\documentclass[12pt,preprint]{aastex}

\shorttitle{Radio galaxies and Emission Line nebulae}

\usepackage{natbib,verbatim}
\bibpunct{(}{)}{;}{a}{}{,}

\begin{document}

\title{WFPC2 LRF Imaging of Emission Line Nebulae in 3CR Radio Galaxies \footnote{Based on observations made with the NASA/ESA Hubble Space Telescope, obtained at the Space Telescope Science Institute, which is operated by the Association of Universities for Research in Astronomy, Inc., under NASA contract NAS 5-26555. These observations are associated with program 5957.}}

\author{G. C. Privon\altaffilmark{1}}

\author{C. P. O'Dea\altaffilmark{2}}

\author{S. A. Baum\altaffilmark{1}}

\author{D. J. Axon\altaffilmark{2}}

\author{P. Kharb\altaffilmark{1}}

\author{C. L. Buchanan\altaffilmark{2}}

\author{W. Sparks\altaffilmark{3}}

\author{M. Chiaberge\altaffilmark{3}}

\altaffiltext{1}{Chester F. Carlson Center for Imaging Science, Rochester Institute of Technology, Rochester, NY 14623}
\altaffiltext{2}{Department of Physics, Rochester Institute of Technology,
    Rochester, NY 14623}
\altaffiltext{3}{Space Telescope Science Institute, Baltimore, MD 21218}

\begin{abstract}
We present HST/WFPC2 Linear Ramp Filter images of high surface brightness emission lines (either 
[OII], [OIII], or H$\alpha$+[NII]) in 80 3CR radio sources. We overlay the emission 
line images on high resolution VLA radio images (eight of which are new reductions 
of archival data) in order to examine the spatial relationship between the optical 
and radio emission. We confirm that the radio and optical emission line structures 
are consistent with weak alignment at low redshift ($z < 0.6$) except in the 
Compact Steep Spectrum (CSS) radio galaxies where both the radio source and the 
emission line nebulae are on galactic scales and strong alignment is seen at all redshifts. 
There are weak trends for the aligned emission line nebulae to be more luminous, 
and for the emission line nebula size to increase with redshift and/or radio power. 
The combination of these results suggests that there is a limited but real capacity 
for the radio source to influence the properties of the emission line nebulae at 
these low redshifts ($z < 0.6$). Our results are consistent with previous suggestions that both 
mechanical and radiant energy are responsible for generating alignment between the 
radio source and emission line gas. 
\end{abstract}


\section{Introduction}
\label{sec:Introduction}

Radio galaxies are an important class of extragalactic objects:  they represent one of the most 
energetic astrophysical phenomena; they may be used as probes of their environments; and they 
are unique probes of the early Universe \citep{McCarthy93}. 
As the 3CR sample of radio galaxies is radio flux-density selected \citep{Bennett62}, 
it provides an unbiased optical sample to study the host galaxies of these radio-loud 
active galactic nuclei (AGN). 
The sample has been well studied at multiple wavelengths and previous ground-based studies 
have investigated the spatial coincidence of optical emission nebulae and the radio 
source \citep{Baum88,Baum89a,McCarthy87,Chambers87}.

\citet{Chambers87} and \citet{McCarthy87} first demonstrated the ``alignment effect'' at high 
redshift ($z \geqslant 0.6$) where the continuum optical and/or IR emission is aligned along the radio axis. An ``alignment effect'' for emission line gas was also shown by \citet{McCarthy87} and \citet{McCarthy95}.  In this paper we focus on the alignment of  the fine scale high surface brightness emission line gas imaged by the Hubble Space Telescope (HST)
with the radio emission. The primary mechanisms for the emission line gas alignment
are thought to be shocks induced by the radio jet and photoionization from the central AGN 
\citep[e.g.,][]{Baum89a,McCarthy93,Best00}.

\citet{McCarthy89} demonstrated the redshift dependence of the alignment effect - 
at $z < 0.1$ there is no alignment, at $0.1 < z < 0.6$ some alignment is seen, 
and for $z > 0.6$ essentially all the powerful radio galaxies show strong alignment.
\citet{Baum89b} found alignment at low $z$ if one compares the emission in the quadrants containing the radio source to quadrants without radio emission. 
Studies which attempt to break the redshift-radio power degeneracy suggest that 
the alignment depends on both redshift and radio power \citep[e.g.,][]{Inskip02b}.
Thus, the alignment depends on both the presence of extended gas in the environment
as well as the ability of the radio source to influence its environment. 

In this paper we examine the relationship of the high surface brightness emission line gas and radio emission over a wide range of redshift ($0.017 < z < 1.406$), focusing on the transition redshift range $z < 0.6$ where the alignment effect starts
to turn on. Using the Wide Field and Planetary Camera 2 (WFPC2) on HST, we obtained 
images with resolutions of 0.$\arcsec$05 -- 0.$\arcsec$1, similar to or better than the 
synthesis imaging resolution obtained with the Very Large Array \footnote{The National 
Radio Astronomy Observatory is a facility of the National Science Foundation operated 
under cooperative agreement by Associated Universities, Inc.} (VLA) at 5 GHz in 
the ``A'' configuration. The higher resolution allows us to probe denser gas in 
the center of these galaxies. In addition, the tunable Linear Ramp Filter (LRF) permits 
narrow band imaging of a selected spectral feature for a wide range of redshifts, 
enabling studies of a specified emission line for a sample such as the 3CR covering 
a large range of redshift.

\citet{Biretta02} conducted an initial study using this data, and published a subset of the data.  Here we expand on their analysis, presenting the complete set of detected images as well as statistical results.

The paper is organized in the following fashion: we discuss the properties of the sample in 
$\S$\ref{sec:Sample}. $\S$\ref{sec:Observations} discusses the observations made and the 
data reduction methods used. In $\S$\ref{sec:Data Analysis} we outline the analysis of the 
data. $\S$\ref{sec:Properties} contains a discussion of our sources and the results of our study. 
In $\S$\ref{sec:Conclusions} we conclude by discussing the implications of our results. 
Appendix \ref{Appendix} contains notes for individual sources.

\section{The Sample}
\label{sec:Sample}

\begin{figure}[h!]
\centering{
\includegraphics[width=8cm]{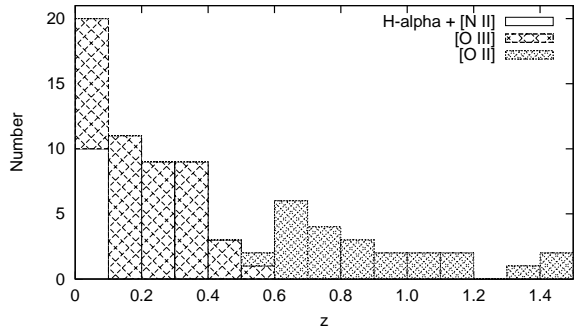}}
\caption{Histogram of the redshift distribution of the sources in our sample.}
\label{redshift_histogram}
\end{figure}

Our sample is drawn from a set of 100 3CR radio galaxies selected from the HST 
3CR imaging survey. Using earlier spectroscopic observations, objects with significant 
emission line flux were selected \citep{Biretta02} for this study to maximize detection. Thus, our results
may not be extendible to objects with very faint emission lines and our observations do not reach uniform surface brightness in the rest-frame of the object. 
Observations were done in 
HST's ``snapshot'' mode, in which targets are randomly selected from the target list 
for observation. 80 sources were observed and constitute our sample, spanning a 
redshift of 0.017 $\leqslant z \leqslant$ 1.406 (See Figure \ref{redshift_histogram}). 
The sample contains 9 FRIs, 58 FRIIs, 12 CSS sources, and 1 source of an unknown morphology. 
22 of the radio galaxies have broad emission lines while 58 have narrow emission lines
in their optical spectrum.

\section{Observations and Data Reduction}
\label{sec:Observations}

\subsection{Radio Observations}

Arcsecond-scale resolution radio maps for the sample objects were obtained 
from a variety of sources (referenced in figure captions). Sources for which 
5 or 8-GHz VLA `A' array maps were not already available were reduced using 
archival data. The cleaning and self-calibration of the raw data was performed 
using the NRAO's AIPS software package. We present new maps for the following 
sources: 3C124, 3C135, 3C284, 3C303.1, 3C341, 3C368, 3C379.1, and 3C382. 
Table \ref{Table:VLA_obs_params} lists the observation parameters for the new 
VLA maps, including the date of observation, frequency and RMS noise of the final map.

\subsection{Optical Observations}

Snapshot images were taken using the LRFs on HST's WFPC2 instrument. There are 3 WF chips and 1 PC chip. The chip on which the LRF images a target varies with the selected wavelength. We have noted the chip upon which a source was imaged in Table \ref{Table:Optical_obs}. Nearby objects ($z \lesssim 0.03$) were imaged in H$\alpha$+[NII]. Objects at intermediate redshifts were imaged in [OIII] 5007 $\AA$ and objects at high redshift ($z > 0.5$) were imaged in [OII] 3727 $\AA$. Table \ref{Table:Optical_obs} also lists which emission line was imaged for each source. All of the LRF observations consisted of a pair of 
300 sec exposures taken in ``snapshot'' mode. The images were calibrated and combined using the CRREJ procedure to remove cosmic rays \citep{Biretta96}. Because the design of the LRF does not allow direct observations of the flat field, the images were flat fielded using the F656N flat.
Note that due to the small size of the WFPC2 pixels, we are sensitive only to high surface brightness 
extended emission line gas.
 
In order to obtain images of only the emission line region (ELR), the continuum contribution to the LRF images was subtracted. This was usually done using an image of the object in the broadband F702W filter. The continuum was scaled to the observed wavelength by assuming a 3000K blackbody spectrum for the continuum \citep{Biretta02}. In some cases the continuum was contaminated by emission lines. \citet{Baum88} discussed the uncertainties in flux calibration due to contamination from emission lines. Using standard line ratios, they found corrections from 1.002 to 1.36 (median 1.09). We have not applied corrections for emission line contamination to our data, but the values given by them indicate the uncertainty due to contamination. Continuum images were not available for four sources, 3C98, 3C273, 3C289, and 3C323.1, and the subtraction could not be performed.  These sources have been left out of the analysis, but their narrow-band LRF images are presented if the source was detected.

\section{Optical Data Analysis}
\label{sec:Data Analysis}

We grouped the emission-line nebulae into 4 broad optical morphological categories: `extended' sources (21 sources), `partially extended' sources (33 sources), slightly or unresolved sources (12 sources), and non-detections (11 sources). Four sources were eliminated due to the lack of a continuum image for subtraction. Sources were classified as extended if the 3-$\sigma$ contour was extended compared with the expected PSF in the continuum subtracted image. `Partially extended' sources are those which feature extensions smaller than 2$\arcsec$ but do not match the PSF. They often resemble an unresolved source, with a small faint extension. We analyzed only the `extended' and `partially extended' sources, as we were able to determine reliable position angles and/or angular sizes for most of these objects. Table \ref{Table:Optical_data} lists the morphological categories for each object.

To examine the spatial coincidence of the ELR and the radio source, the radio and optical images for the `extended' sources were combined into overlays. In general, the images were registered using the radio core (for the radio sources), and the centroid of the host galaxy (obtained from the continuum images). One `extended' source (3C305) has a known dust lane, making centroiding on the continuum image unreliable. We attempted to determine a center through ellipse fitting to the outer portion of the galaxy, but this was unsuccessful due to the close companion galaxy. The registration in the image presented was done using centroiding on the continuum image. The overlays are shown in Figures \ref{3C46-montage} - \ref{3C381-overlay} along with the continuum and emission-line images for each source.

Position angles (PAs) for the `extended' sources were measured using the emission within the 3-$\sigma$ contour, with the PA passing through the nucleus of the host galaxy in the direction of the longest extent of the emission line nebulae. PAs for the partially extended sources were determined using IRAF's `ellipse' task. Radio PAs were taken 
from the literature (if available), or measured from hotspot-to-hotspot for FRIIs or across the longest
extent for FRIs on the radio maps. The measurement of the difference in the PAs is independent of
the spatial alignment of the radio and optical images as the position angles are independently determined from the images. 

A measure of alignment between the ELR and the radio source was determined by taking the absolute 
value of the difference in position angles of the radio source and the ELR. 
If $\Delta$PA $\leqslant$ 
30 degrees, the ELR and radio source were considered aligned. In some cases the ELR displayed 
a position angle for the inner regions which differed from the large scale position angle. 
In most cases, the usage of an ``inner PA'' did not cause misaligned objects to be moved 
into the aligned category, so only the large-scale PA was used for the purposes of our analysis.

Emission line fluxes were measured using IRAF's `apphot' set of tasks. Luminosities and sizes 
were calculated using a cosmology of H$_{0}$=71 km s$^{-1}$, $\Omega$$_{M}$=0.27, and 
$\Omega$$_{vac}$=0.73. Values were obtained using an online cosmology calculator \citep{Wright06}. In order to compare the luminosities from each source, the measured 
fluxes were converted to H$\alpha$ fluxes using line ratios derived from observations of 
narrow-line radio galaxies \citep{Koski78}. The ratios of [OIII], [OII], and 
H$\alpha$+[NII] to H$\alpha$ were 3, 1.23, and 2.08 respectively.

\section{Properties of the Nebulae}
\label{sec:Properties}

Here we discuss the properties of the nebulae and their relationship to the radio source. 
Individual source descriptions are provided in Appendix \ref{Appendix}.  In Table \ref{Table:Radio_data} we provide morphological information about the radio structure, including classifications, position angles, and angular sizes.  We also present luminosities for each source.  Table \ref{Table:Optical_data} contains the morphological information for the continuum-subtracted emission line images.  This includes position angles, angular sizes, and fluxes.  In addition, this table contains the V-band magnitudes for the host galaxies and the position angle of the galaxy.  

The detection statistics varied depending on the emission line observed.  Nine of ten sources (90\%) imaged in H$\alpha$+[NII] were detected. Forty-three of forty-seven sources (91\%) imaged in [OIII] were detected, while fifteen of twenty-three (65\%) [OII] detections were made.

The statistical significance of the observed relationships discussed in this section are given in Table 
\ref{Table:Correlations}. The Kendall tau value is a measure of the correlation between 
the specified measurements. The Spearman's rho value is a similar measure, describing 
how likely the two measurements are to increase or decrease concordantly. Statistical 
values for non-correlations are given in Table \ref{Table:Non-Correlations}.

\subsection{Alignment Effect}
\label{ssec:alignment}

\begin{figure}[h!]
\centering{
\includegraphics[width=8cm]{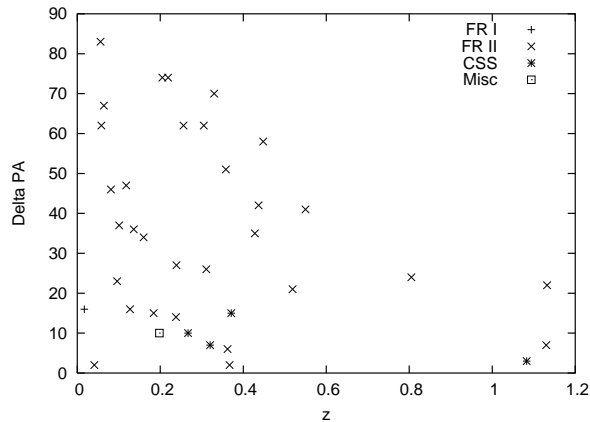}}
\caption{Relative alignment of the radio source and ELR as a function of redshift.
No correlation was present with the Kendall tau test giving a non-correlation probability
of 0.0913 and Spearman's rho test giving a probability of 0.1540 of a correlation
not being present.}
\label{delta_pa-redshift}
\end{figure}

We find that 19 of 37 sources for which a PA could be measured showed 
alignment between the high surface brightness ELR and the radio source to within 30 degrees (Figure \ref{delta_pa-redshift}). 
We have 4 sources at $z > 0.6$ which all show strong alignment. In the remaining 33 
sources at $z < 0.6$,  15 show alignment. The median values of the position angle difference
in bins of redshift do not decrease significantly as a function of $z$ out to 0.6. 
However, the upper envelope of the position angle differences  does seem to decrease 
with redshift. 
We clearly see a weaker alignment effect at low redshift than is seen at the high redshifts 
$z > 0.6$. 
Our high resolution HST data show that the lack of strong alignment 
at $z < 0.6$ is not due to a lack of spatial resolution. 

We investigated the possibility of the misaligned objects being predominantly broad 
line radio galaxies (BLRGs). According to the unified scheme \citep{Urry95}, BLRGs are 
not expected to show alignment as the jet would be pointing almost along the line of sight. 
However, BLRGs make up only 11 of the sources with measured position angle differences,
and several of the BLRGS show alignment (3C234, 3C268.3, 3C277.1)\footnote{3C268.3 and 
3C277.1 are also CSS sources.}.
Thus, orientation of the radio jet close to the line of sight is not responsible for the 
misalignment at $z < 0.6$.

A potential concern is that much of the low redshift observations ($0.03 < z < 0.5$) are
obtained using the [OIII] line which is higher  ionization than the [OII] and
H$\alpha$+[NII]. If the high ionization gas has very different morphology and 
position angle than the lower excitation gas, 
this might reduce the strength of the alignment at low redshift. However, observations
of radio galaxies in which both [OII] and [OIII] are obtained show that the 
estimated large scale position angles are consistent \citep[e.g.][]{Tilak05}.

Another possible bias is the extent of the ELRs for different emission lines. Studies of single objects in multiple emission lines indicate typical ratios in the extended emission of $\sim$1 - 6 for [O III]/[O II] and $\sim$1 - 4 for [O III]/H$\alpha$ \citep[e.g.][]{Tilak05,Solorzano03,Robinson00}. Thus, as the transition between lines is made, the ELR should be smaller for [O II] than for [O III]. As will be discussed in $\S$ \ref{ssec:elr-size_l}, the opposite is found.

Over the redshift range of our sample (0.017 $\leqslant z \leqslant$ 1.406), the expected
(1+z)$^4$ dimming of surface brightness is a factor of $\sim$30 - though the details will
depend on the actual surface brightness profiles. Thus, both the transition from [OIII] to
[OII]  and the (1+z)$^4$  effect should result in apparently smaller ELR at higher redshift. 
Despite this, we see emission line size increasing with redshift (see $\S$ \ref{ssec:elr-size_l}), 
suggesting that this result is real. 

In order to investigate the possible effect of an instrumental sensitivity selection
effect on the detection of alignment, we redshifted the faintest level in the emission
line images to see what would be detected if located at $z=0.5$ (just prior to
the onset of strong alignment). None of the misaligned sources showed any alignment
when viewed using these modified levels. This indicates that the alignment effect is
not due to a selection effect of the sensitivity of the instrument, but rather to
some other inherent attribute of the source and its environment. 

\subsubsection{Comparison with Ground Based Data}
\label{ssec:groundbased}

\begin{figure}[h!]
\includegraphics[width=0.5\textwidth]{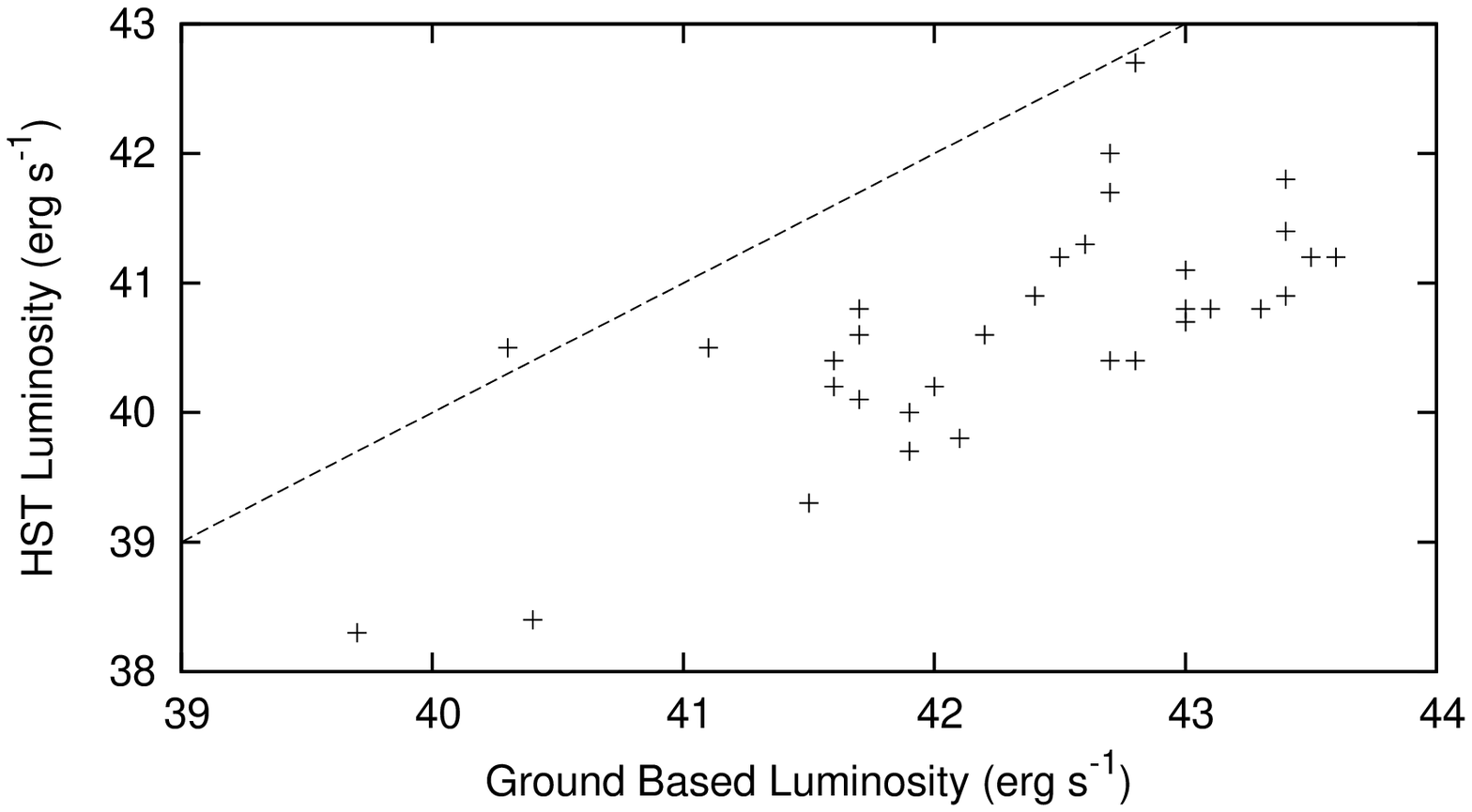}
\includegraphics[width=0.5\textwidth]{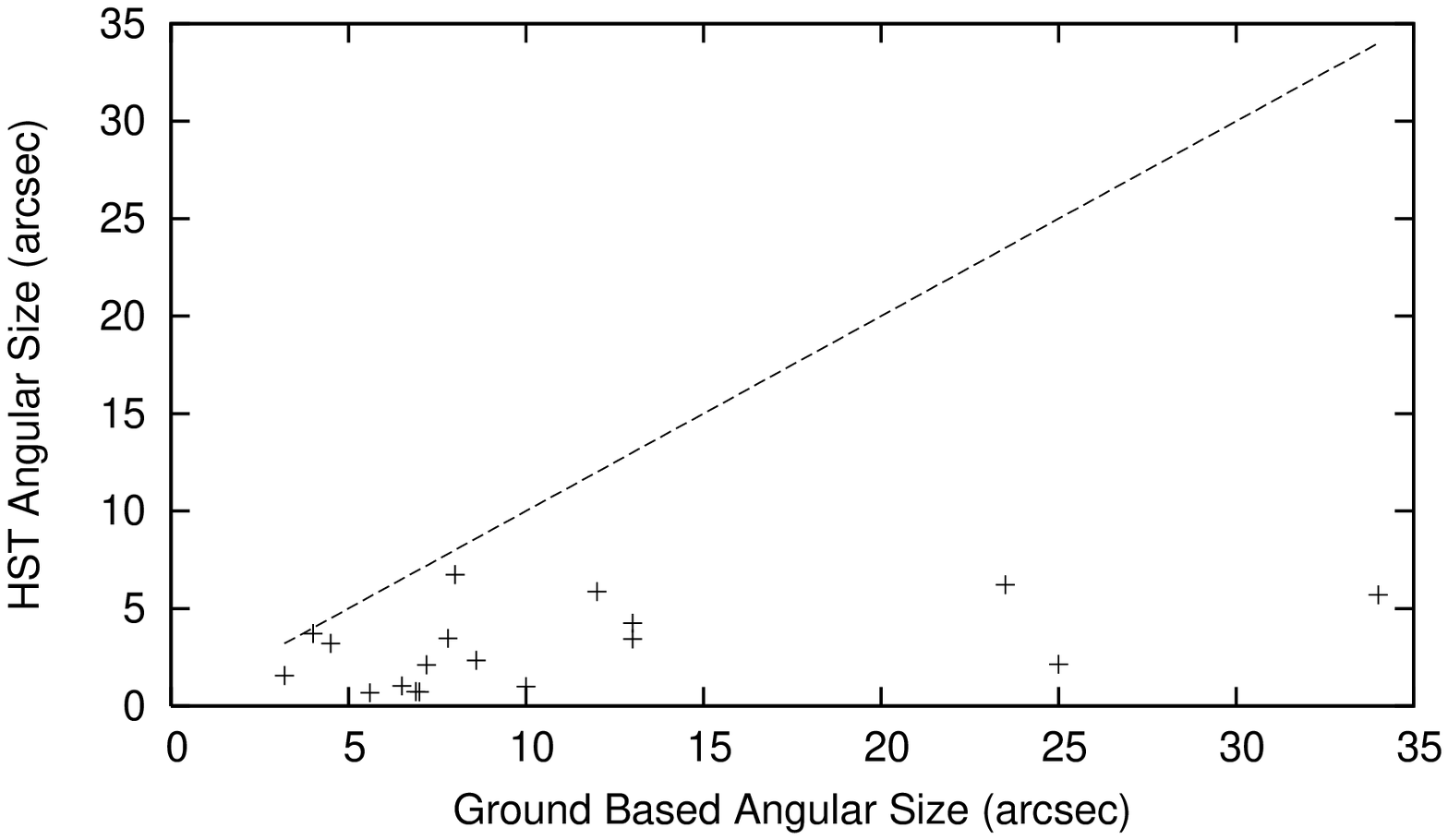}
\includegraphics[width=0.5\textwidth]{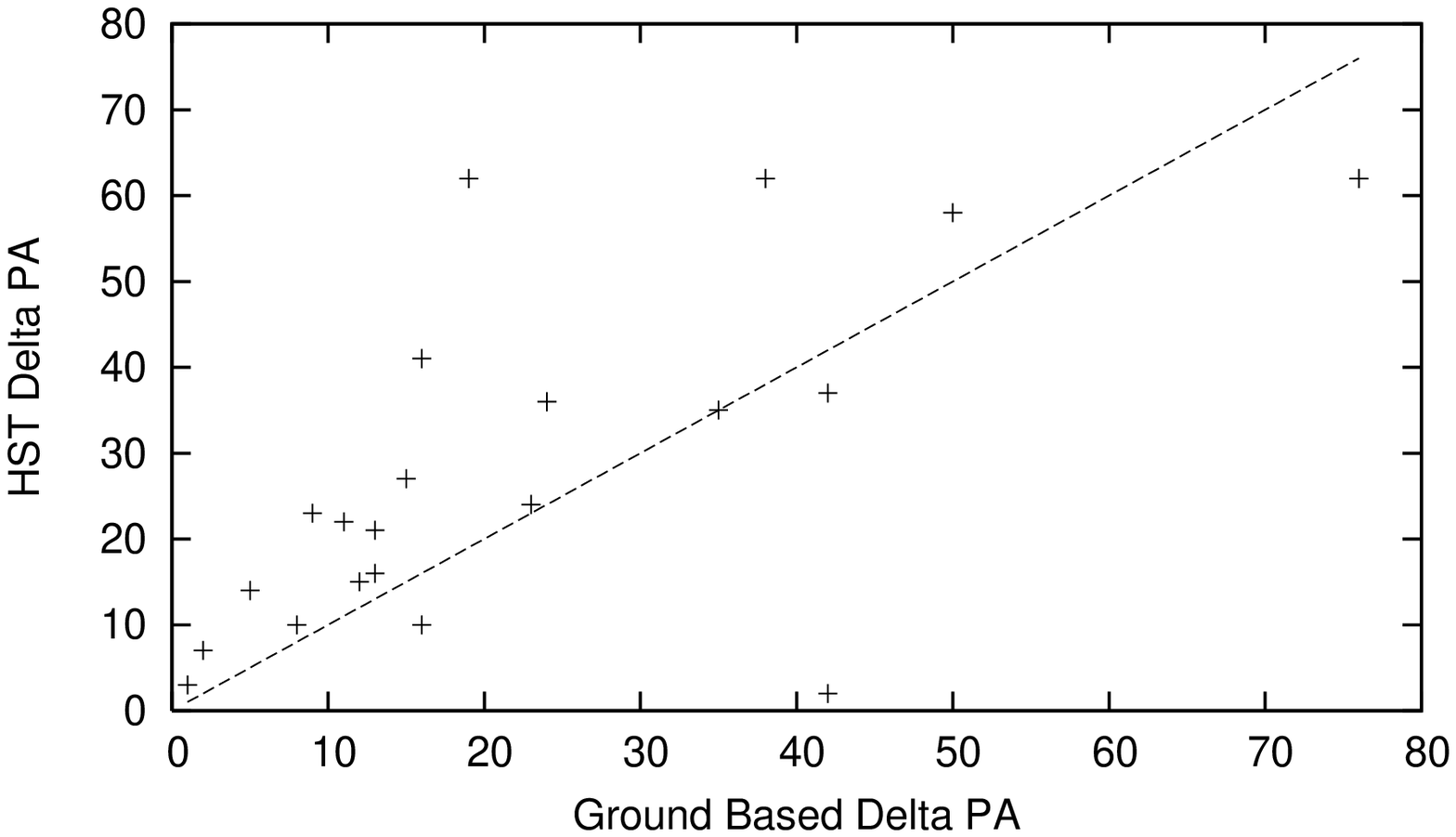}
\caption{Top Left: Comparison of the HST luminosities from our data with ground based data from 
previous studies \citep{Baum89a,McCarthy95}. The HST luminosities are much lower due to 
the lower surface brightness sensitivity of WFPC2. Top Right: Comparison of the angular sizes 
in our HST data with ground based data. The extended emission is of low surface brightness 
and we do not detect the full extent of the ELR as seen from the ground. Bottom Left: 
A comparison of the Delta PA derived from the HST data and ground based data. While many 
of the values are quite similar, some are significantly different. The line in all plots is y=x.}
\label{fig:hst-ground}
\end{figure}

The emission line images from HST were compared with previous studies using ground based 
imaging (Figure \ref{fig:hst-ground}). Morphologies and position angles were generally consistent between 
the two sets consistent with the hypothesis that the nebulae position angles do not
exhibit large changes in position angle as a function of size scale. 
However, because of the  lower surface brightness sensitivity of the HST/WFPC2 camera,
the HST images show only the brighter inner regions of the nebulae. Thus, the HST
measured sizes and emission line fluxes tend to be smaller than those measured
on ground based images. 

\begin{figure}[h!]
\includegraphics[width=.5\textwidth]{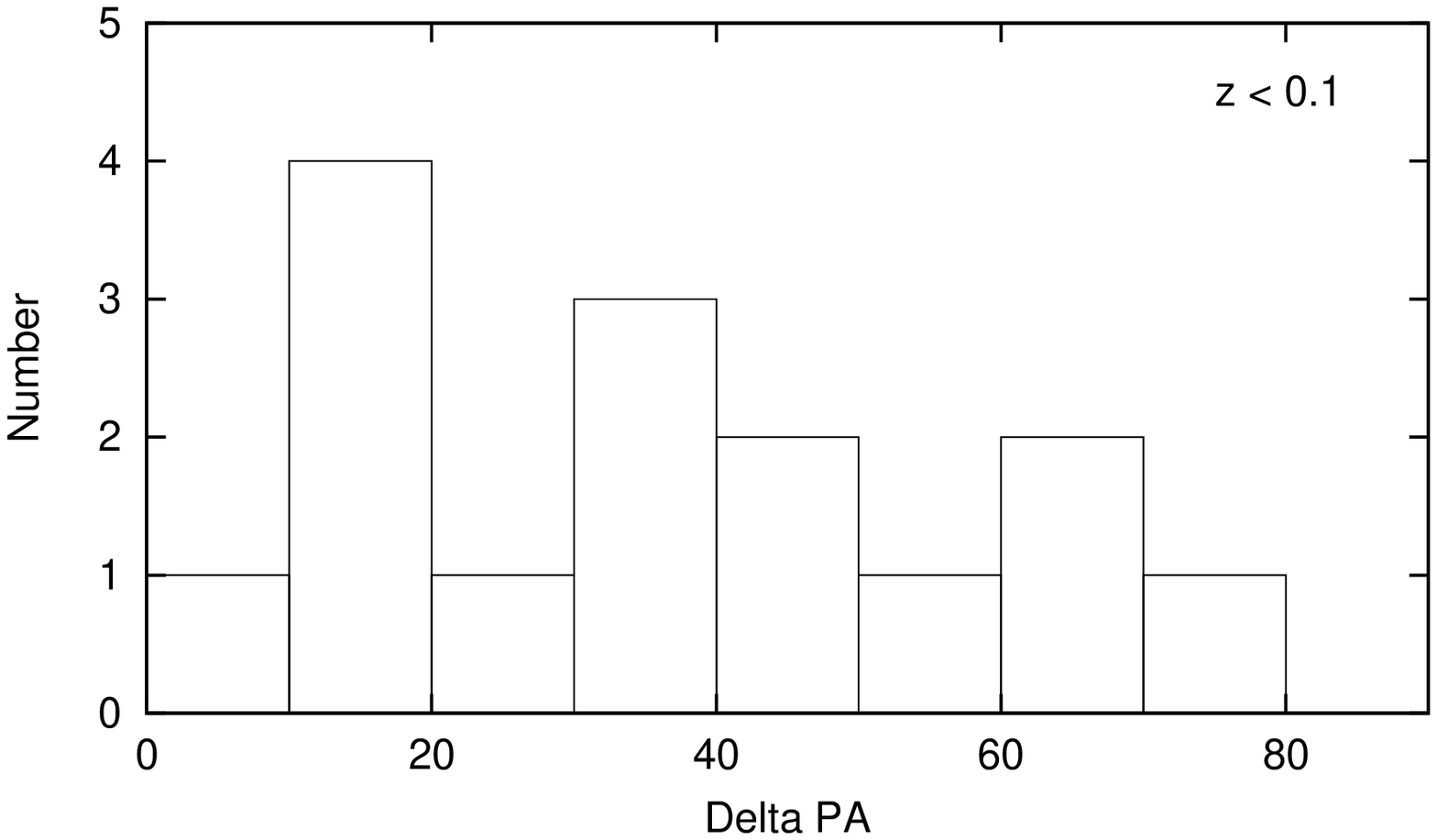}
\includegraphics[width=.5\textwidth]{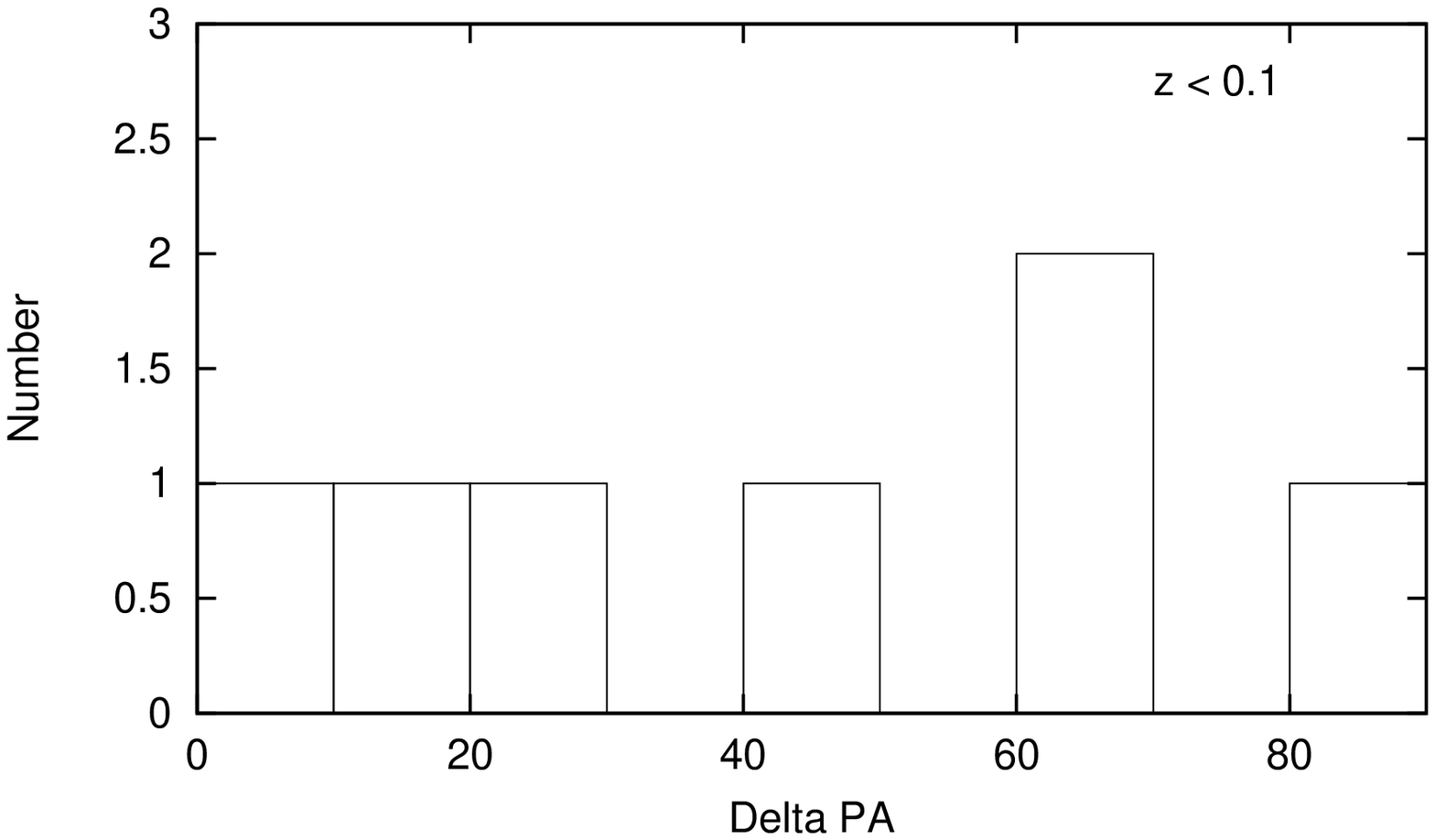}\\
\includegraphics[width=.5\textwidth]{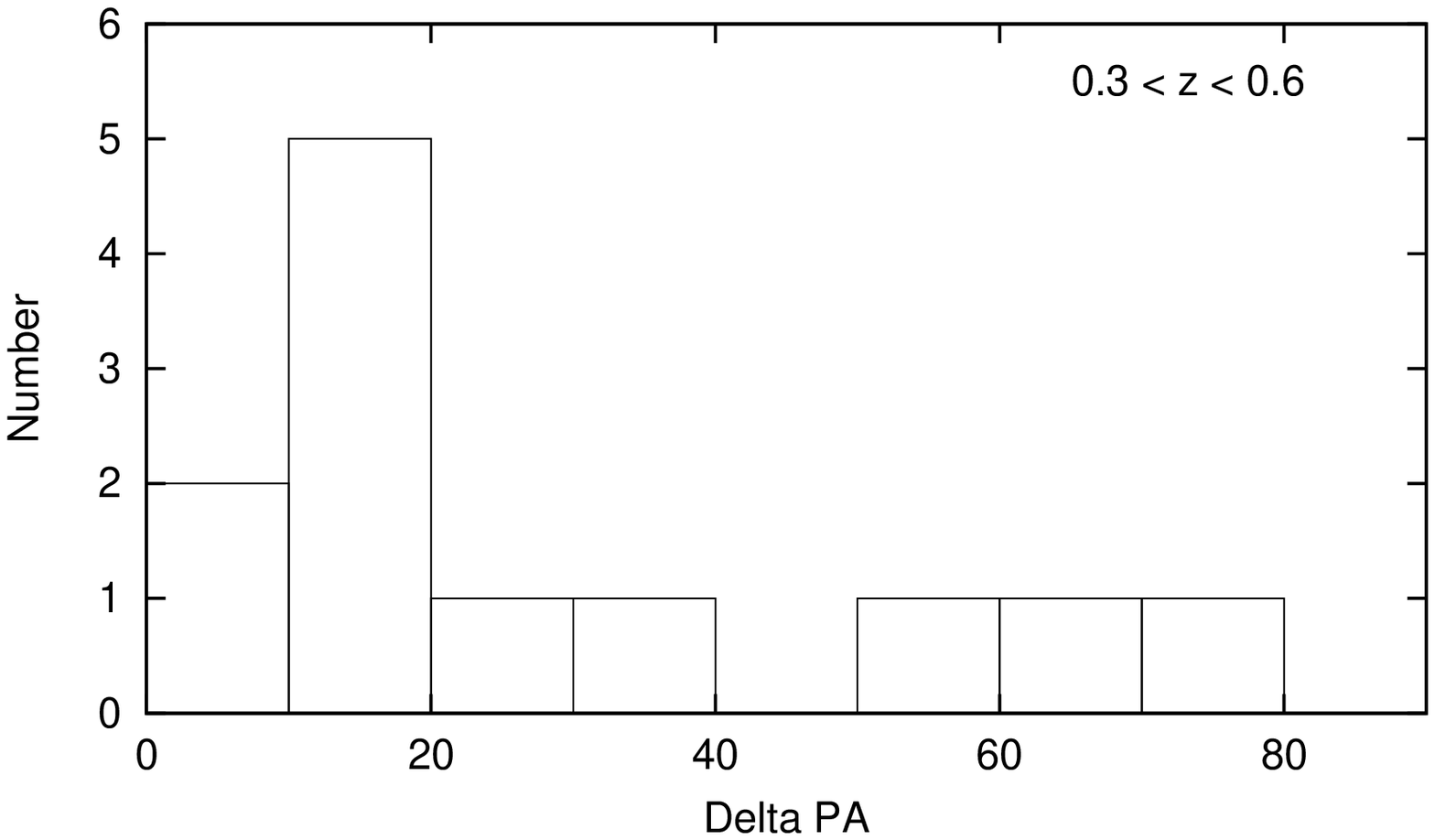}
\includegraphics[width=.5\textwidth]{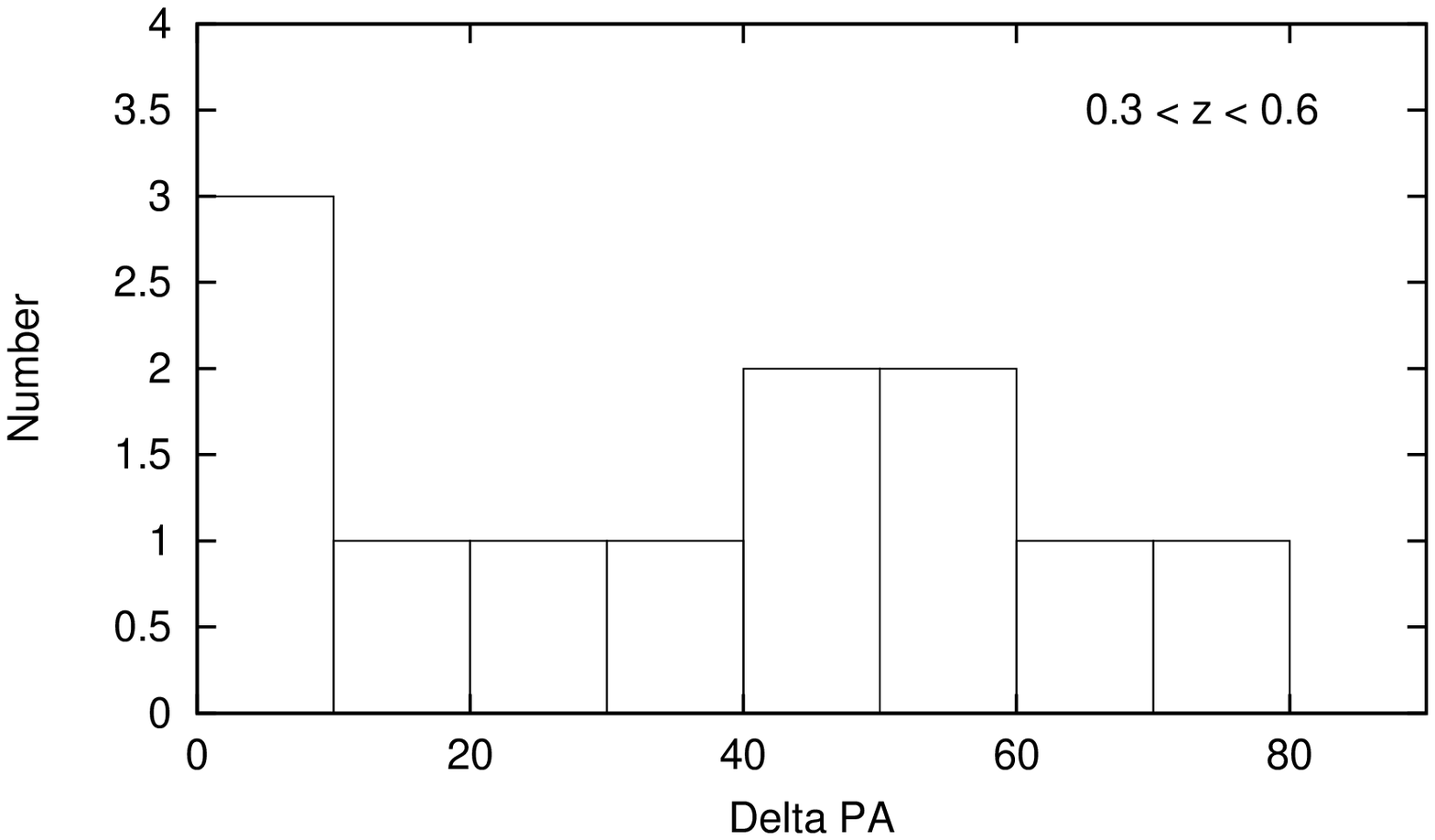}\\
\includegraphics[width=.5\textwidth]{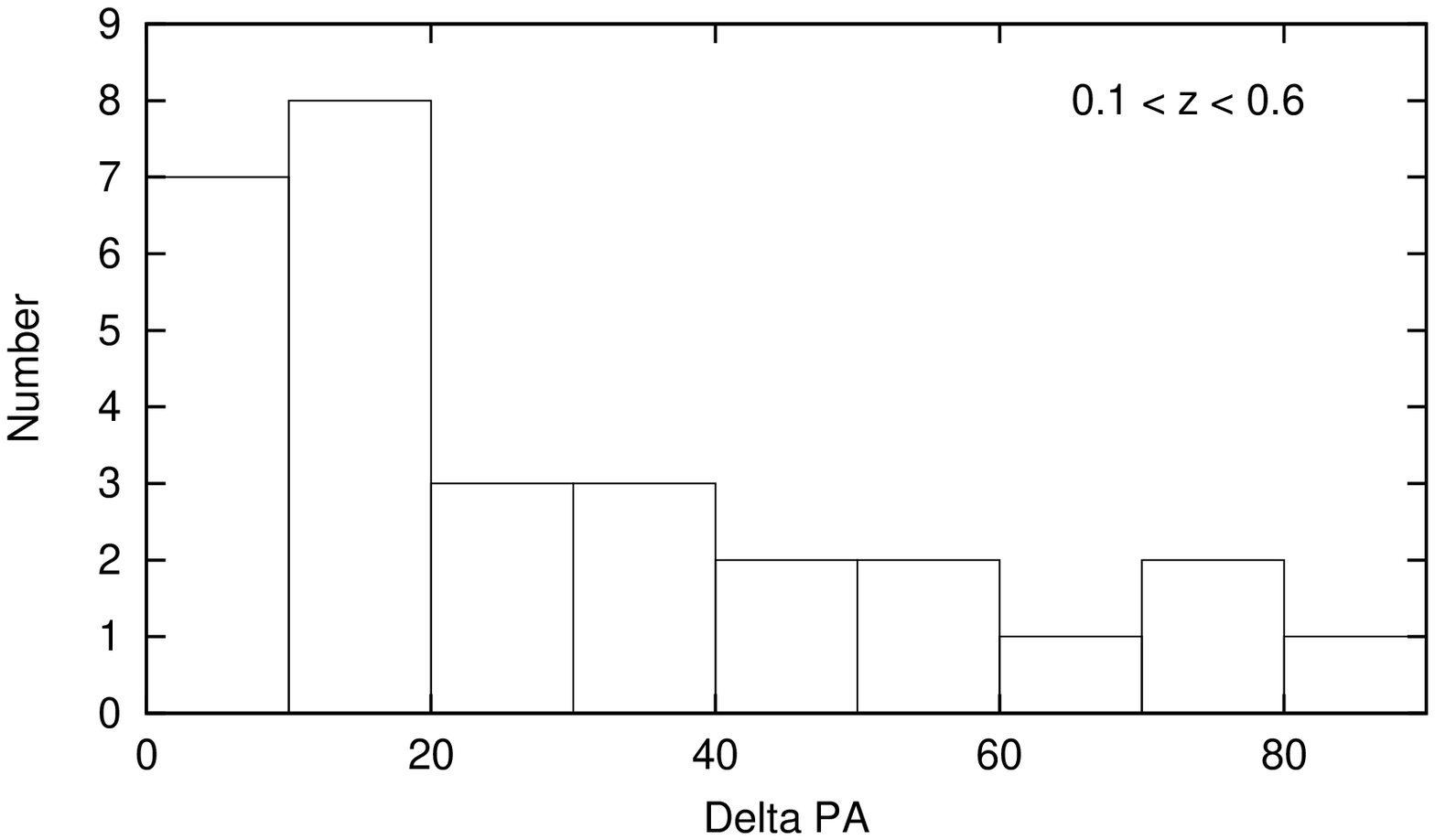}
\includegraphics[width=.5\textwidth]{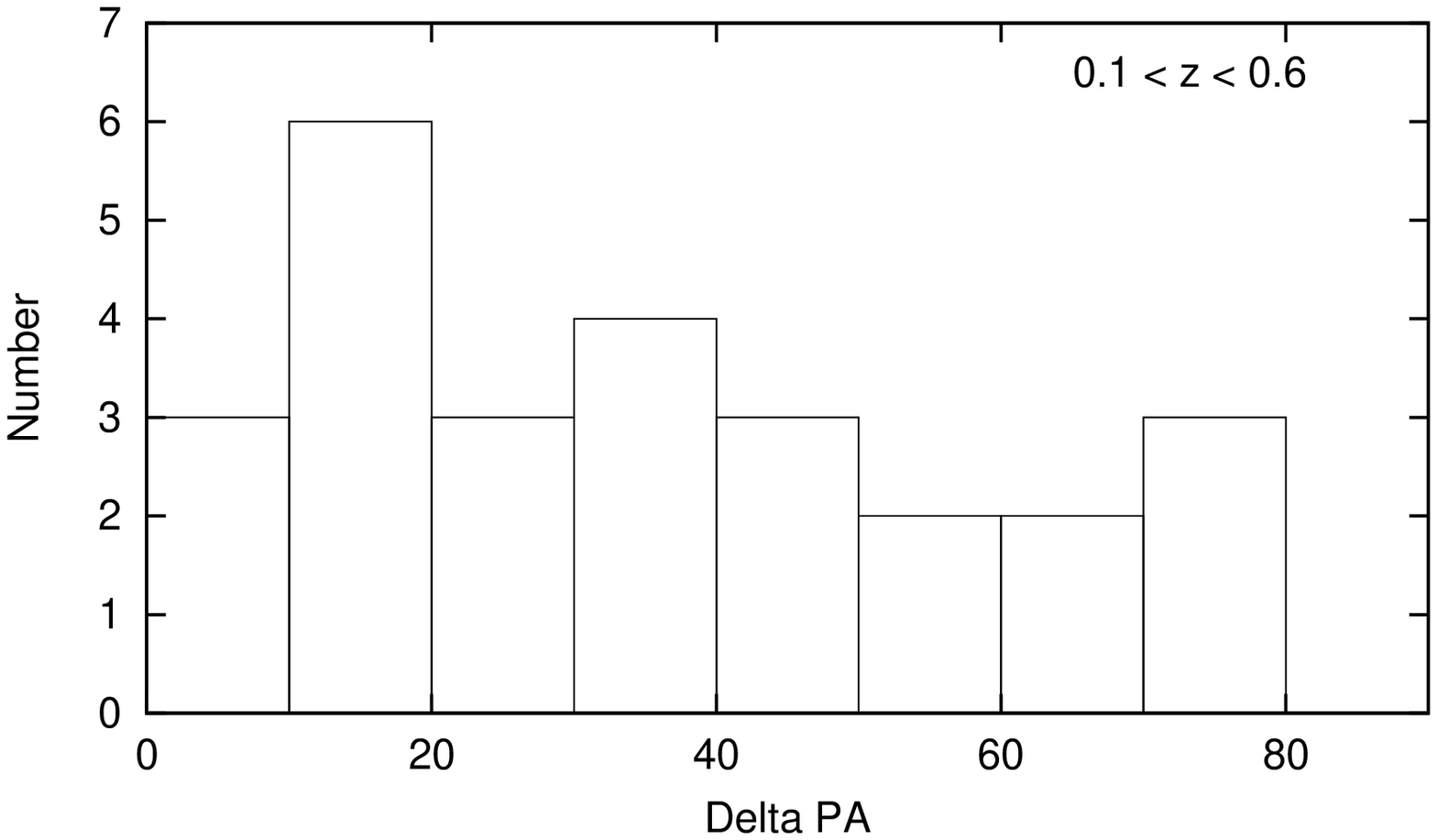}\\
\caption{These figures show histograms of relative alignment between the emission line 
region and the radio source for various redshift ranges. The left column plots ground-based data compiled 
in \citet{McCarthy95} while the right column shows our HST/WFPC2 data.  
The top two plots are for $z < 0.1$. The middle row is for $0.3 < z < 0.6$ 
and the bottom row is for $0.1 < z < 0.6$.}
\label{all_delta_pas}
\end{figure}

Figure \ref{all_delta_pas} shows a comparison of the distribution of position angle differences from
ground-based data \citep{Baum88,Rigler92,McCarthy95} with our HST data. 
There is a suggestion that the HST data shows an alignment effect which 
is weaker than seen in the ground-based data at the same redshifts. The 
Kolmogorov-Smirnov (KS) test 
indicates that the distribution of position angle differences in the HST data is
consistent with a uniform distribution of angles, while the ground based data are
not (see Table \ref{Table:KS-results}). 
If larger samples confirm this result, this may be due to the HST data mainly being 
sensitive to the higher surface brightness, more compact emission, suggesting that the 
alignment effect is mainly produced in fainter, more extended emission line gas. 

\subsection{Relative Sizes of the Emission Line Region and the Radio Source}

\begin{figure}[h!]
\includegraphics[width=8cm]{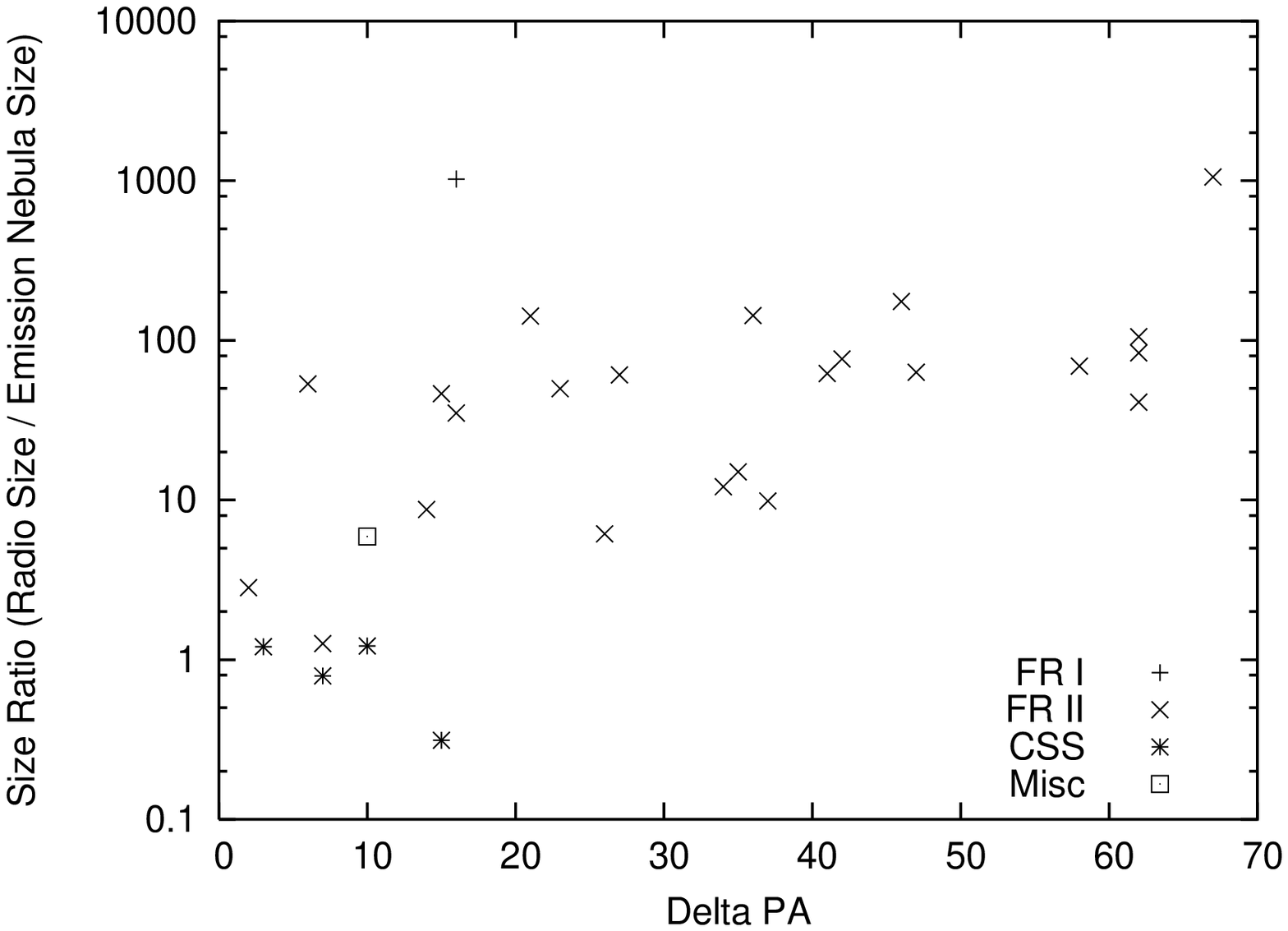}
\hfill
\includegraphics[width=8cm]{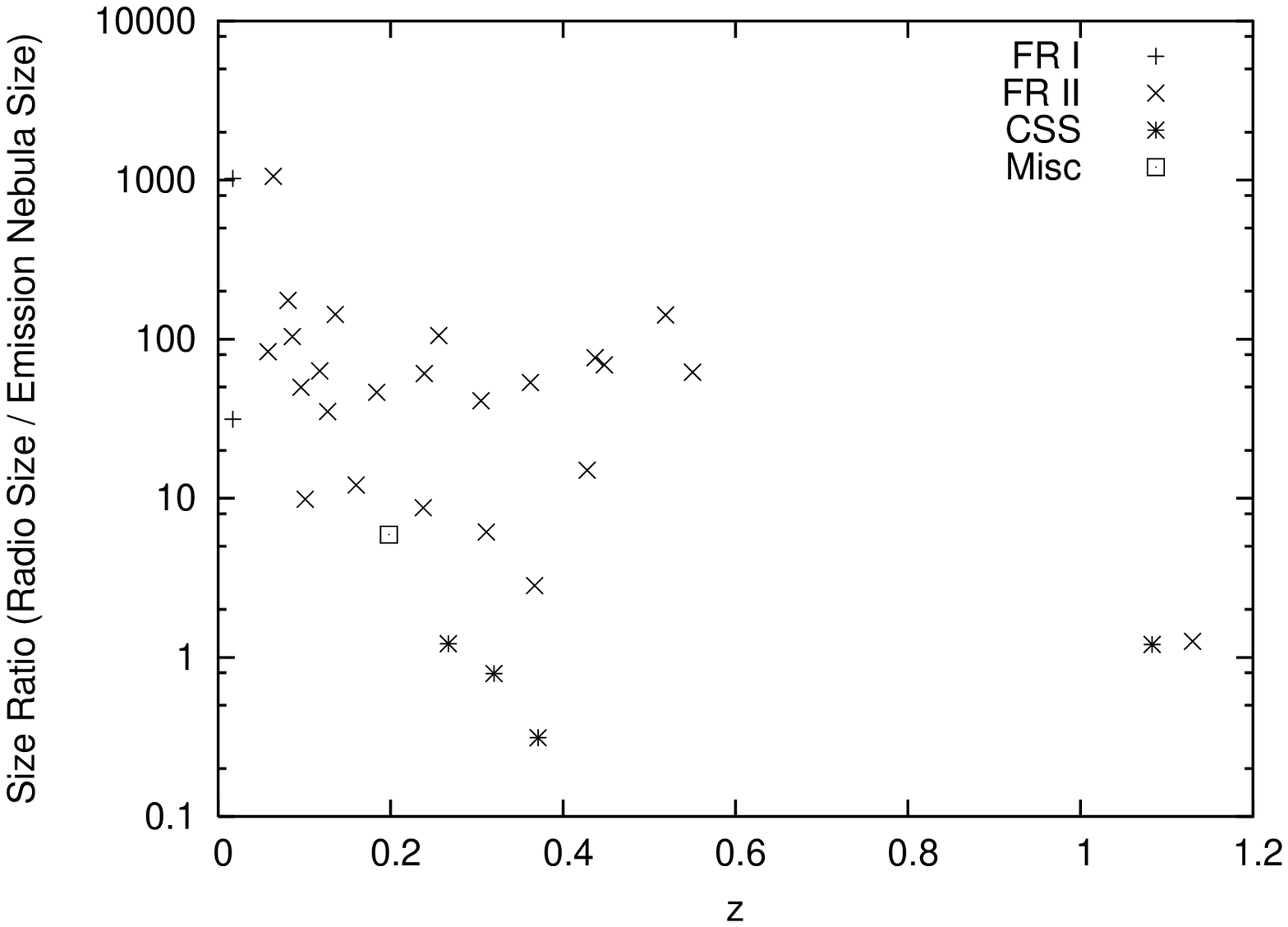}
\caption{Left: Plot of the calculated size ratio between the radio source and the emission line region as a function of alignment between the two. The sources with size ratios of roughly one have optical morphologies which are closely aligned with the radio source. Both Kendall tau and Spearman's rho tests indicate the presence of a correlation with non-correlation probabilities of 0.0001 and 0.0003 respectively. Right: Size ratio as a function of $z$. The sources with size ratios near unity are found at both high and low redshifts.}
\label{size_ratio-delta_pa}
\end{figure}

\begin{figure}[h!]
\centering{
\includegraphics[width=8cm]{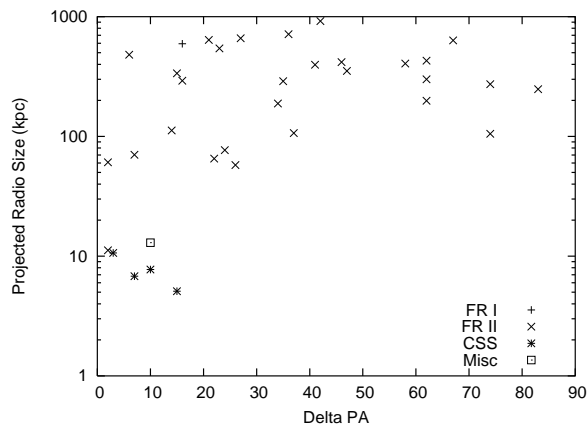}}
\caption{Plot of the projected radio size as a function of the relative alignment between the radio source and the emission line nebulae. All of the small sources are aligned while only some of the large sources are aligned.} 
\label{delta_pa-radio_size}
\end{figure}

We computed the `size ratio' comparing the relative projected linear sizes of the 
radio source and the high surface brightness emission line region. The linear sizes for FR II and CSS 
sources were measured from hotspot to hotspot, and the linear sizes for FR I 
sources were measured across the overall extent. Linear sizes for the emission 
line regions were measured roughly along the long axis (if such an axis existed). 
Comparing the relative sizes of the radio source and ELR with the relative alignment 
indicates that sources with similar radio and ELR sizes are more frequently aligned than 
sources in which the ELR is much smaller than the radio source (Figure 
\ref{size_ratio-delta_pa}). This is true at all redshifts (Figure \ref{size_ratio-delta_pa}).
We also note that all six of the small sources (D $<$ 15 kpc, i.e., CSS) are well aligned (Figure 
\ref{delta_pa-radio_size}). All six of these have size ratios between $\sim$0.3 and $\sim$ 6. 
These results suggest that the alignment at low redshift results when the source is 
smaller/younger and of a similar size as the nebula.  
This is consistent with suggestions that the alignment is due to 
mechanical energy input from the radio source \citep[e.g.,][]{deVries99,Axon00,Best00,Inskip02a,Inskip02b}.   
At low redshifts, suitable gas clouds are likely to exist only on the scale of the host 
galaxy. 

In addition, 13 of the 29 large sources (size greater than 50 kpc) also show alignment 
to within 30$^{\circ}$. These sources are much bigger than their emission line nebulae
(factors of 10-100).  Some of this alignment may be produced by chance if the position 
angles differences in at least some low $z$ sources are randomly distributed.  However,
some of the alignment may be real since we suspect the effect is starting
to turn on at these redshifts. 
Diagnostic line ratios and kinematics indicate that shocks are important when
the radio source size is comparable to the aligned emission line nebula size and that
photoionization dominates when the radio source is much larger than the
nebula \citep[e.g.,][]{Best00, Inskip02a, Moy02, ODea02, Labiano05}. 
Thus, in these large sources, the alignment may be produced by
photoionization from the central AGN along the ``ionization cone".  
Recombination times for warm gas
(T $\sim$10$^{4}$ K) are roughly 10$^{3}$ years \citep{Osterbrock89} while the lifetimes 
of the radio sources are roughly 10$^{7}$ -- 10$^{8}$ years, indicating the ionizing 
mechanism must be on-going throughout the lifetime of the source.

\subsection{Emission Line Nebula Size and Luminosity}
\label{ssec:elr-size_l}

\begin{figure}[h!]
\centering{
\includegraphics[width=8cm]{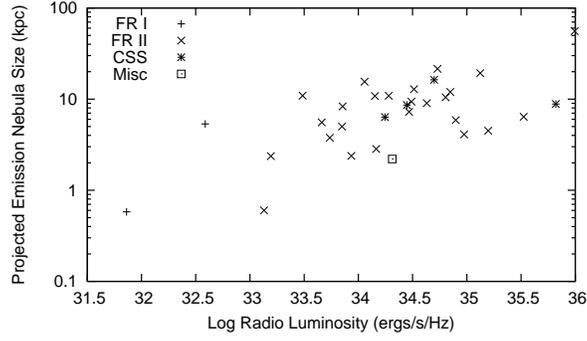}}
\caption{Plot of projected emission line nebulae size against logarithm of the radio luminosity. 
The probabilities of this being a non-correlation are 0.0039 (Kendall tau) and 0.0067 
(Spearman's rho) for the whole sample and are 0.0119 (Kendall tau) and 0.0172 (Spearman's rho) 
for $z < 0.6$.}
\label{emission_size-radio_luminosity}
\end{figure}

\begin{figure}[h!]
\centering{
\includegraphics[width=8cm]{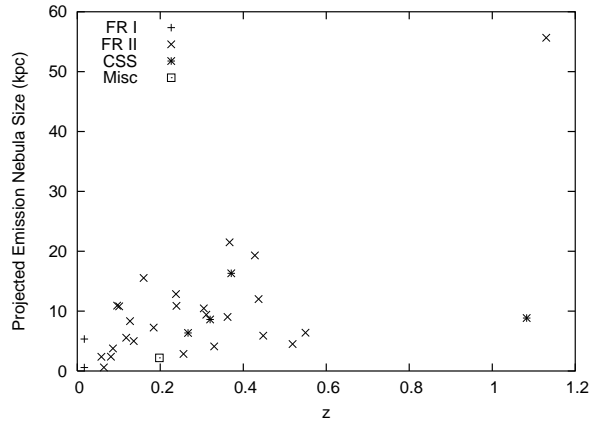}}
\caption{Plot of the projected emission line nebulae size as a function of redshift. The 
probabilities of this being a non-correlation are 0.0050 (Kendall tau) and 0.0055 
(Spearman's rho) for the whole sample and are 0.0152 (Kendall tau) and 0.0136 (Spearman's rho) 
for $z < 0.6$.}
\label{emission_size-redshift}
\end{figure}

The size of the high surface brightness ELR increases with increasing radio power (Figure 
\ref{emission_size-radio_luminosity}) and/or with increasing redshift 
(Figure \ref{emission_size-redshift}).
Both correlations are somewhat weaker for just the $z < 0.6$ objects.
At higher redshifts, dense gas clouds are found on increasingly larger scales 
culminating in the large Ly-$\alpha$ halos surrounding powerful radio galaxies 
\citep{Ojik97, Reuland03}. 
Of course, the radio luminosity will also increase with redshift in flux density limited 
samples, meaning the correlations may be related.
Studies which attempt to break the redshift-radio power degeneracy suggest that 
the alignment depends on both redshift and radio power \citep[e.g.,][]{Inskip02b}.
The more powerful radio sources can provide both  more mechanical energy and more ionizing
photons \citep{Baum89a, Rawlings91}. Thus both the presence of gas on larger scales as well
as the larger energy input from the radio source could result in the emission line
size increasing with redshift and radio power. 

\begin{figure}[h!]
\includegraphics[angle=270,width=8cm]{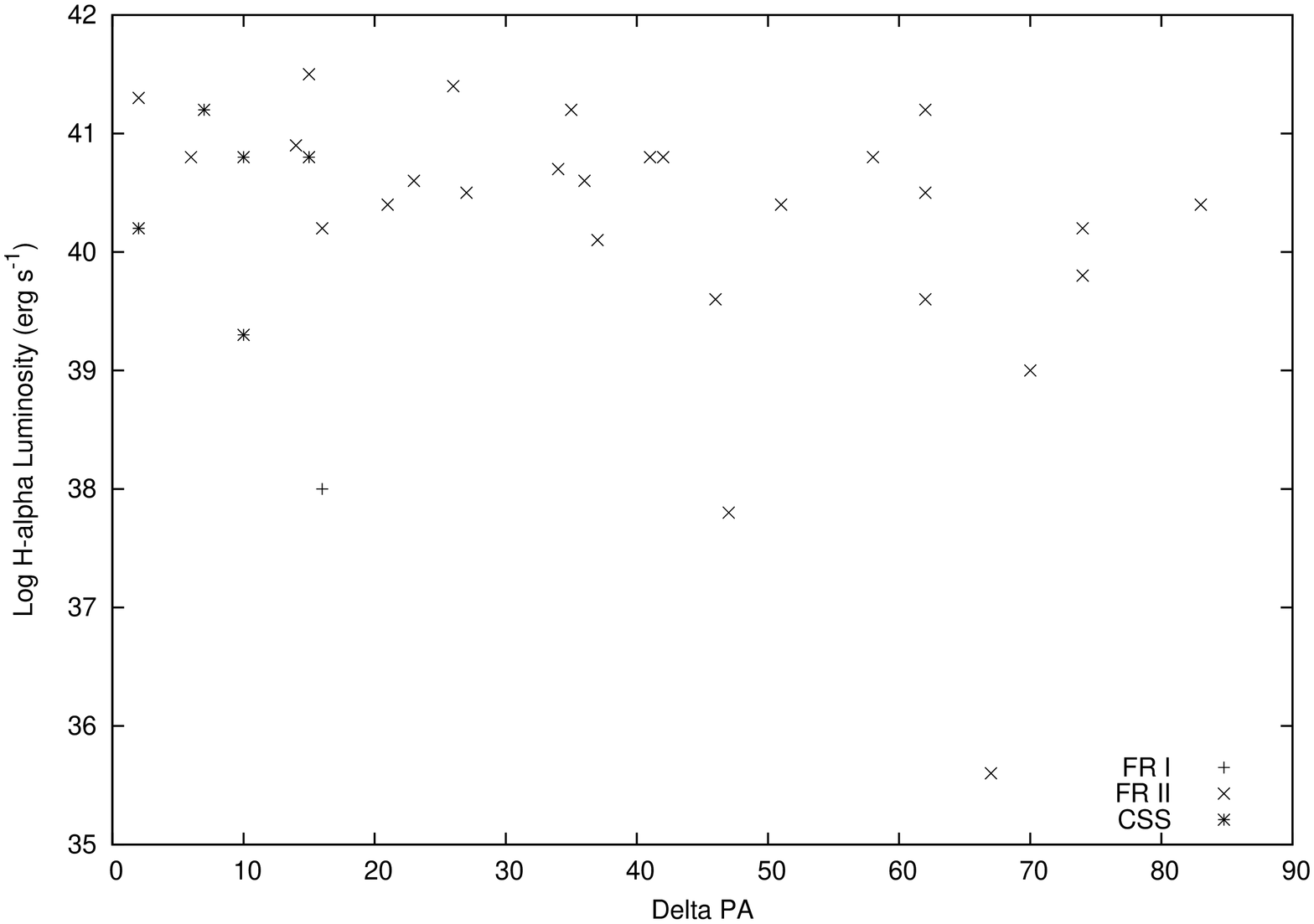}
\hfill
\includegraphics[angle=270,width=8cm]{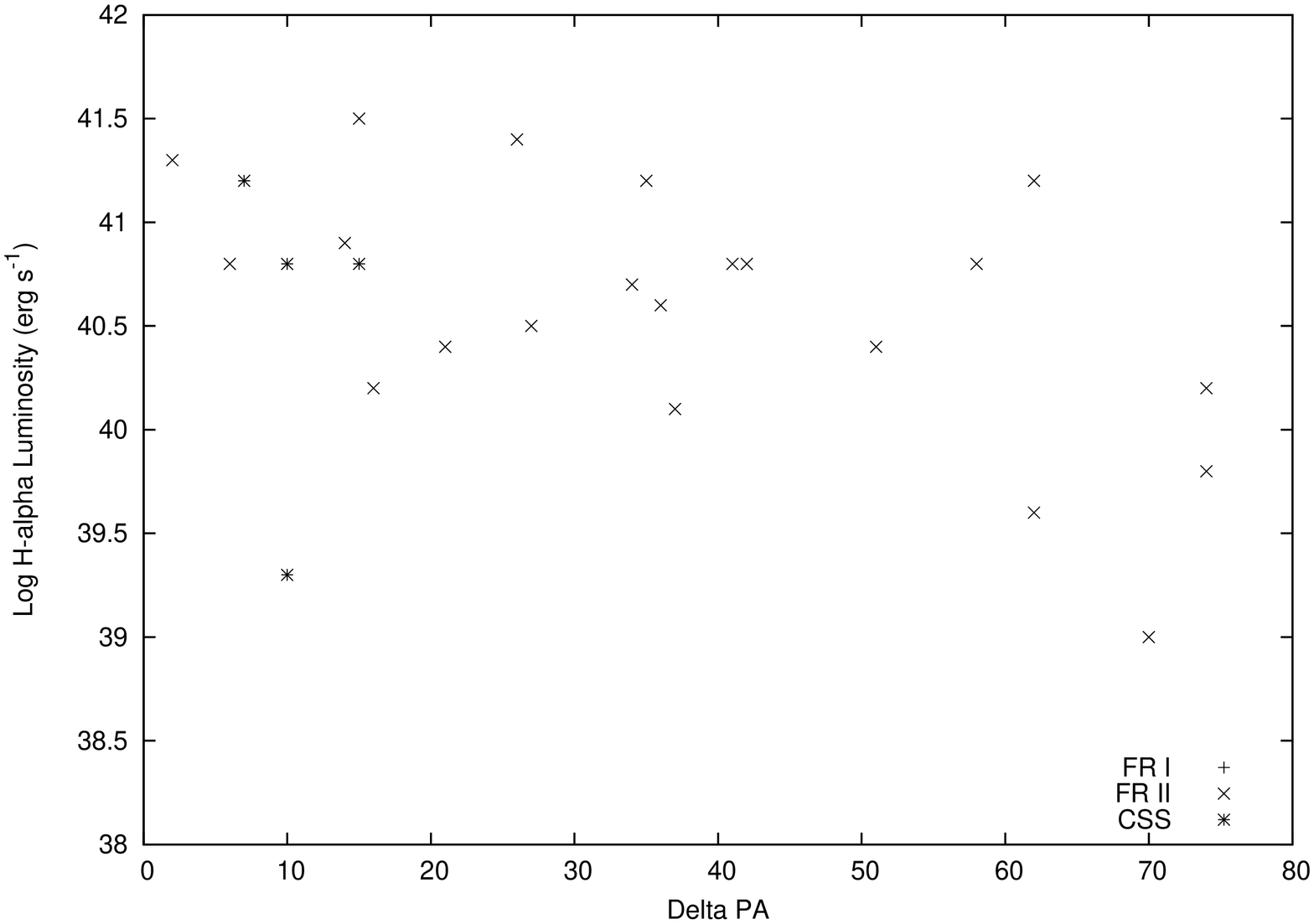}
\caption{Left: Plot of logarithm of the emission line nebulae luminosity against the relative 
alignment between the radio source and the emission line nebulae. The probabilities of 
this being a non-correlation are 0.0083 (Kendall tau) and 0.0134 (Spearman's rho) for 
the whole sample and are 0.0356 (Kendall tau) and 0.0473 (Spearman's rho) 
for $z < 0.6$.
Right: Same plot as that on the left, but for sources with $0.1 < z < 0.6$ only. The probability 
of this being a non-correlation is 0.0429 (Kendall tau). }
\label{emission_luminosity-delta_pa}
\end{figure}

Comparing the luminosity of the ELR with the relative alignment (Figure 
\ref{emission_luminosity-delta_pa}) indicates aligned sources have more luminous 
ELRs. The correlation becomes weaker for the redshift range $z < 0.6$. We find that
the correlation is further weakened if we also exclude $z < 0.1 $
(Figure \ref{emission_luminosity-delta_pa}). 
The existence of the (weak) correlation at $z < 0.6$ suggests that the radio source
provides additional energy to the emission line gas along the direction of
the radio axis and thus, there is a `real' though weak alignment effect at these 
low redshifts. 

\section{Conclusions}
\label{sec:Conclusions}

We present WFPC2 LRF images of selected emission lines in a sample of 3CR radio galaxies with 
bright emission lines. 
We detect clumpy, high surface brightness emission line gas with a wide range of 
morphologies. We examine the properties of the emission line gas and its relationship to
the radio source, and in particular the nature of the alignment effect at $z < 0.6$ where
the effect is beginning to become important. 

We find there is a weak alignment effect in our low $z$ sample ($z < 0.6$). The effect is clearly
weaker than seen at high redshifts and is also somewhat weaker than seen at similar redshifts in
ground-based data. This suggests that the alignment effect is dominated by the more
extended lower surface brightness emission line gas which falls below our detection
threshold in WFPC2. 


The alignment at low redshift is strong when the radio source and the nebula are both of 
order galactic scales (e.g., CSS sources). This implies there is mechanical energy input from 
the radio source as it propagates through the ambient gas in the host galaxy 
\citep[e.g.,][]{deVries99,Axon00}.

There is a weak trend for the nebulae in the aligned sources (at all radio/optical size ratios)
to have higher line luminosity.
This suggests that the radio source preferentially provides energy to gas which is along
the radio source axis. This is likely to be predominantly mechanical energy when the
radio source and nebulae have similar sizes, and predominantly photoionization when the
radio source is much larger than the nebula 
\citep[e.g.,][]{Best00,Inskip02a,Moy02,ODea02,Labiano05}. 

There is a weak trend for the size of the high surface brightness emission line nebulae 
to increase with  
increasing radio power and/or redshift over the range $z < 0.6$. This may be due to both the 
increasing presence of  gas on large scales and the ability of the more
powerful radio sources  to provide additional mechanical energy and/or ionizing photons
as a function of redshift.  

Thus, there is evidence for a modest excess of aligned sources at low redshift
($z < 0.6$). 
The alignment is strong when the radio source and emission line nebula are both on
galactic scales. There are also weak trends for the aligned emission line
nebulae to be more luminous, and for the emission line size to increase with redshift
and radio power. The combination of these results  suggests that 
there is a limited but real capacity for the radio source to influence the properties
of the emission line nebulae at these low redshifts.

\acknowledgments
Support for program 5957  was provided by NASA through a grant from the Space Telescope Science Institute, which is operated by the Association of Universities for Research in Astronomy, Inc., under NASA contract NAS 5-26555. 
This research has made use of the NASA/IPAC Extragalactic Database (NED) which is operated 
by the Jet Propulsion Laboratory, California Institute of Technology, under contract with 
the National Aeronautics and Space Administration. This research has also made use of 
NASA's Astrophysics Data System Bibliographic Services. We are grateful to the referee
for comments which helped to clarify the presentation of the results. 


{\it Facilities:} \facility{HST (WFPC2)}, \facility{VLA (NRAO)} 


\clearpage

\begin{figure}
\rotatebox{-90}{\includegraphics[scale=.8]{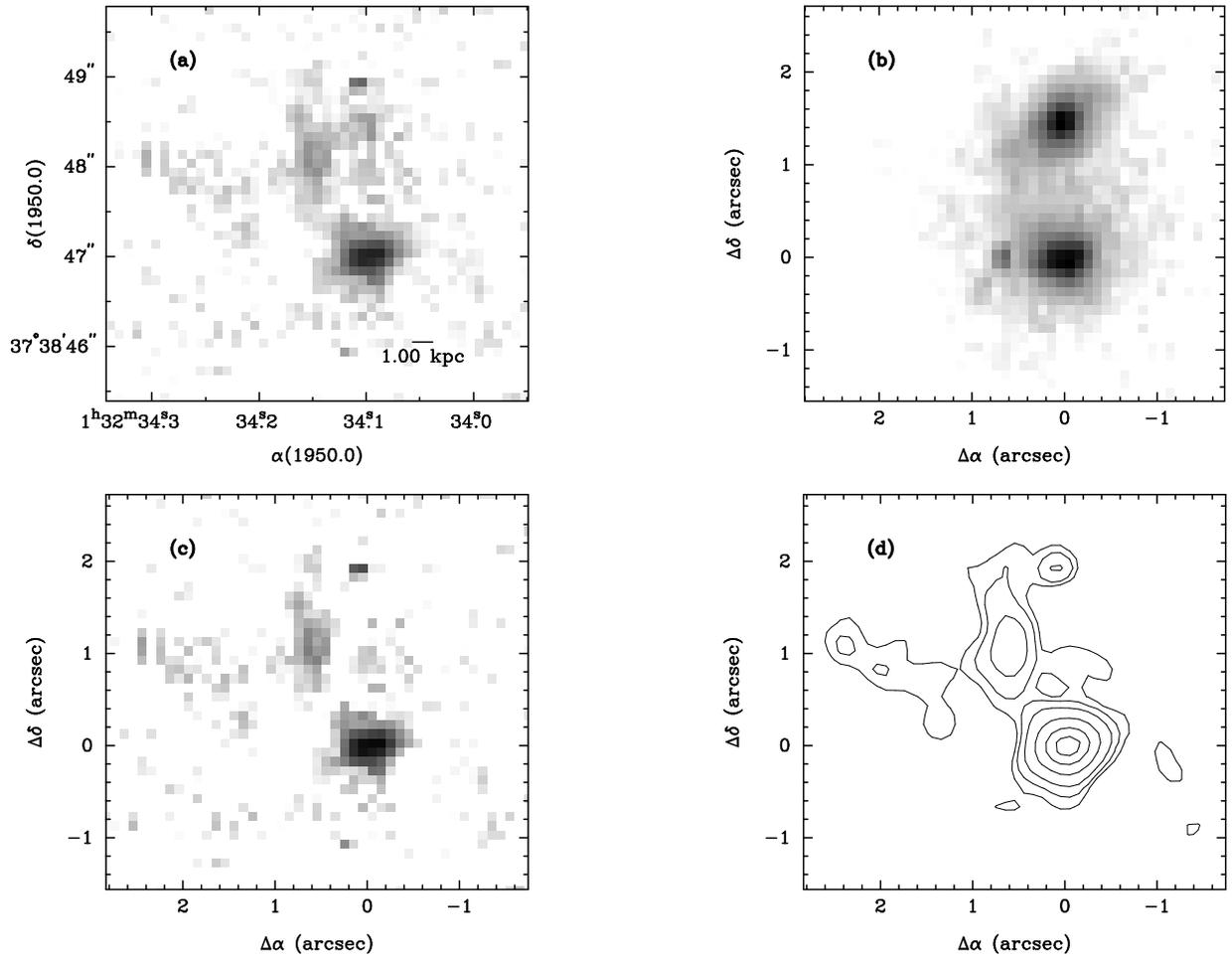}}
\caption{3C46 (Optical Montage) - Starting at the upper left, going clockwise: LRF Image; Broadband Image; Continuum Subtracted image [Contours]; Continuum Subtracted Image [Grey scale]}
\label{3C46-montage}
\end{figure}

\begin{figure}
\includegraphics[scale=.5]{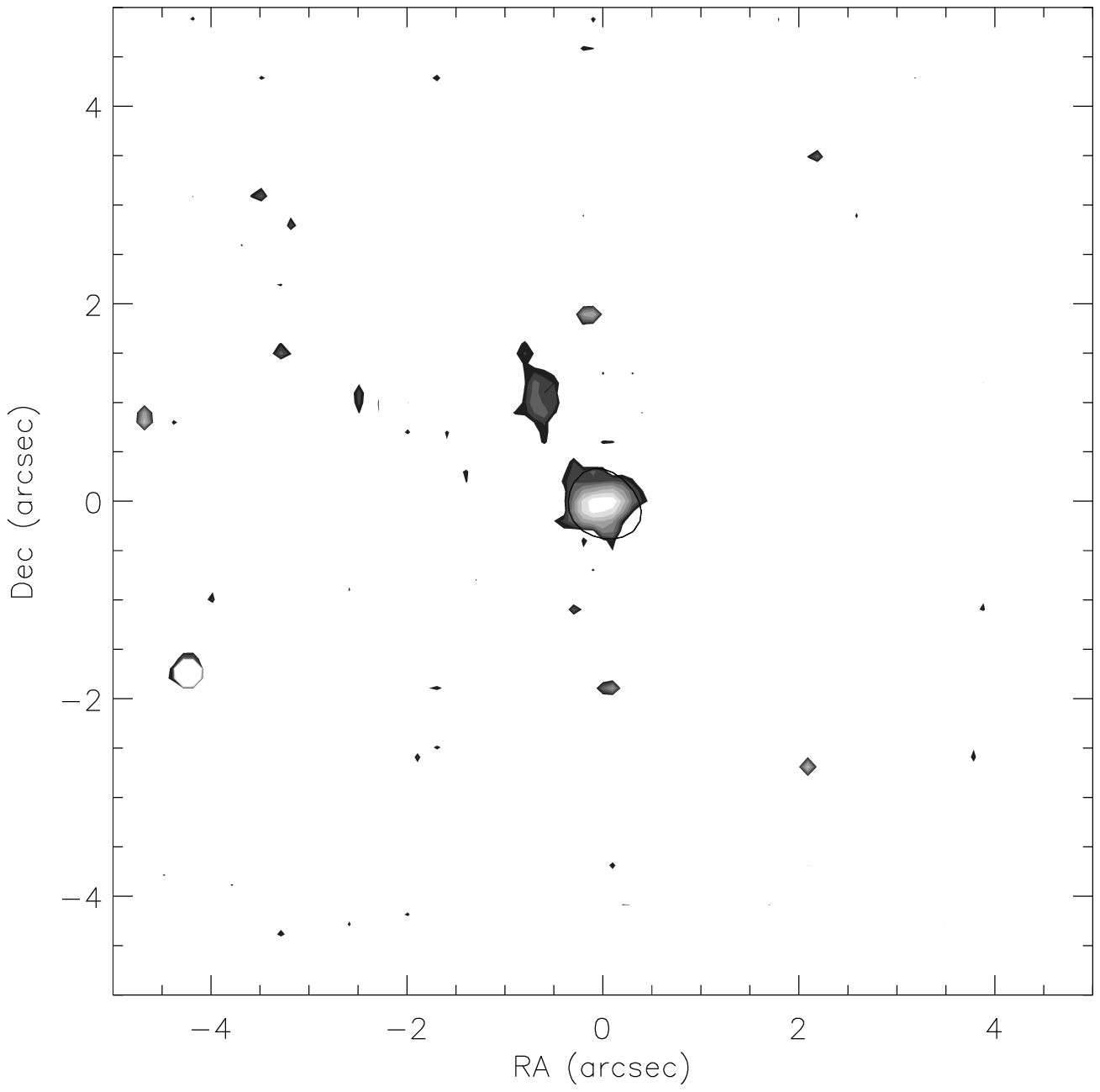}
\hfill
\includegraphics[scale=.5]{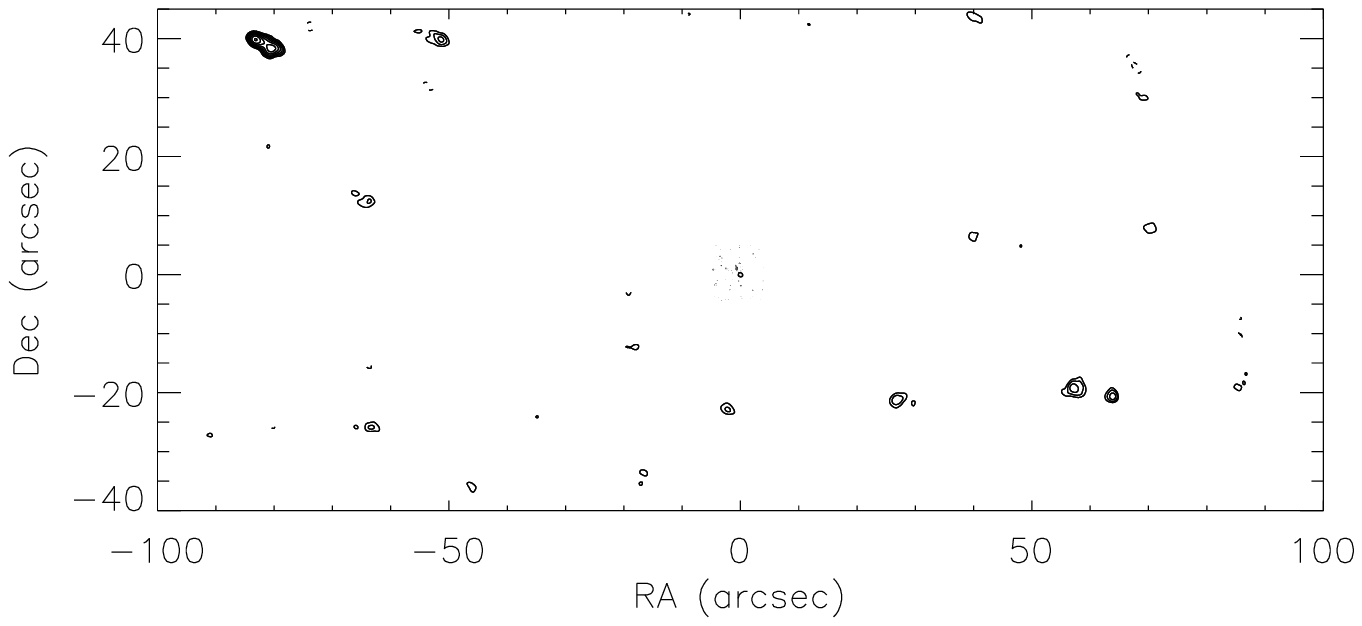}
\caption{3C46 (Radio/Optical Overlay) - Radio is shown in contours. Optical is shown in grey scale. Left: Closeup of the core. Right: View of the overall radio source. Radio levels are (1.215, 1.719, 2.431, 3.437, 4.861, 6.875, 9.722, 13.750, 19.445, 27.499) mJy. Optical levels are (56.5, 79.9, 113.0, 159.9, 159.9, 226.1, 319.7, 452.2, 639.5) * {$10^{-14}$} erg $s^{-1}$ $cm^{-2}$. Both images have the same levels. Radio map from \citet{Neff95}.}
\label{3C46-overlay}
\end{figure}

\clearpage

\begin{figure}
\rotatebox{-90}{\includegraphics[scale=.8]{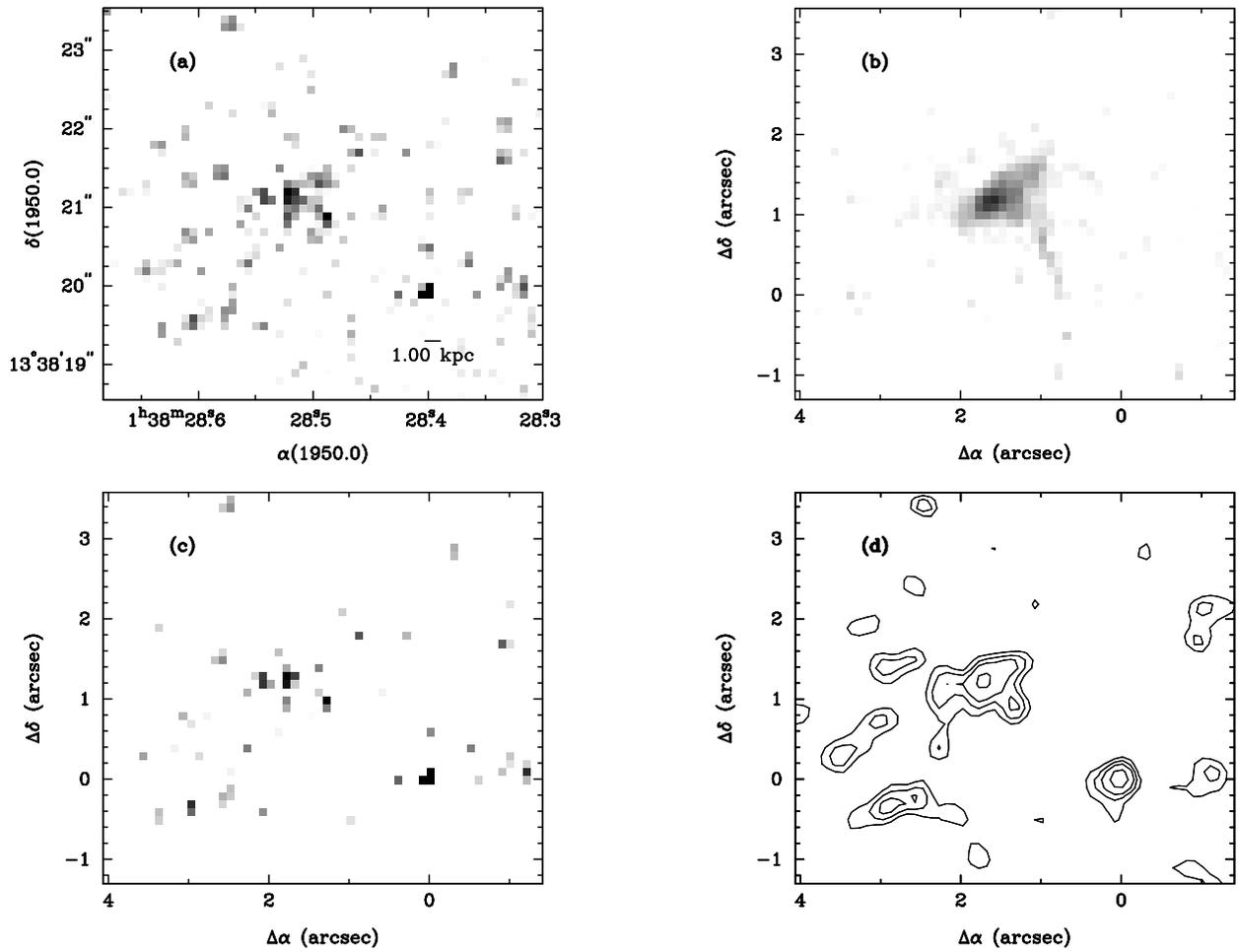}}
\caption{3C49 (Optical Montage) - Starting at the upper left, going clockwise: LRF Image; Broadband Image; Continuum Subtracted image [Contours]; Continuum Subtracted Image [Grey scale]}
\label{3C49-montage}
\end{figure}

\begin{figure}
\includegraphics[scale=.5]{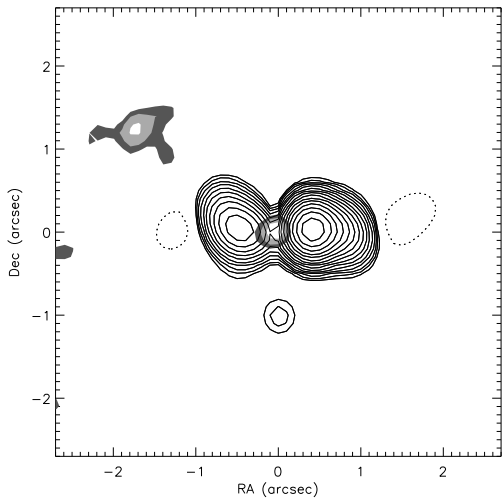}
\caption{3C49 (Radio/Optical Overlay) - Radio is shown in contours. Optical is shown in grey scale. Radio levels are (3.555, 5.028, 7.110, 10.055, 14.220, 20.110, 28.440, 40.220, 56.880, 80.440, 113.760, 160.881, 227.520, 321.762, 455.040) mJy. Optical levels are (18.2, 25.7, 33.3) * {$10^{-14}$} erg $s^{-1}$ $cm^{-2}$ The radio map is from \citet{Neff95}.}
\label{3C49-overlay}
\end{figure}

\clearpage

\begin{figure}
\rotatebox{-90}{\includegraphics[scale=.8]{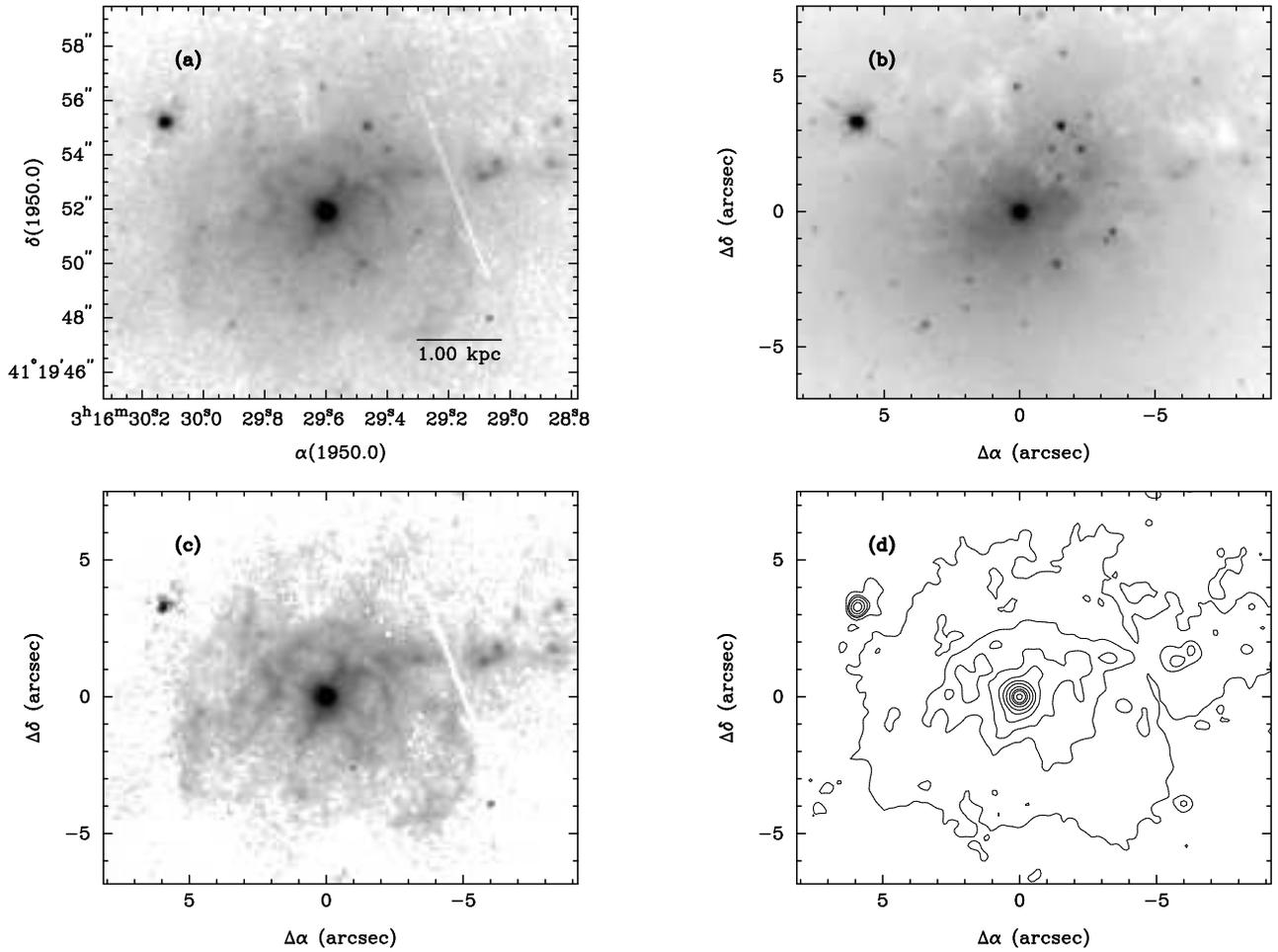}}
\caption{3C84(Optical Montage) - Starting at the upper left, going clockwise: LRF Image; Broadband Image; Continuum Subtracted image [Contours]; Continuum Subtracted Image [Grey scale]}
\label{3C84-montage}
\end{figure}

\begin{figure}
\includegraphics[scale=.5]{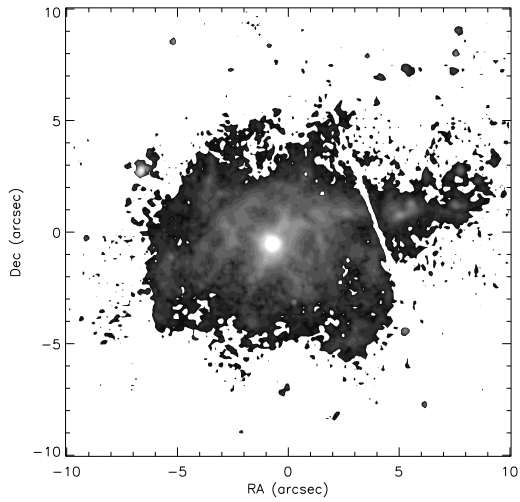}
\caption{3C84 (Emission Line Only) - Optical is shown in grey scale. Optical levels are (48.6, 68.7, 97.1, 137.3, 194.2, 274.7, 388.5, 549.4, 777.0, 1098.8, 1553.9, 2197.6, 3108.0, 4395.2, 6215.8, 8790.4) * {$10^{-14}$} erg $s^{-1}$ $cm^{-2}$ (Note: In Radio maps 3C84 is highly core dominated. A worthwhile 5 or 8 GHz radio map could not be located, so only the emission line image is presented.)}
\label{3C84-overlay}
\end{figure}

\clearpage

\begin{figure}
\includegraphics[scale=0.5]{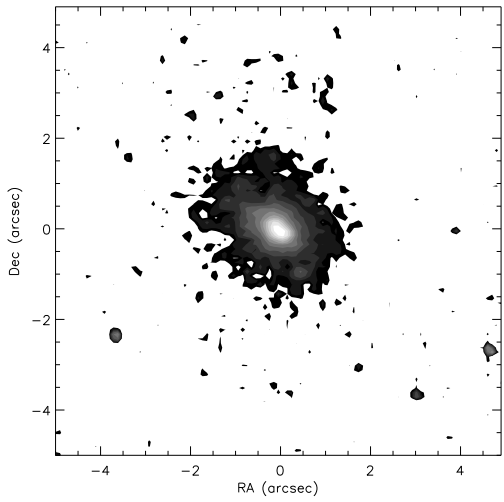}
\caption{3C98 (Narrow band Image) - This is the narrow-band optical (LRF) image. The levels are (44.5, 62.9, 89.0, 125.8, 177.9, 251.6, 355.8, 503.2, 711.7, 1006.5, 1423.4, 2012.9) {$10^{-14}$} erg $s^{-1}$ $cm^{-2}$.}
\label{3C98}
\end{figure}

\clearpage

\begin{figure}
\rotatebox{-90}{\includegraphics[scale=.8]{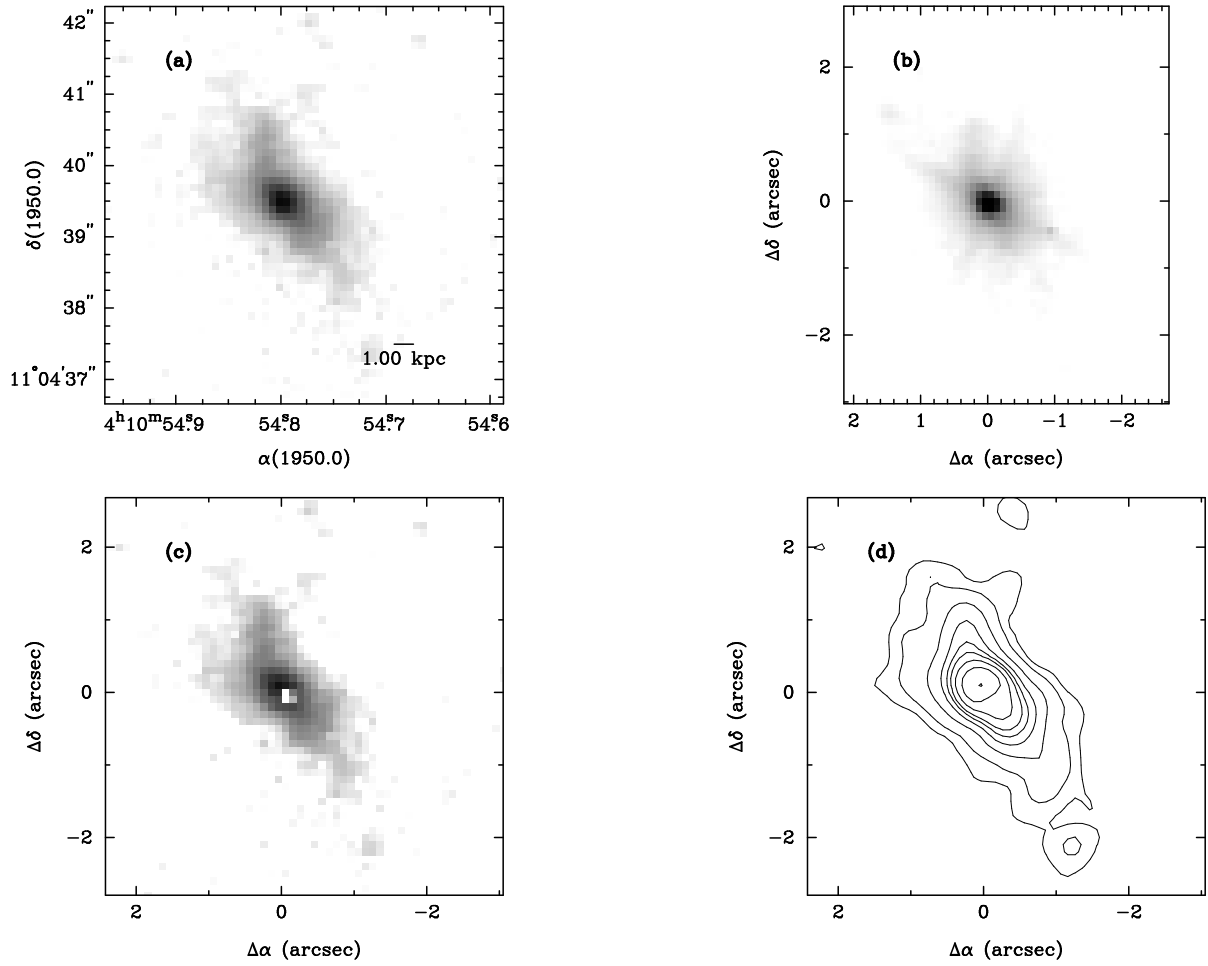}}
\caption{3C109(Optical Montage) - Starting at the upper left, going clockwise: LRF Image; Broadband Image; Continuum Subtracted image [Contours]; Continuum Subtracted Image [Grey scale]}
\label{3C109-montage}
\end{figure}

\begin{figure}
\includegraphics[scale=.5]{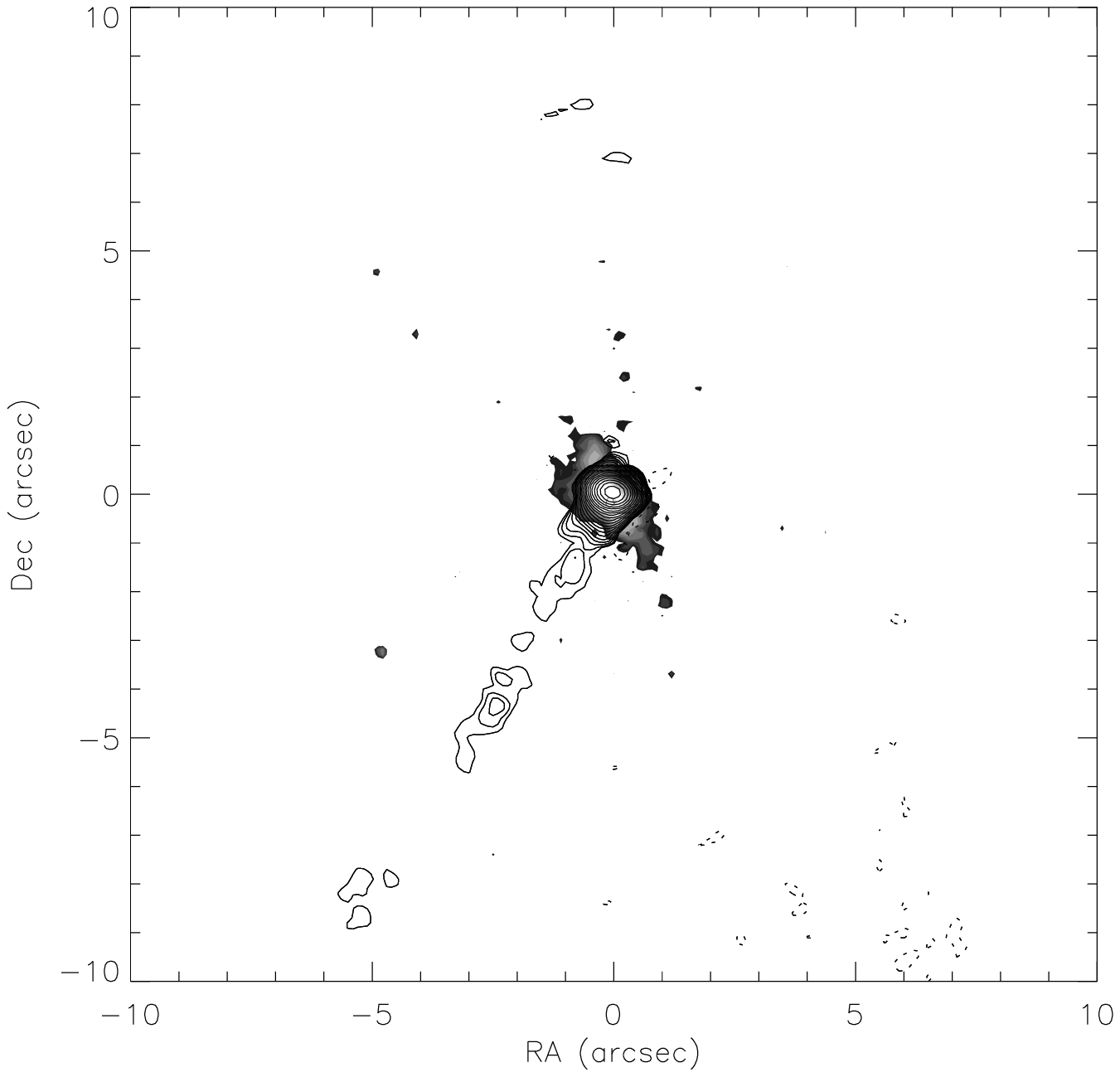}
\hfill
\includegraphics[scale=.5]{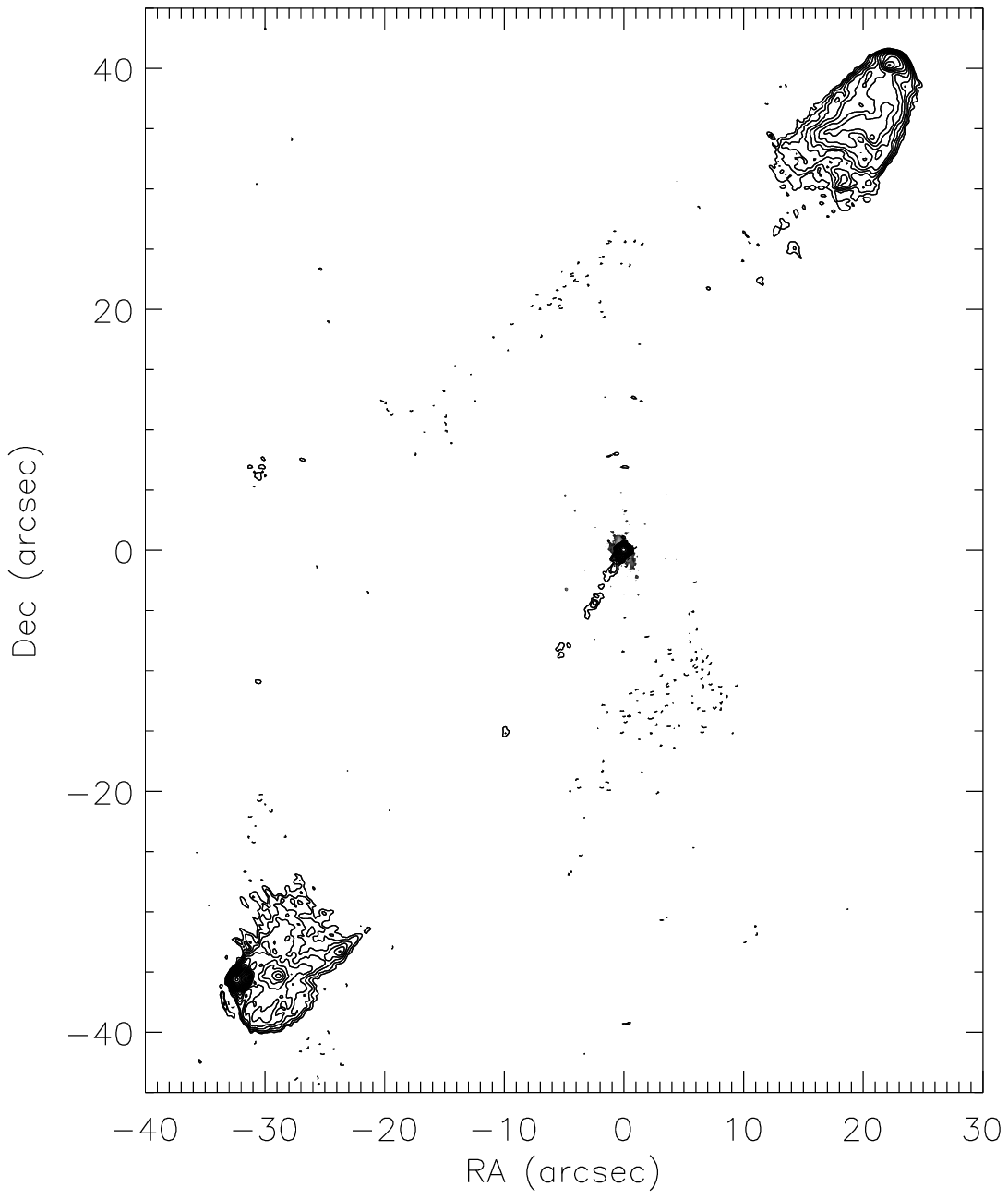}
\caption{3C109 (Radio/Optical Overlay) - Radio is shown in contours. Optical is shown in grey scale. Left: Closeup of the core. Right: View of the overall radio source. Radio levels are (0.199, 0.281, 0.398, 0.563, 0.796, 1.125, 1.591, 2.250, 3.182, 4.501, 6.365, 9.001, 12.730, 18.002, 25.459, 36.005, 50.918, 72.009, 101.837, 144.019) mJy. Optical levels are (46.7, 66.0, 93.3, 132.0, 186.7, 264.0, 373.4, 528.0, 749.8) * {$10^{-14}$} erg $s^{-1}$ $cm^{-2}$. Both images have the same levels.}
\label{3C109-overlay}
\end{figure}

\clearpage

\begin{figure}
\rotatebox{-90}{\includegraphics[scale=.8]{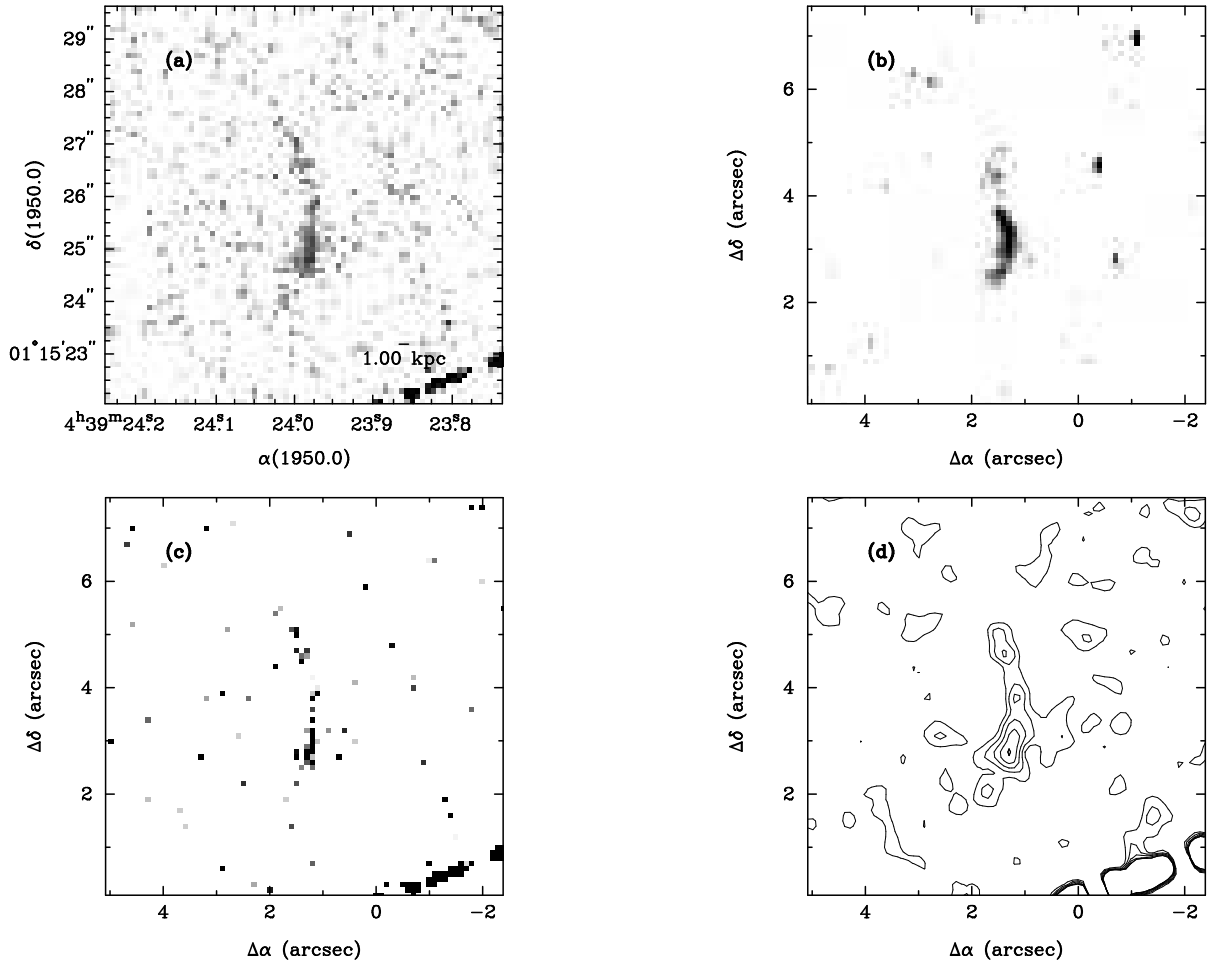}}
\caption{3C124 (Optical Montage) - Starting at the upper left, going clockwise: LRF Image; Broadband Image; Continuum Subtracted image [Contours]; Continuum Subtracted Image [Grey scale]}
\label{3C124-montage}
\end{figure}

\begin{figure}
\includegraphics[scale=.5]{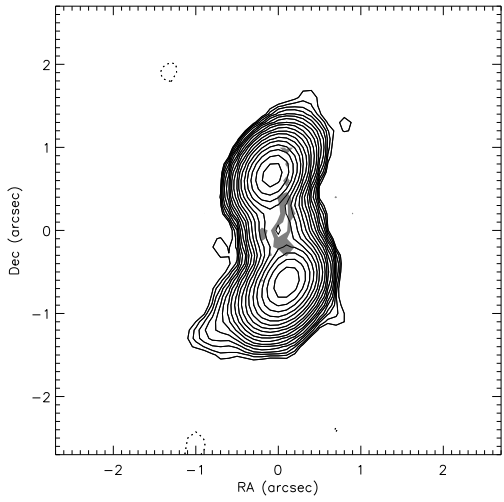}
\caption{3C124 (Radio/Optical Overlay) - Radio is shown in contours. Optical is shown in grey scale. Radio levels are (0.182, 0.258, 0.365, 0.516, 0.730, 1.032, 1.459, 2.064, 2.918, 4.127, 5.837) mJy. Optical levels are (64.6, 91.4) * {$10^{-14}$} erg $s^{-1}$ $cm^{-2}$.}
\label{3C124-overlay}
\end{figure}

\clearpage

\begin{figure}
\rotatebox{-90}{\includegraphics[scale=.8]{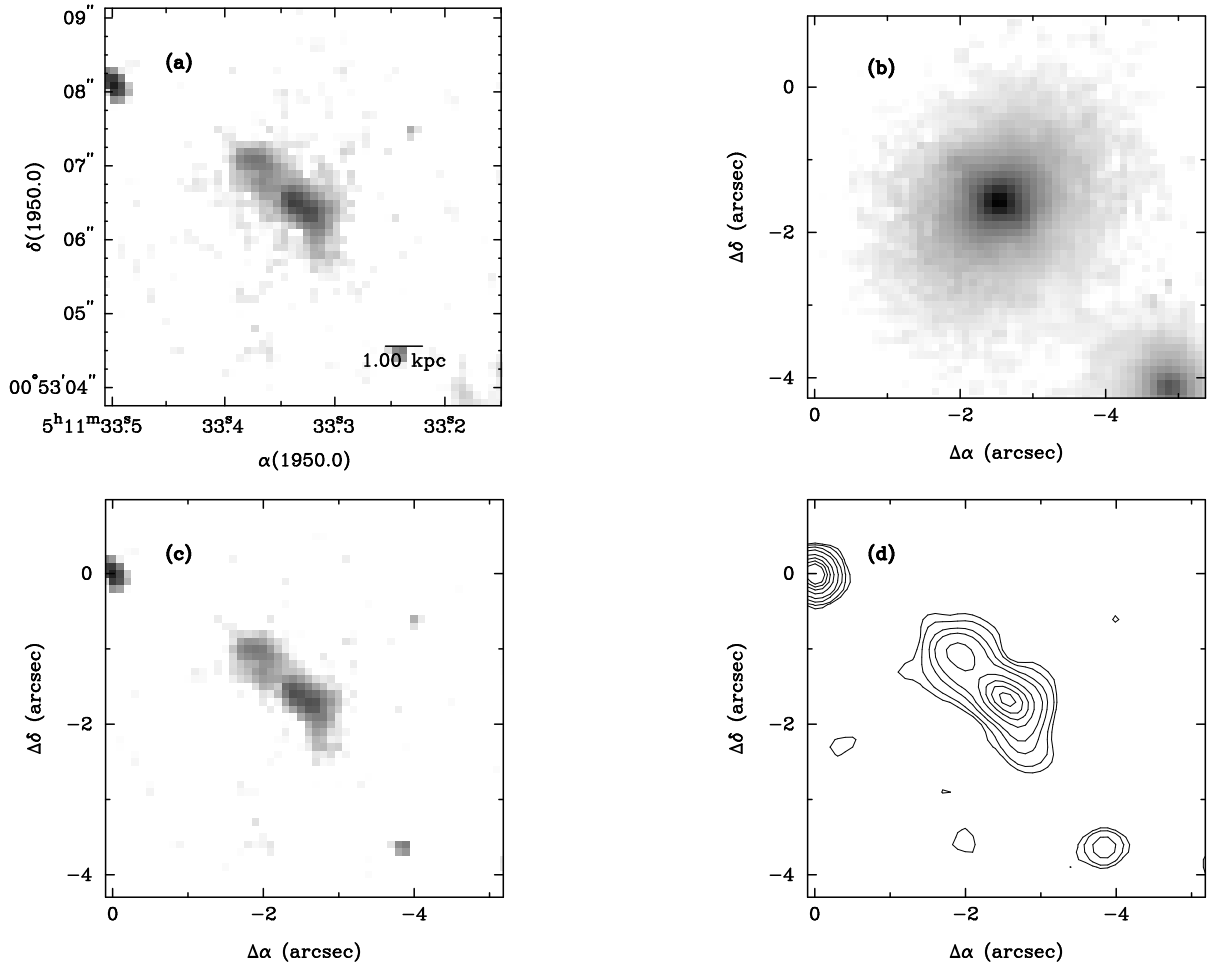}}
\caption{3C135 (Optical Montage) - Starting at the upper left, going clockwise: LRF Image; Broadband Image; Continuum Subtracted image [Contours]; Continuum Subtracted Image [Grey scale]}
\label{3C135-montage}
\end{figure}

\begin{figure}
\includegraphics[scale=.5]{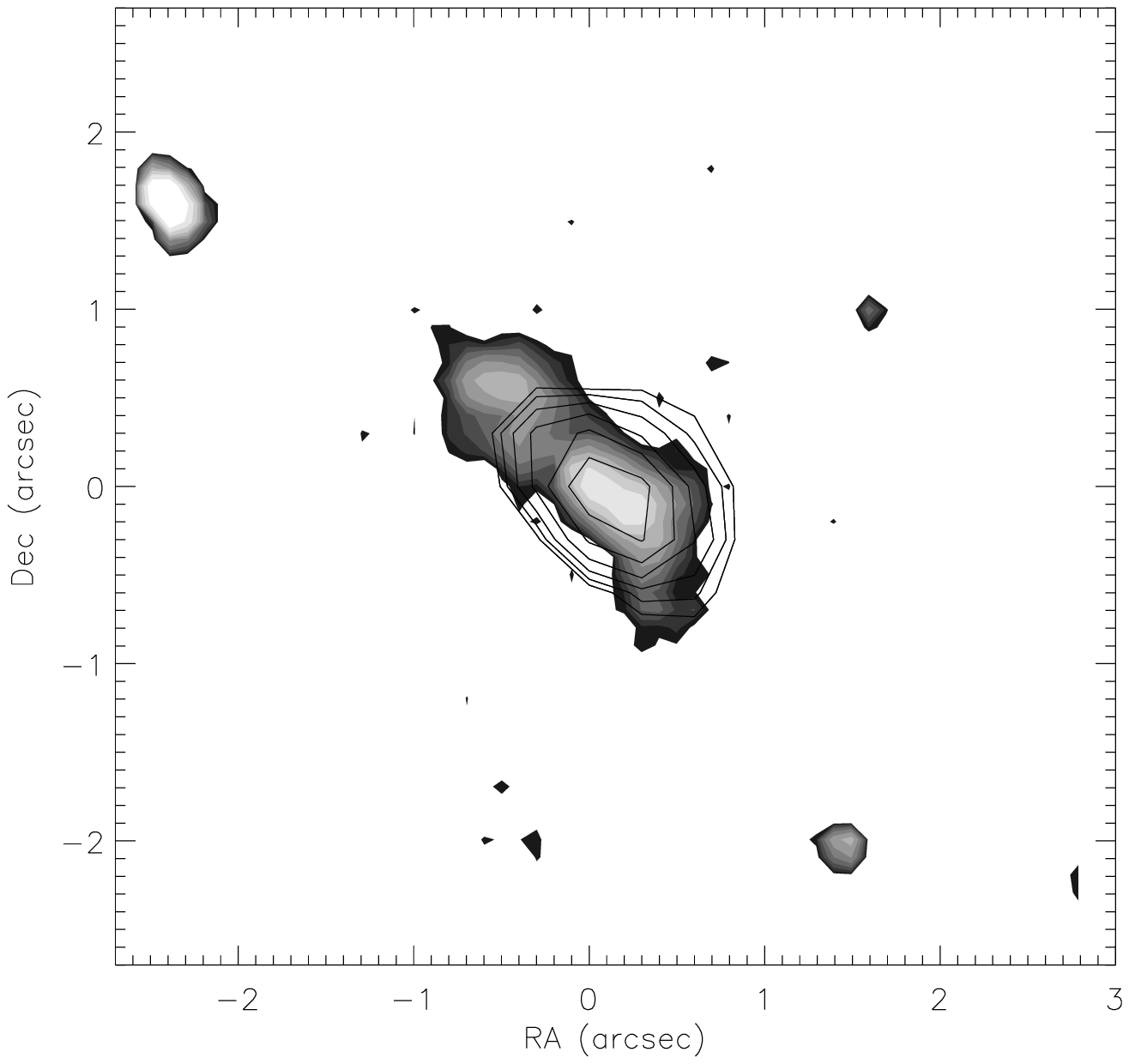}
\hfill
\includegraphics[scale=.5]{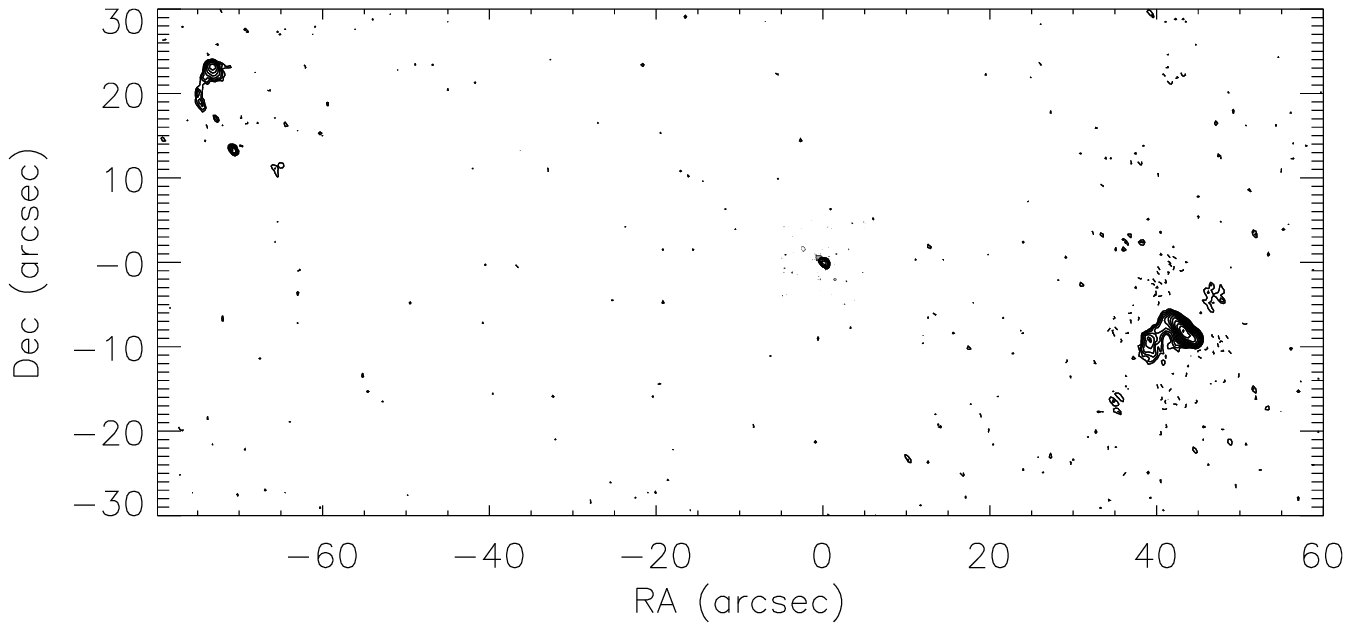}
\caption{3C135 (Radio/Optical Overlay) - Radio is shown in contours. Optical is shown in grey scale. Left: Closeup of the core. Right: View of the overall radio source. Radio levels are (0.151, 0.214, 0.303, 0.429, 0.606, 0.857, 1.212, 1.714, 2.424, 3.428, 4.848, 6.856) mJy. Optical levels are (96.0, 135.9, 192.2, 271.8, 384.3, 543.5, 768.6, 1087.0, 1537.3, 2174.0) * {$10^{-14}$} erg $s^{-1}$ $cm^{-2}$. Both images have the same levels.}
\label{3C135-overlay}
\end{figure}

\clearpage

\begin{figure}
\rotatebox{-90}{\includegraphics[scale=.8]{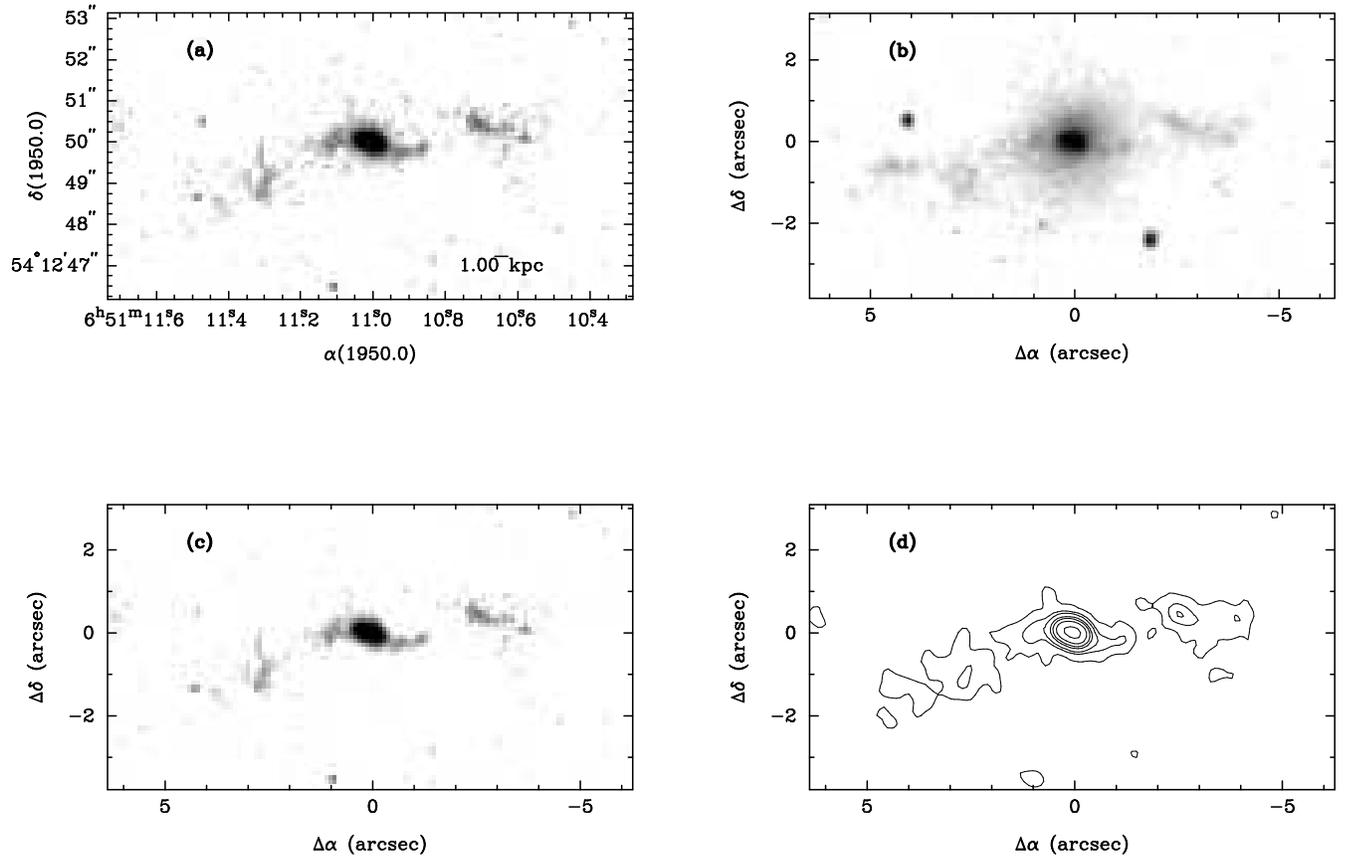}}
\caption{3C171 (Optical Montage) - Starting at the upper left, going clockwise: LRF Image; Broadband Image; Continuum Subtracted image [Contours]; Continuum Subtracted Image [Grey scale]}
\label{3C171-montage}
\end{figure}

\begin{figure}
\includegraphics[scale=.5]{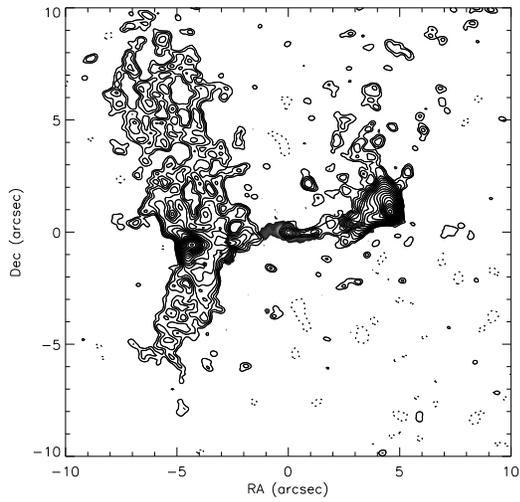}
\caption{3C171 (Radio/Optical Overlay) - Radio is shown in contours. Optical is shown in grey scale. Radio levels are (0.164, 0.231, 0.327, 0.463, 0.654, 0.925, 1.309, 1.851, 2.618, 3.702, 5.235, 7.404, 10.470, 14.807, 20.941, 29.615, 41.882, 59.230, 83.763) mJy. Optical levels are (52.4, 74.0, 104.7, 148.1, 209.4, 296.2, 418.9, 592.4, 837.7, 1184.7, 1675.4, 2369.4) * {$10^{-14}$} erg $s^{-1}$ $cm^{-2}$. The radio map courtesy of David Floyd (private communication).}
\label{3C171-overlay}
\end{figure}

\clearpage

\begin{figure}
\rotatebox{-90}{\includegraphics[scale=.8]{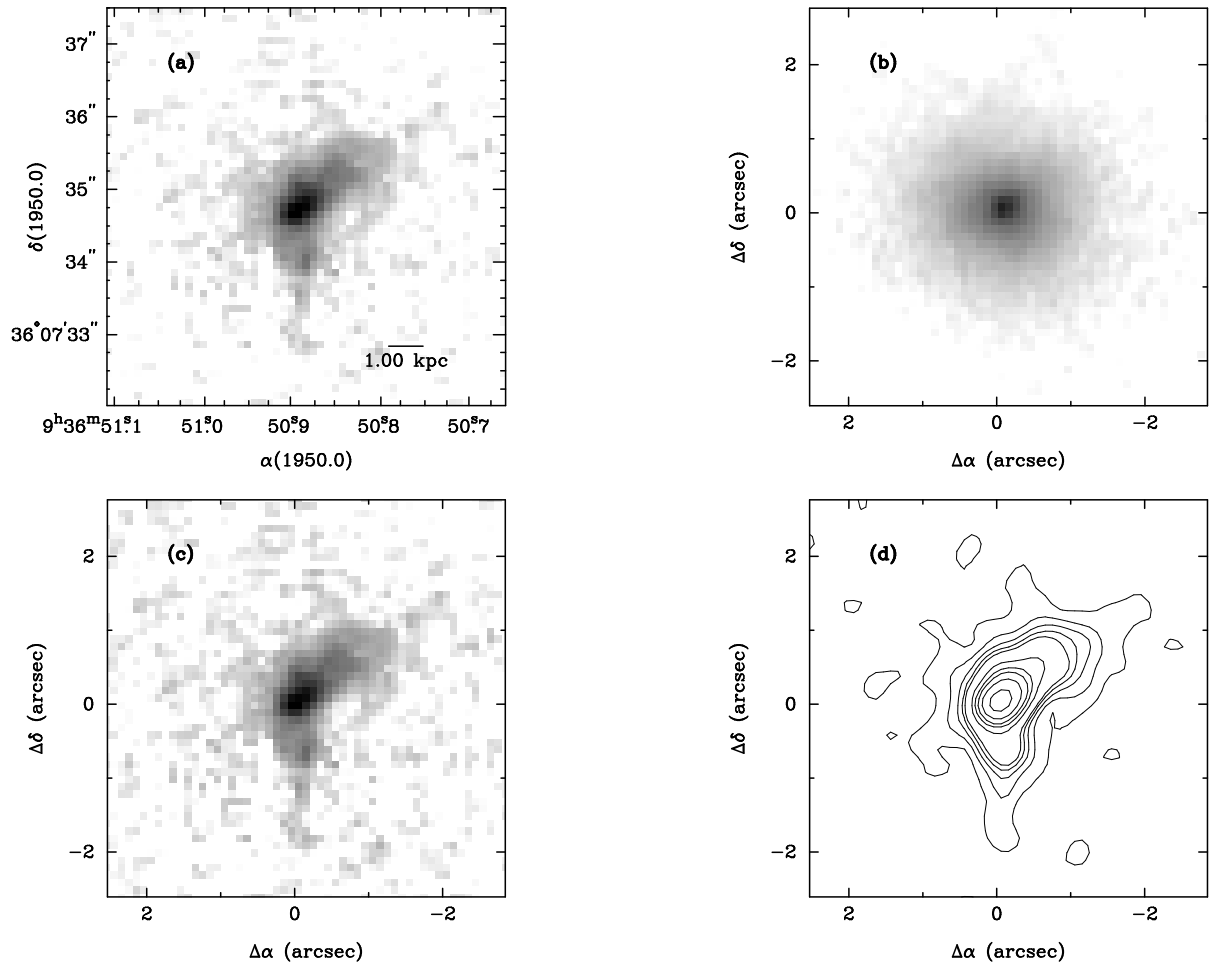}}
\caption{3C223 (Optical Montage) - Starting at the upper left, going clockwise: LRF Image; Broadband Image; Continuum Subtracted image [Contours]; Continuum Subtracted Image [Grey scale]}
\label{3C223-montage}
\end{figure}

\begin{figure}
\includegraphics[scale=.5]{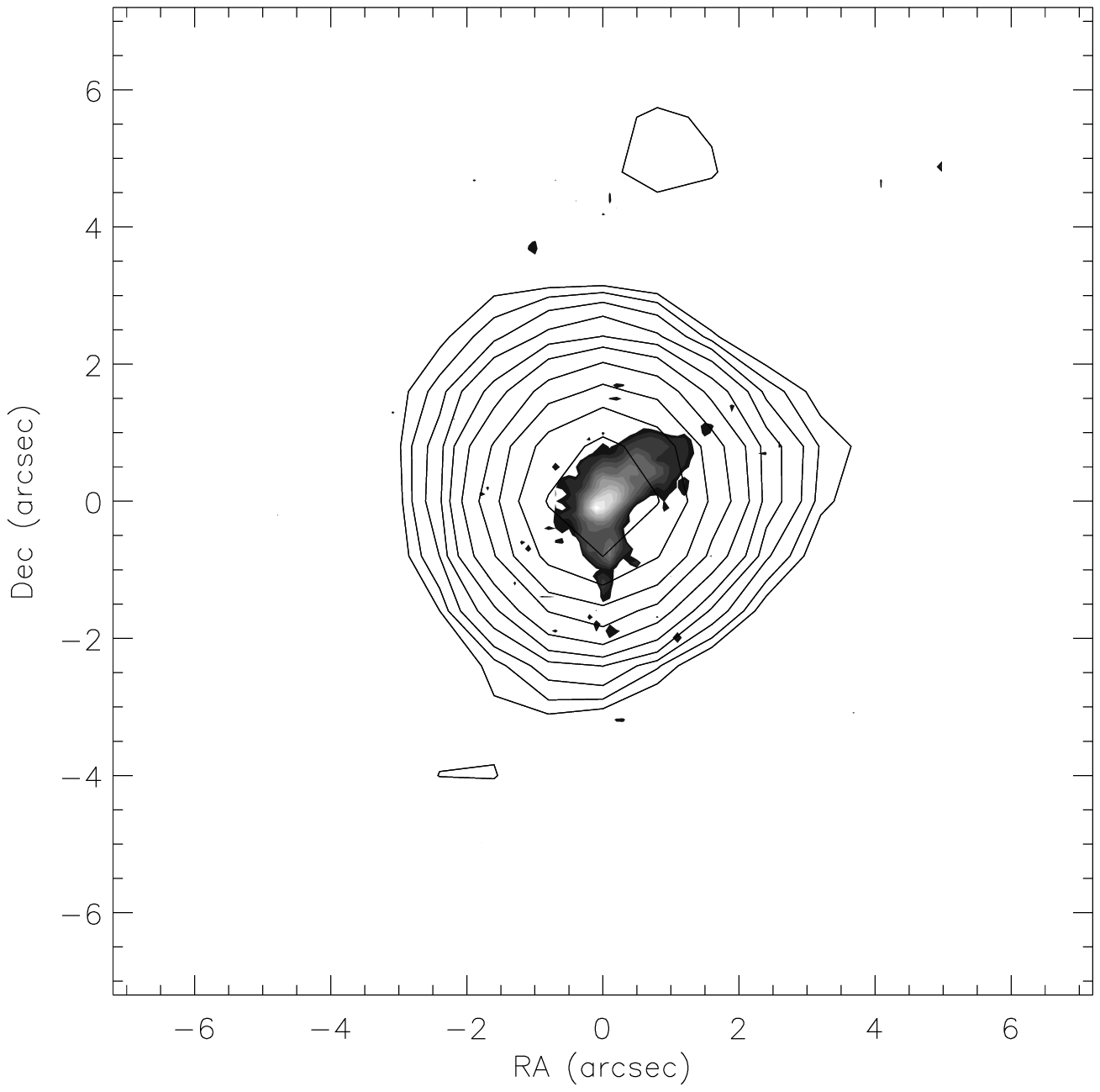}
\hfill
\includegraphics[scale=.5]{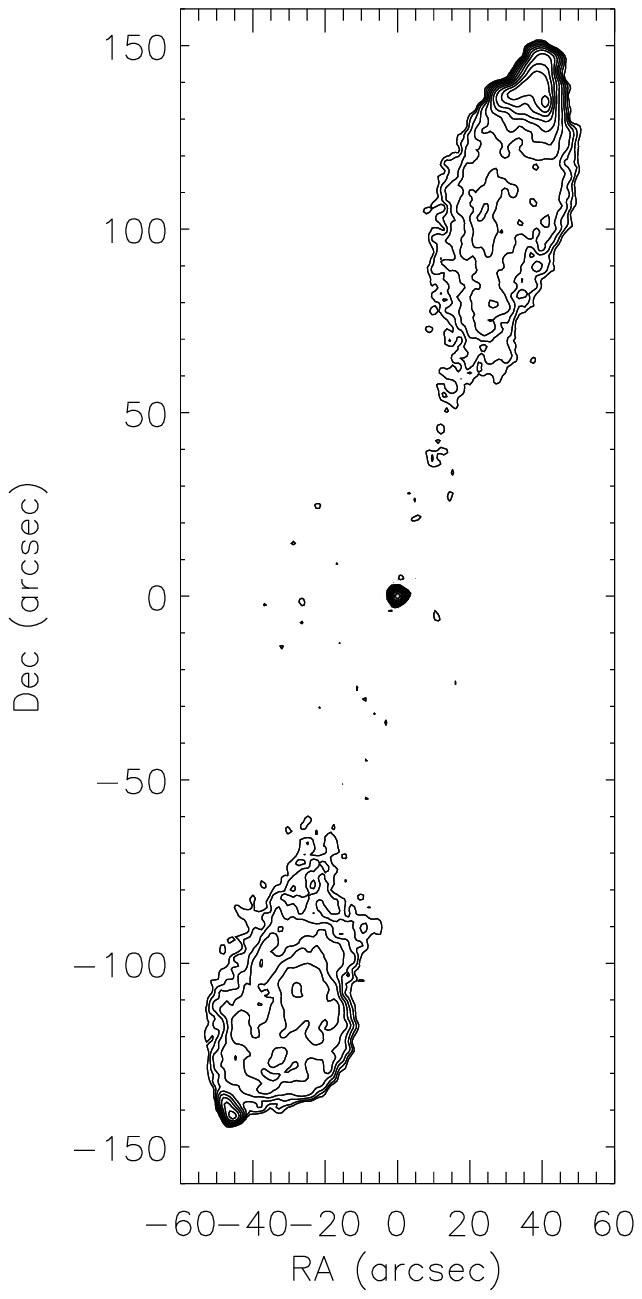}
\caption{3C223 (Radio/Optical Overlay) - Radio is shown in contours. Optical is shown in grey scale. Left: Closeup of the core. Right: View of the overall radio source. Radio levels are (0.258, 0.365, 0.516, 0.730, 1.032, 1.459, 2.064, 2.919, 4.128, 5.838, 8.256) mJy. Optical levels are (100.8, 142.6, 201.7, 285.2, 403.3, 570.4, 806.7, 1140.8, 1613.3, 2281.6, 3226.7, 4563.3, 6453.5) * {$10^{-14}$} erg $s^{-1}$ $cm^{-2}$. Both images have the same levels. The radio map courtesy of David Floyd (private communication).}
\label{3C223-overlay}
\end{figure}

\clearpage

\begin{figure}
\rotatebox{-90}{\includegraphics[scale=.8]{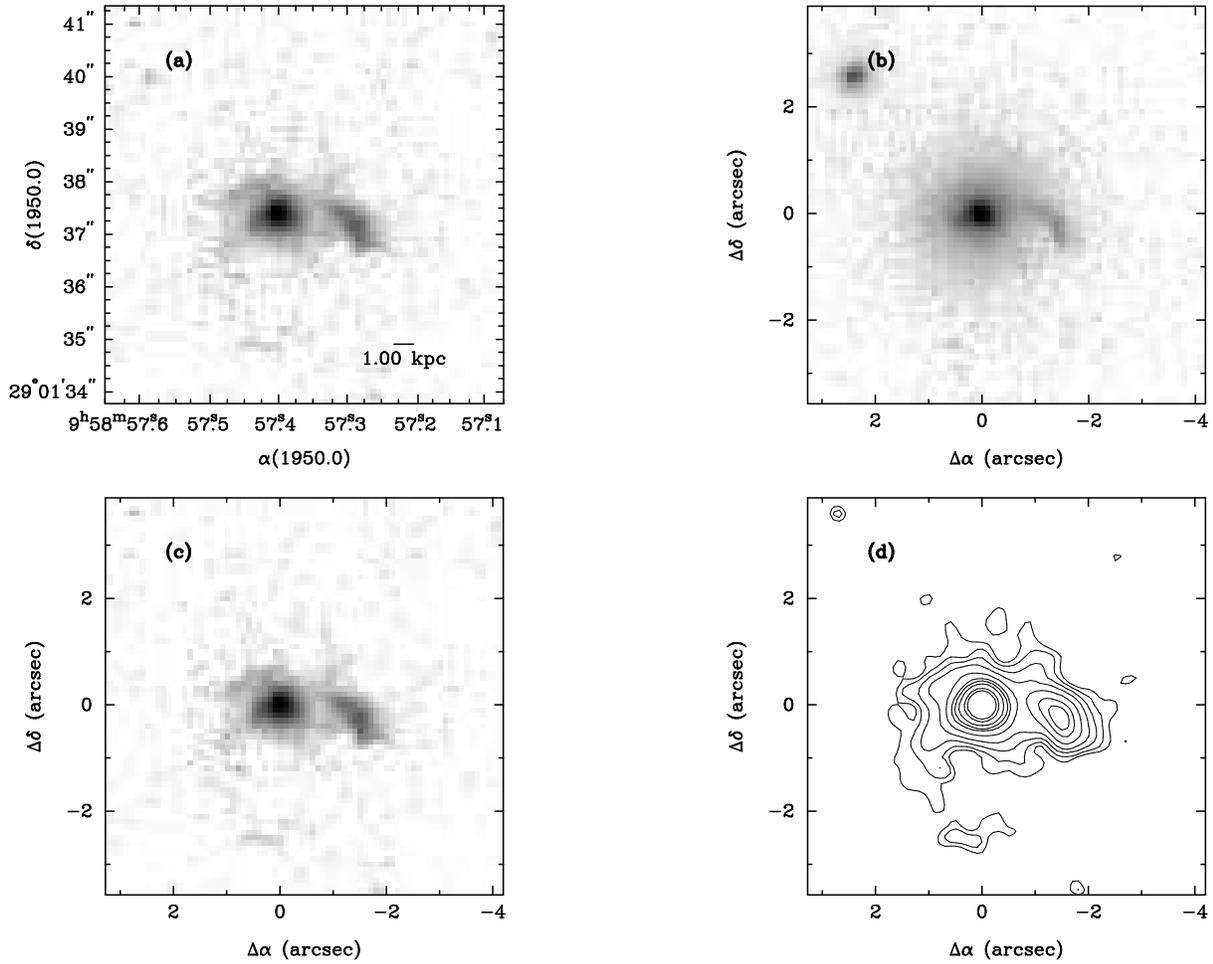}}
\caption{3C234 (Optical Montage) - Starting at the upper left, going clockwise: LRF Image; Broadband Image; Continuum Subtracted image [Contours]; Continuum Subtracted Image [Grey scale]}
\label{3C234-montage}
\end{figure}

\begin{figure}
\includegraphics[scale=.5]{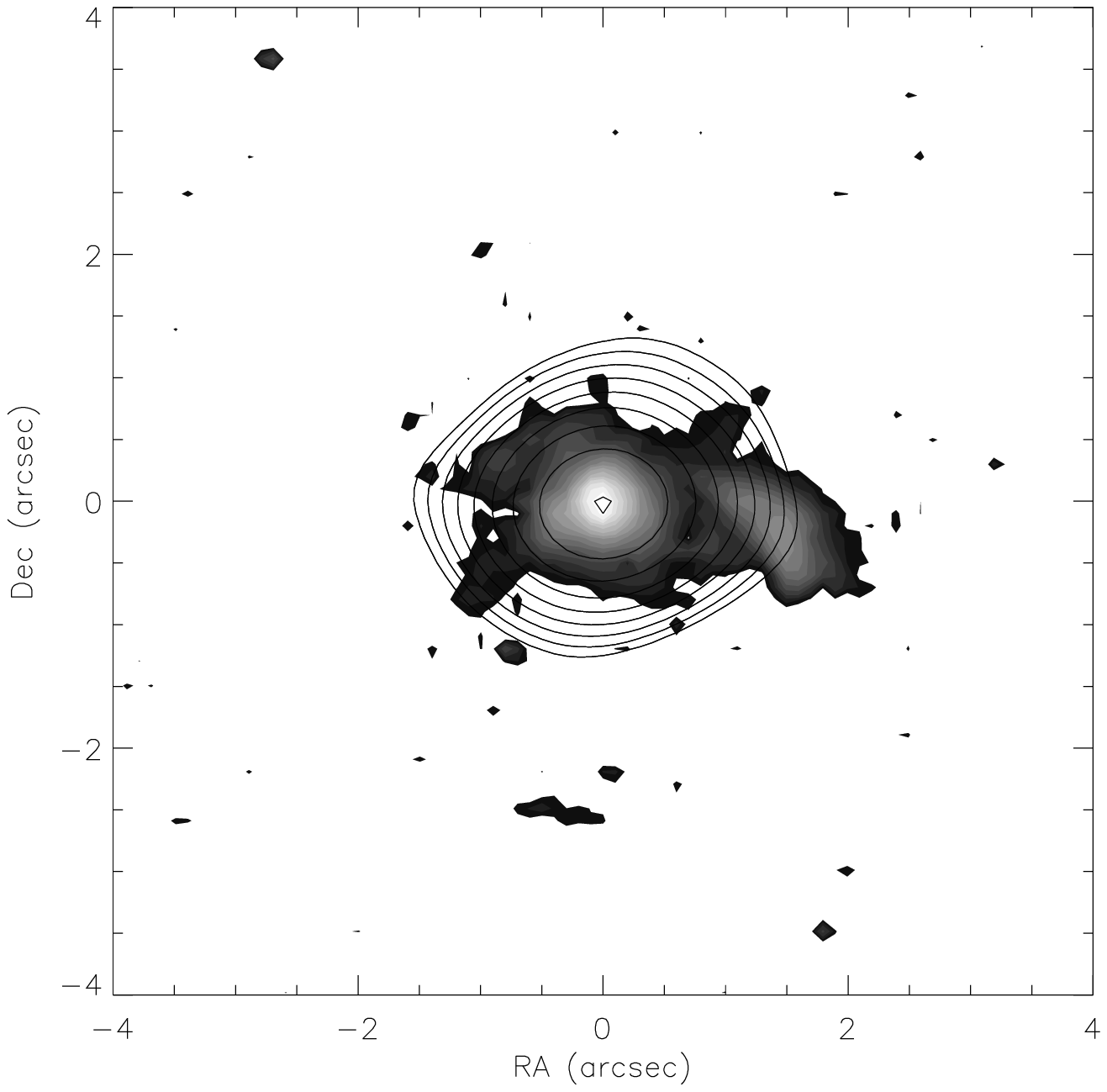}
\hfill
\includegraphics[scale=.5]{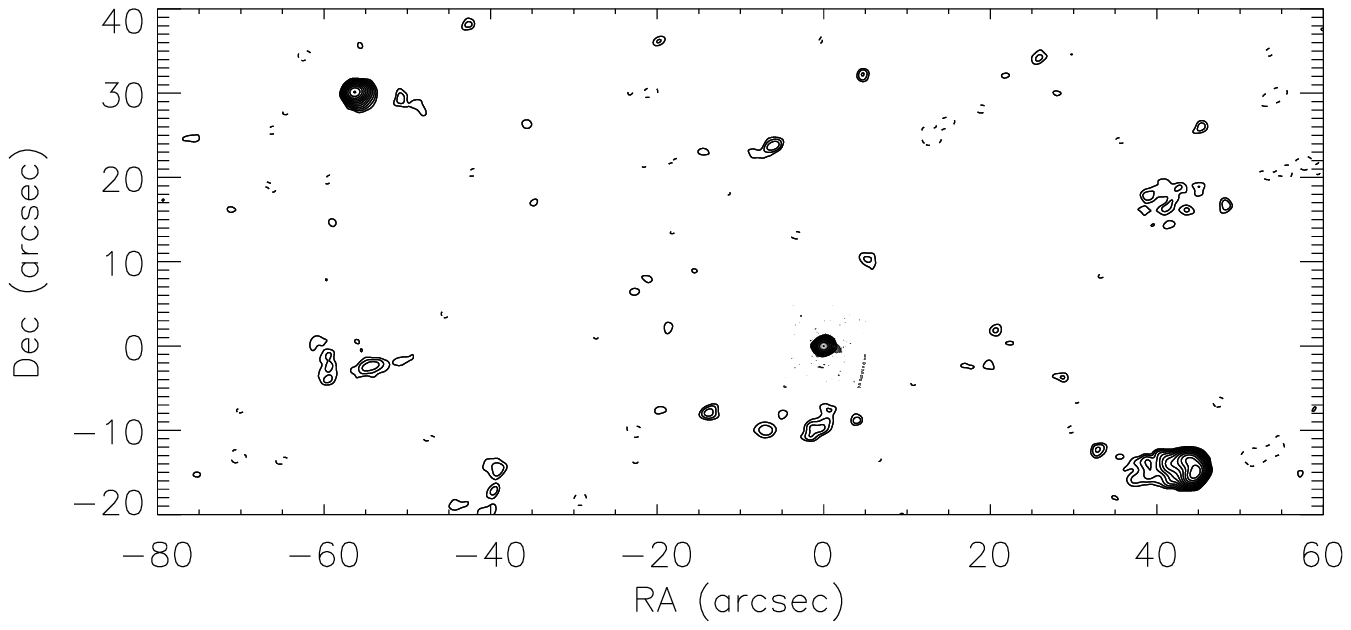}
\caption{3C234 (Radio/Optical Overlay) - Radio is shown in contours. Optical is shown in grey scale. Left: Closeup of the core. Right: View of the overall radio source. Radio levels are (2.226, 3.148, 4.451, 6.295, 8.903, 12.590, 17.806, 25.181, 35.611, 50.362, 71.222, 100.724, 142.445, 201.447, 284.890, 402.895, 569.779) mJy. Optical levels are (134.6, 190.4, 269.3, 380.8, 538.6, 761.7, 1077.1, 1523.3, 2154.3, 3046.6, 4308.6, 6093.3, 8617.2, 12186.6, 17234.5, 24373.2, 34468.9) * {$10^{-14}$} erg $s^{-1}$ $cm^{-2}$. Both images have the same levels. Radio map from \citet{Neff95}.}
\label{3C234-overlay}
\end{figure}

\clearpage

\begin{figure}
\rotatebox{-90}{\includegraphics[scale=.8]{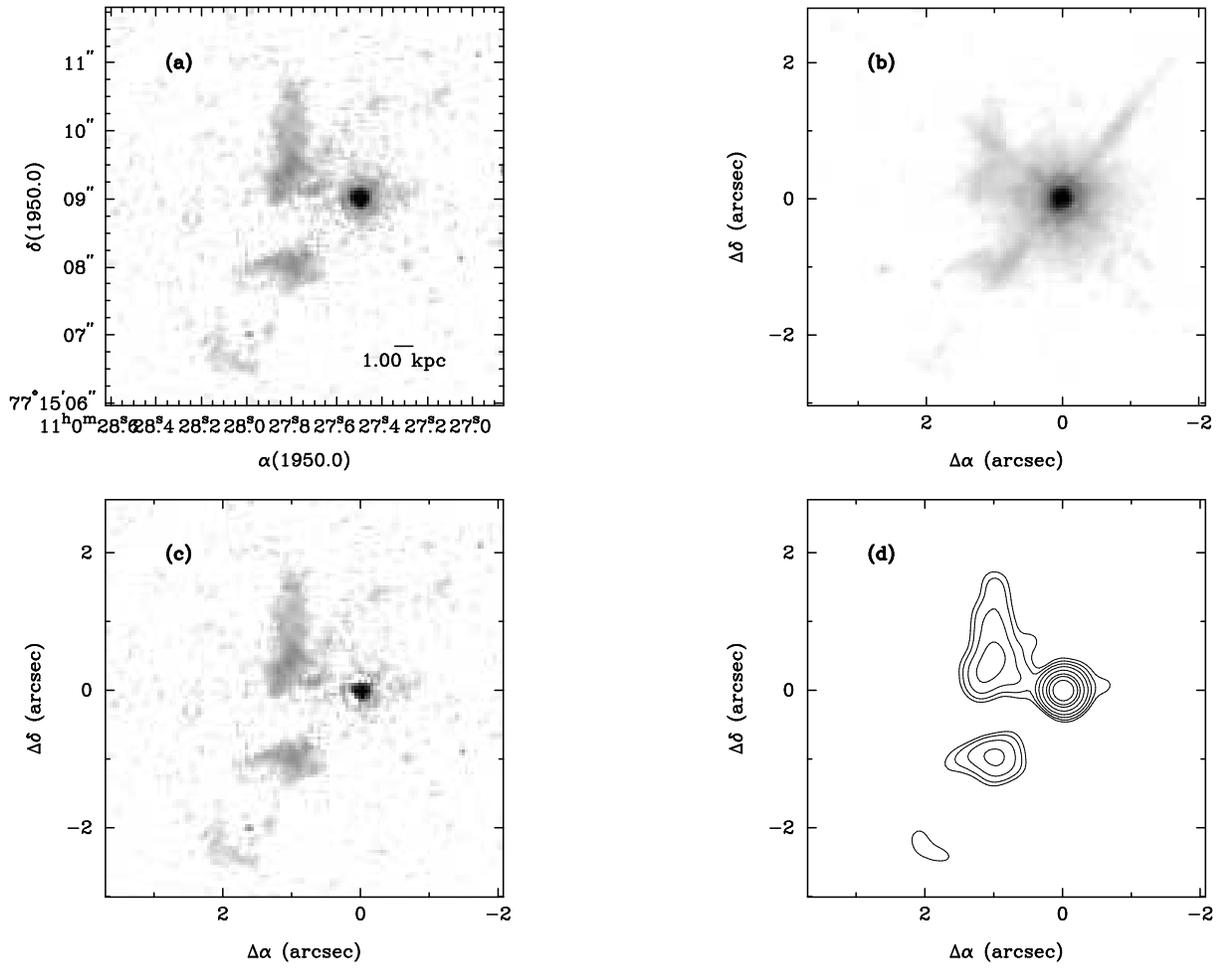}}
\caption{3C249.1 (Optical Montage) - Starting at the upper left, going clockwise: LRF Image; Broadband Image; Continuum Subtracted image [Contours]; Continuum Subtracted Image [Grey scale]}
\label{3C249.1-montage}
\end{figure}

\begin{figure}
\includegraphics[scale=.5]{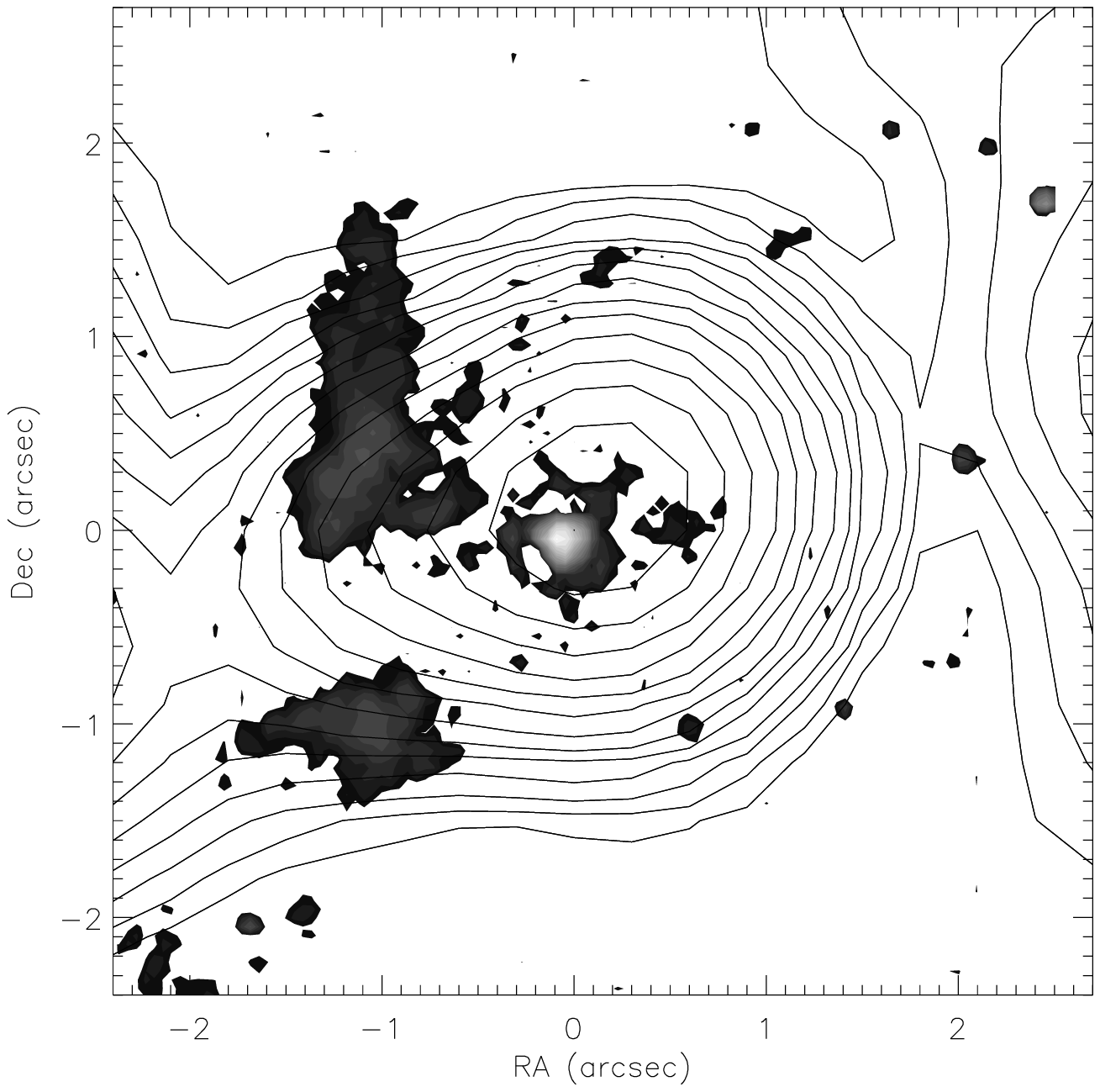}
\hfill
\includegraphics[scale=.5]{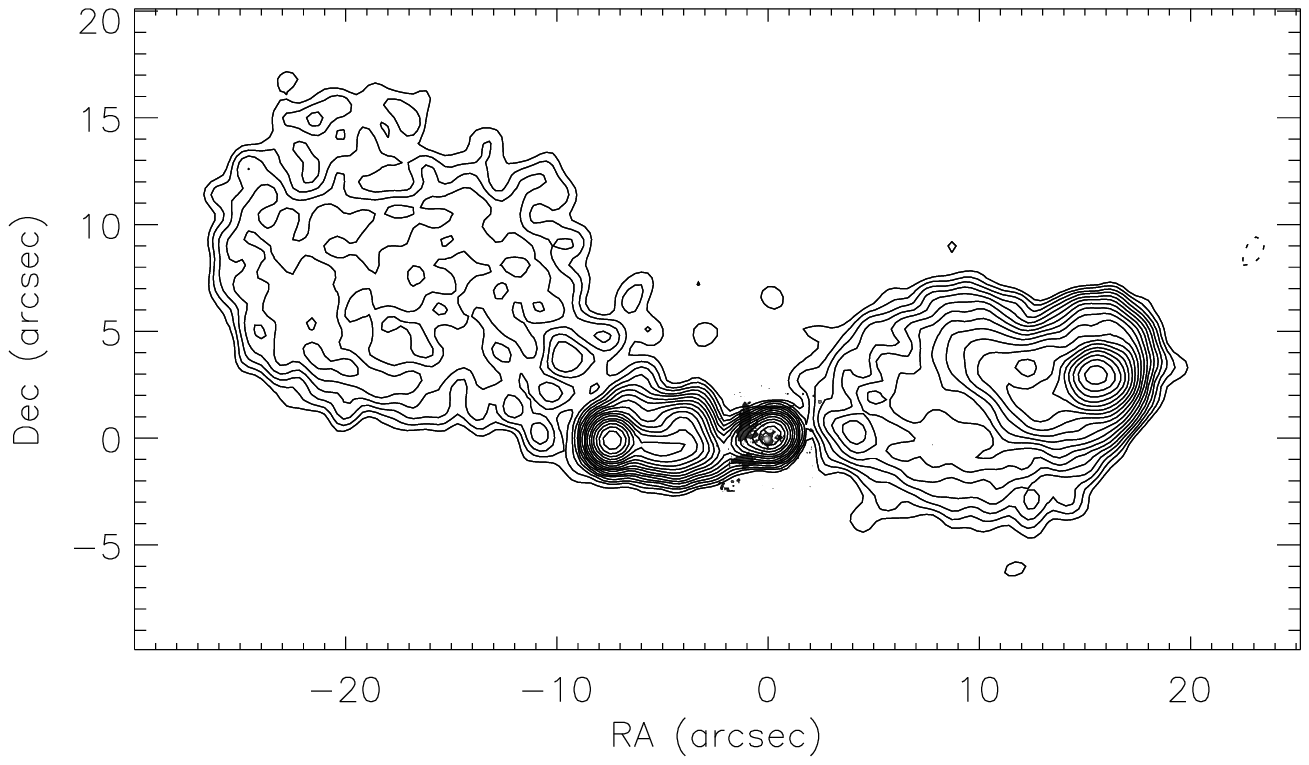}
\caption{3C249.1 (Radio/Optical Overlay) - Radio is shown in contours. Optical is shown in grey scale. Left: Closeup of the core. Right: View of the overall radio source. Radio levels are (0.674, 0.953, 1.348, 1.907, 2.696, 3.813, 5.393, 7.627, 10.786, 15.253, 21.571, 30.506, 43.142, 61.013, 86.285, 122.025, 172.570, 244.050) mJy. Optical levels are (184.0, 260.2, 368.0, 520.4, 735.9, 1040.7, 1471.9, 2081.5, 2943.7, 4163.0, 33304.3, 47099.4, 66608.6, 94198.7) * {$10^{-14}$} erg $s^{-1}$ $cm^{-2}$. Both images have the same levels. Radio map from J. P. Laing (Image obtained from the 3CRR Atlas (http://www.jb.man.ac.uk/atlas).).}
\label{3C249.1-overlay}
\end{figure}

\clearpage

\begin{figure}
\rotatebox{-90}{\includegraphics[scale=.8]{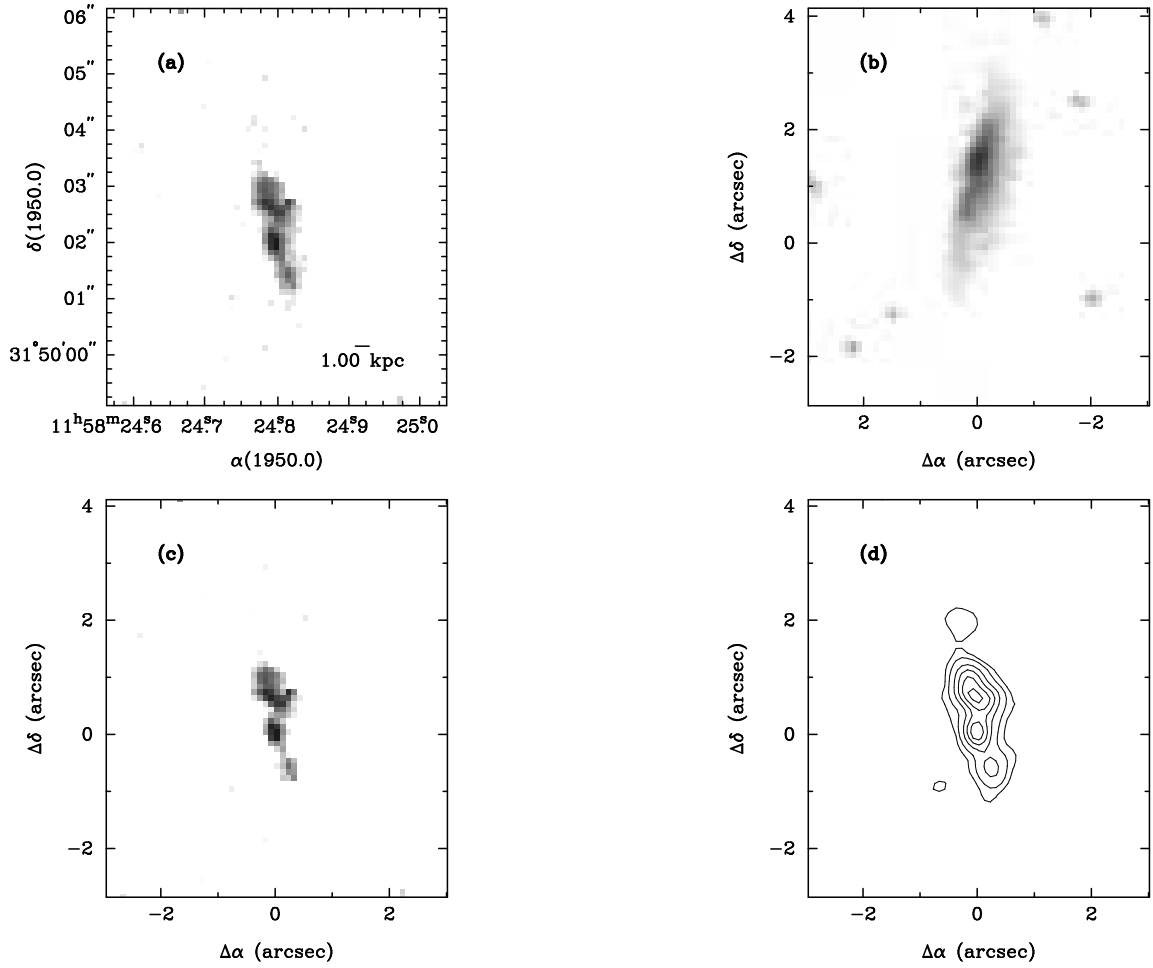}}
\caption{3C268.2 (Optical Montage) - Starting at the upper left, going clockwise: LRF Image; Broadband Image; Continuum Subtracted image [Contours]; Continuum Subtracted Image [Grey scale]}
\label{3C268.2-montage}
\end{figure}

\begin{figure}
\includegraphics[scale=.5]{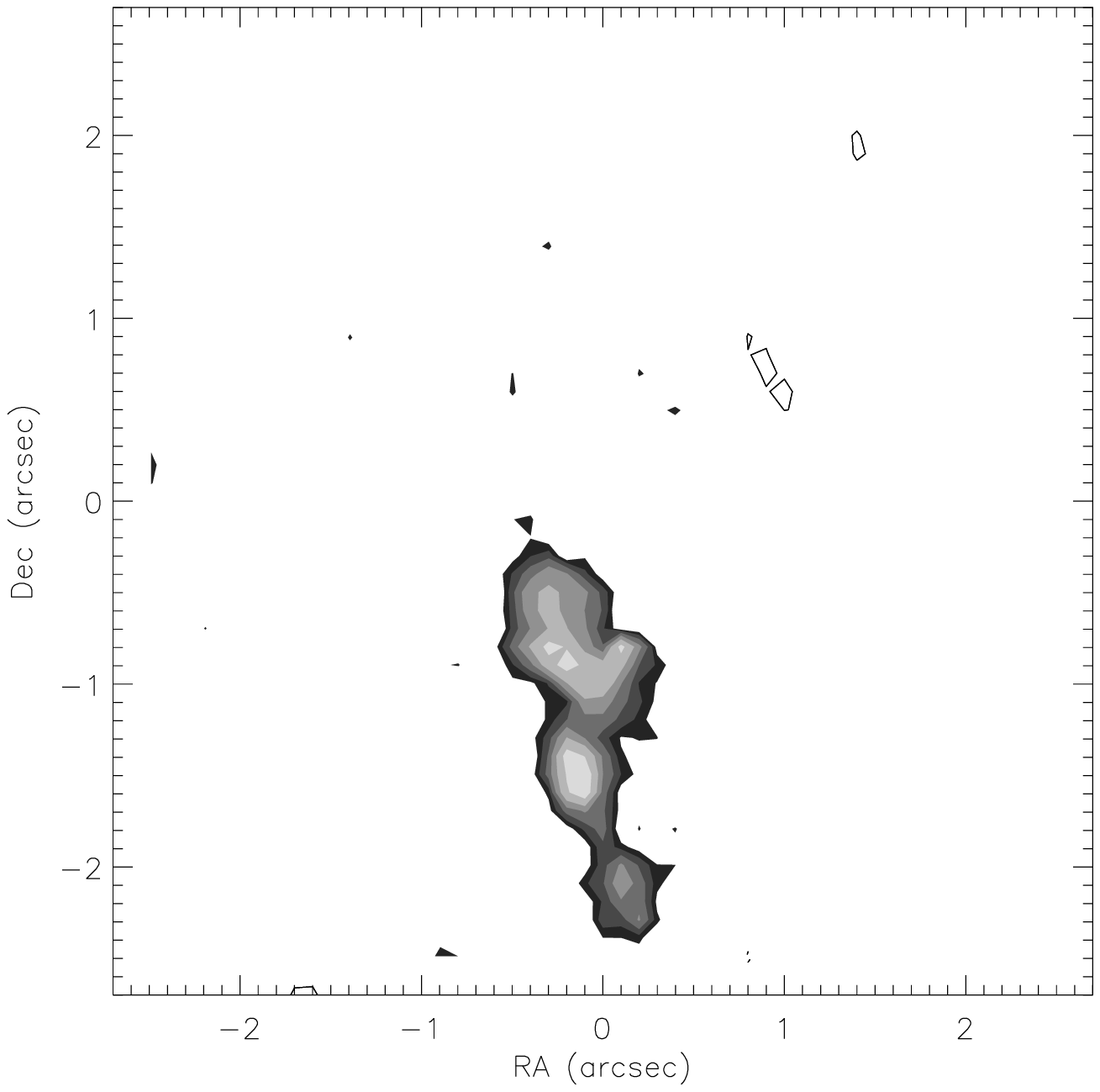}
\hfill
\includegraphics[scale=.5]{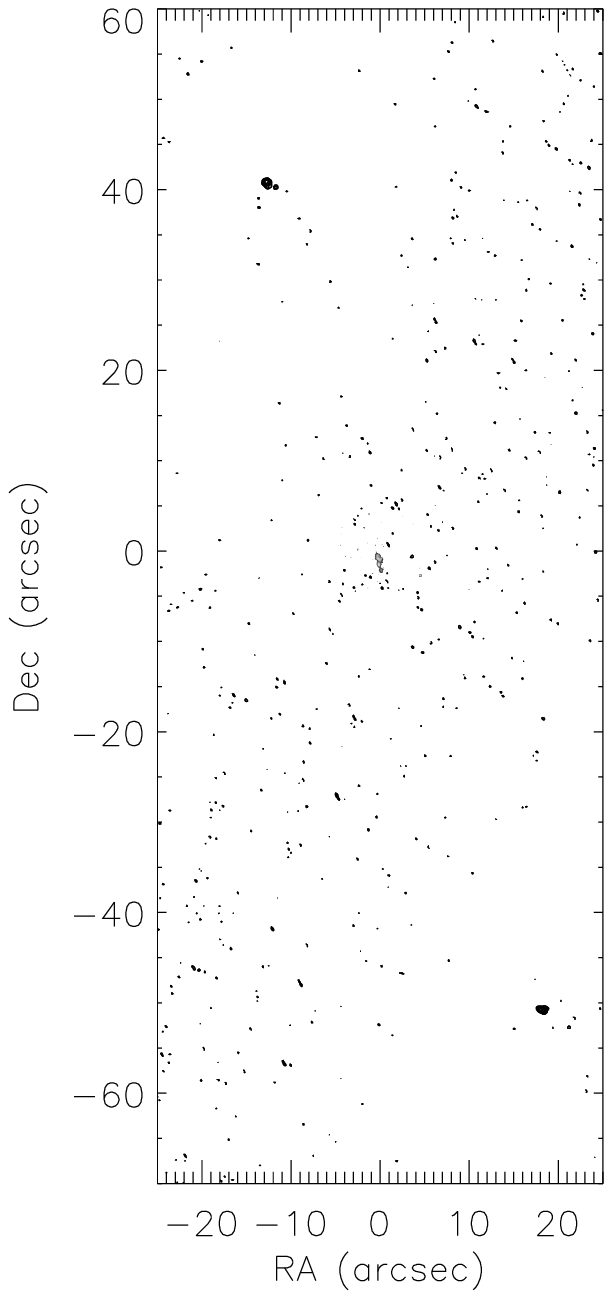}
\caption{3C268.2 (Radio/Optical Overlay) - Radio is shown in contours. Optical is shown in grey scale. Left: Closeup of the core. Right: View of the overall radio source. Radio levels are (1.212, 1.714, 2.424, 3.428, 4.848, 6.856, 9.696, 13.712) mJy. Optical levels are (62.8, 88.8, 125.6, 177.6, 251.2, 355.2, 502.3) * {$10^{-14}$} erg $s^{-1}$ $cm^{-2}$. Both images have the same levels. The radio map is from \citet{Neff95}.}
\label{3C268.2-overlay}
\end{figure}

\clearpage

\begin{figure}
\rotatebox{-90}{\includegraphics[scale=.8]{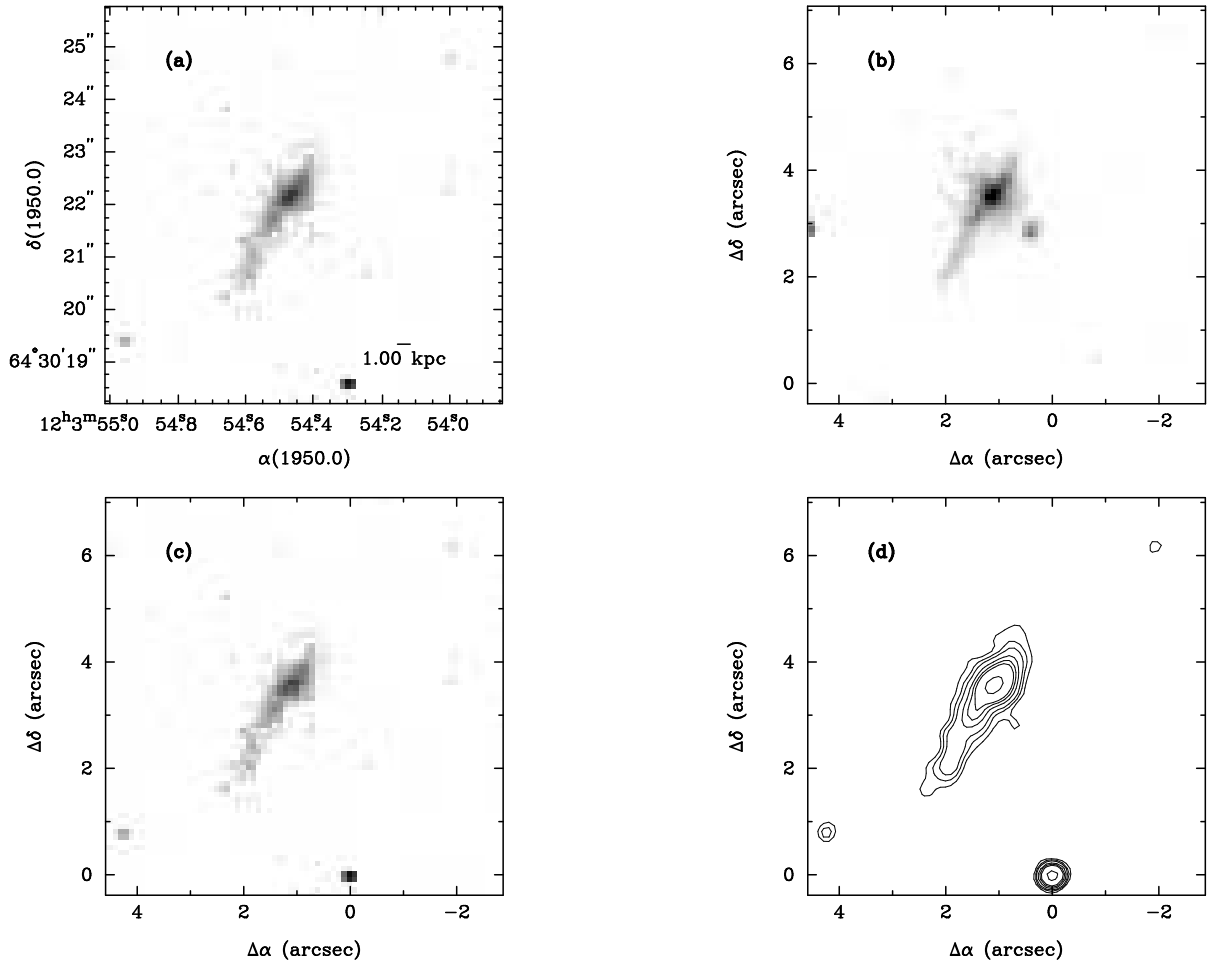}}
\caption{3C268.3 (Optical Montage) - Starting at the upper left, going clockwise: LRF Image; Broadband Image; Continuum Subtracted image [Contours]; Continuum Subtracted Image [Grey scale]}
\label{3C268.3-montage}
\end{figure}

\begin{figure}
\includegraphics[scale=.5]{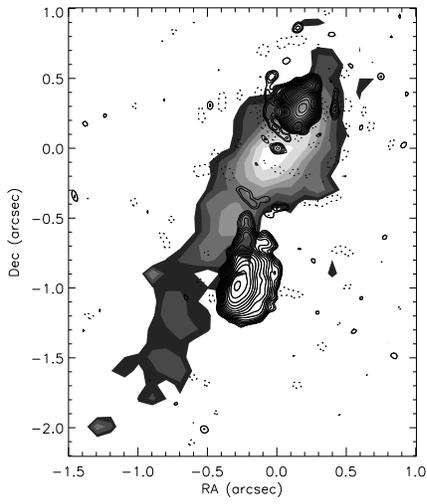}
\caption{3C268.3 (Radio/Optical Overlay) - Radio is shown in contours. Optical is shown in grey scale. Radio levels are (0.193, 0.272, 0.385, 0.545, 0.770, 1.090, 1.541, 2.179, 3.082, 4.358, 6.163, 8.716, 12.326, 17.432, 24.653, 34.864, 49.306, 69.729, 98.611, 139.457) mJy. Optical levels are (51.9, 73.4, 103.9, 146.9, 207.7, 293.8, 415.4) * {$10^{-14}$} erg $s^{-1}$ $cm^{-2}$. The radio map is from \citet{Ludke98}.}
\label{3C268.3-overlay}
\end{figure}

\clearpage

\begin{figure}
\includegraphics[scale=0.5]{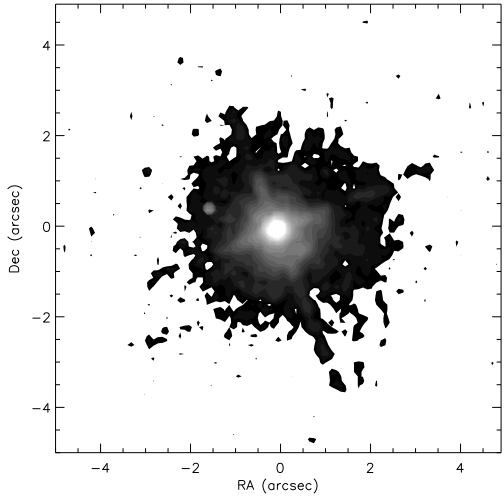}
\caption{3C273 (Narrow band Image). Optical levels are (110.8, 156.6, 221.5, 313.3, 443.0, 626.6, 886.1, 1253.1, 1772.2, 2506.3, 3544.4, 5012.5, 7088.8, 10025.0, 14177.5, 20050.1 28355.1, 40100.2, 56710.2, 80200.4) * {$10^{-14}$} erg $s^{-1}$ $cm^{-2}$.}
\label{3C273}
\end{figure}

\clearpage

\begin{figure}
\rotatebox{-90}{\includegraphics[scale=.8]{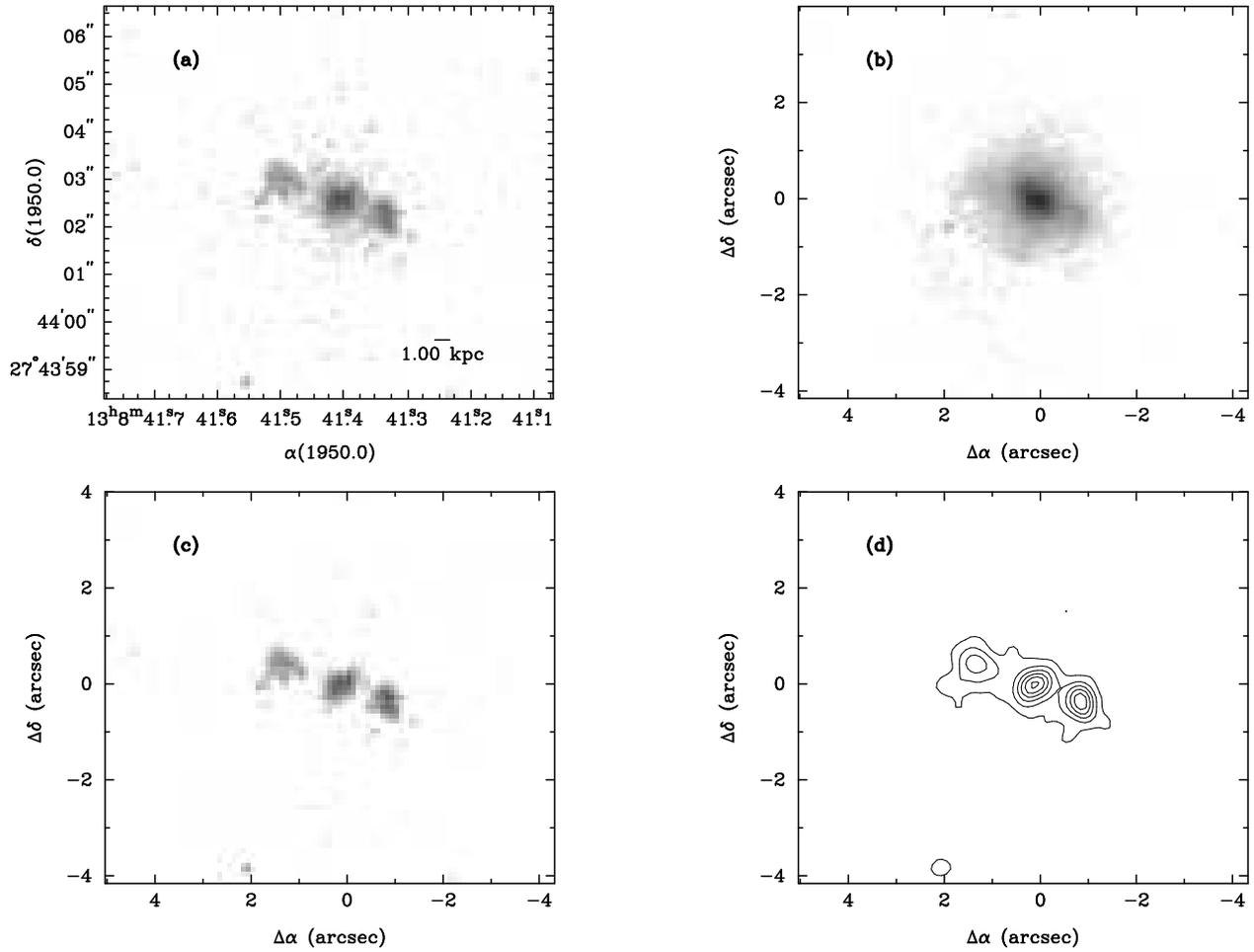}}
\caption{3C284 (Optical Montage) - Starting at the upper left, going clockwise: LRF Image; Broadband Image; Continuum Subtracted image [Contours]; Continuum Subtracted Image [Grey scale]}
\label{3C284-montage}
\end{figure}

\begin{figure}
\includegraphics[scale=.5]{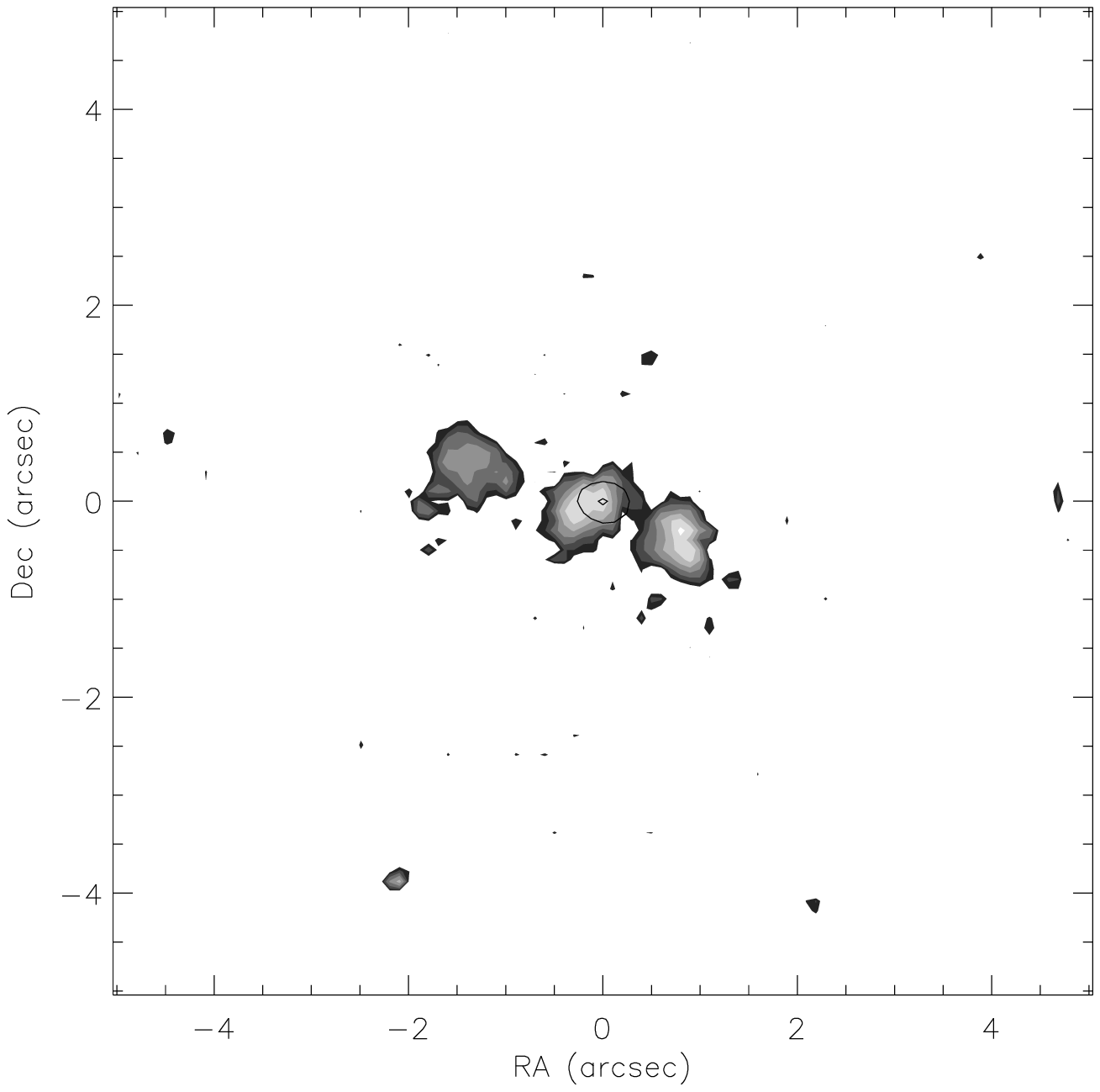}
\hfill
\includegraphics[scale=.5]{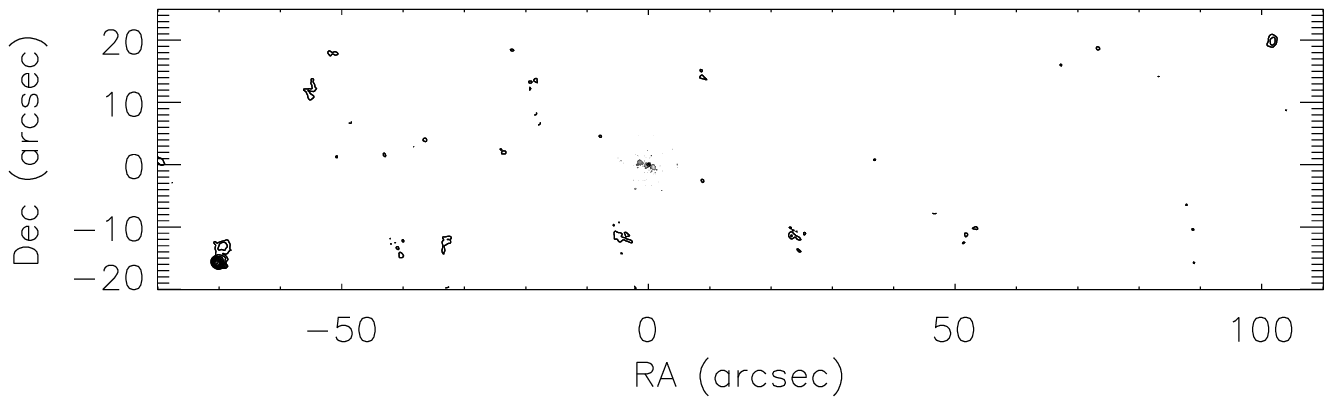}
\caption{3C284 (Radio/Optical Overlay) - Radio is shown in contours. Optical is shown in grey scale. Left: Closeup of the core. Right: View of the overall radio source. Radio levels are (1.796, 2.540, 3.593, 5.081, 7.186, 10.162, 14.371, 20.324, 28.742) mJy. Optical levels are (49.1, 69.5, 98.3, 139.0, 196.5, 277.9, 393.0) * {$10^{-14}$} erg $s^{-1}$ $cm^{-2}$. Both images have the same levels.}
\label{3C284-overlay}
\end{figure}

\clearpage

\begin{figure}
\rotatebox{-90}{\includegraphics[scale=.8]{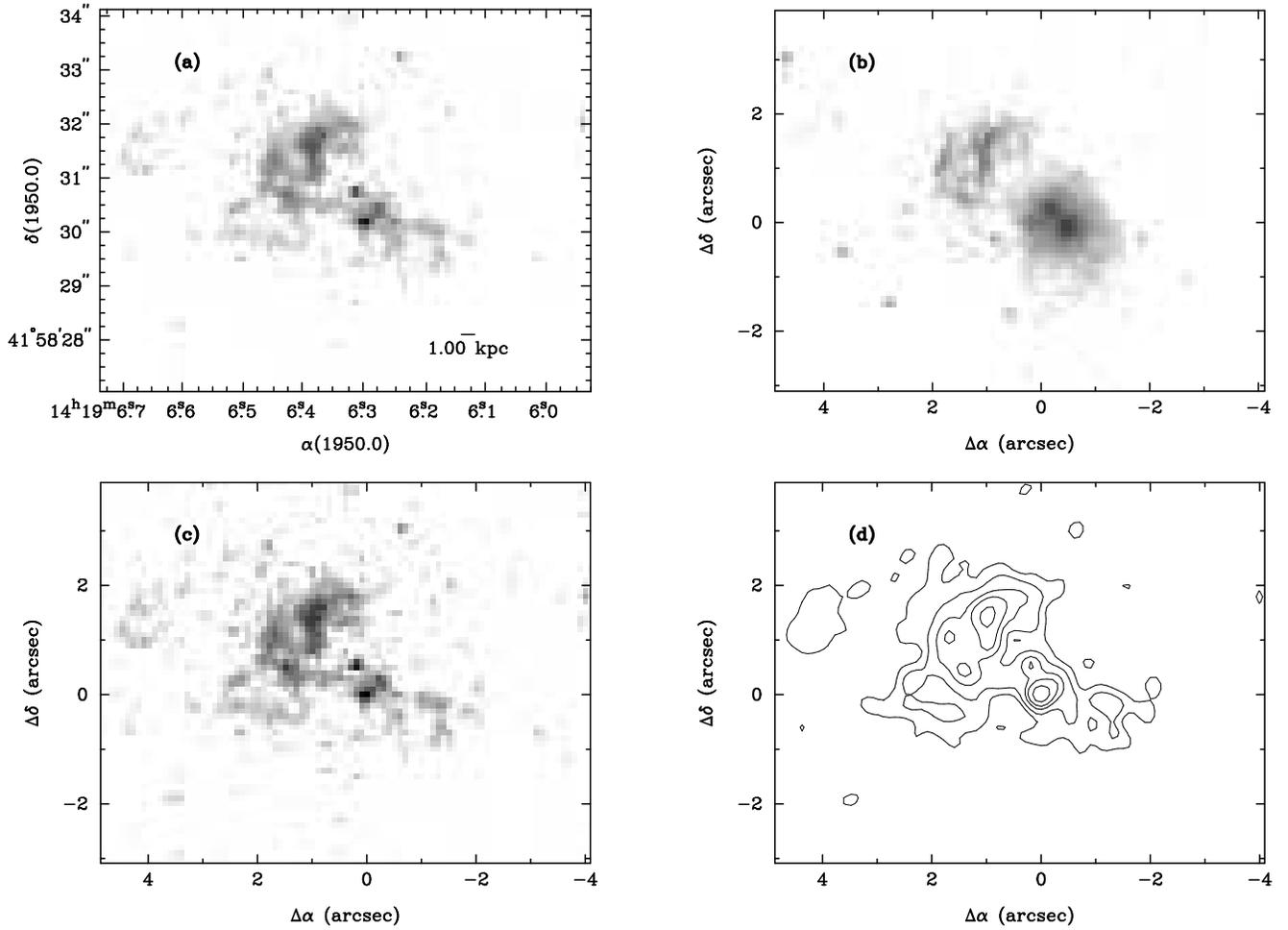}}
\caption{3C299(Optical Montage) - Starting at the upper left, going clockwise: LRF Image; Broadband Image; Continuum Subtracted image [Contours]; Continuum Subtracted Image [Grey scale]}
\label{3C299-montage}
\end{figure}

\begin{figure}
\includegraphics[scale=.5]{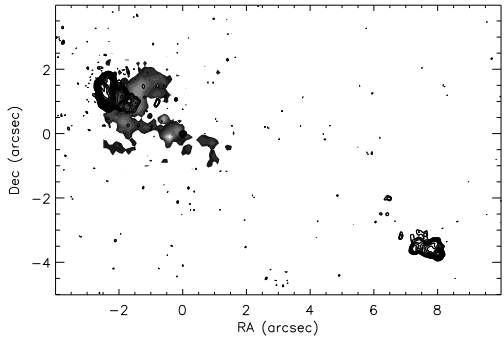}
\caption{3C299 (Radio/Optical Overlay) - Radio is shown in contours. Optical is shown in grey scale. Radio levels are (0.430, 0.608, 0.860, 1.216, 1.720, 2.432, 3.439, 4.864, 6.878, 9.728, 13.757, 19.455, 27.514, 38.910, 55.027, 77.820, 110.054, 155.640, 220.109, 311.281) mJy. Optical levels are (52.6, 74.3, 105.1, 148.7, 210.3, 297.4, 420.6, 594.8, 841.1, 1189.5, 1682.3) * {$10^{-14}$} erg $s^{-1}$ $cm^{-2}$. The radio map was provided by J. P. Laing (Image obtained from the 3CRR Atlas (http://www.jb.man.ac.uk/atlas).). }
\label{3C299-overlay}
\end{figure}

\clearpage

\begin{figure}
\rotatebox{-90}{\includegraphics[scale=.8]{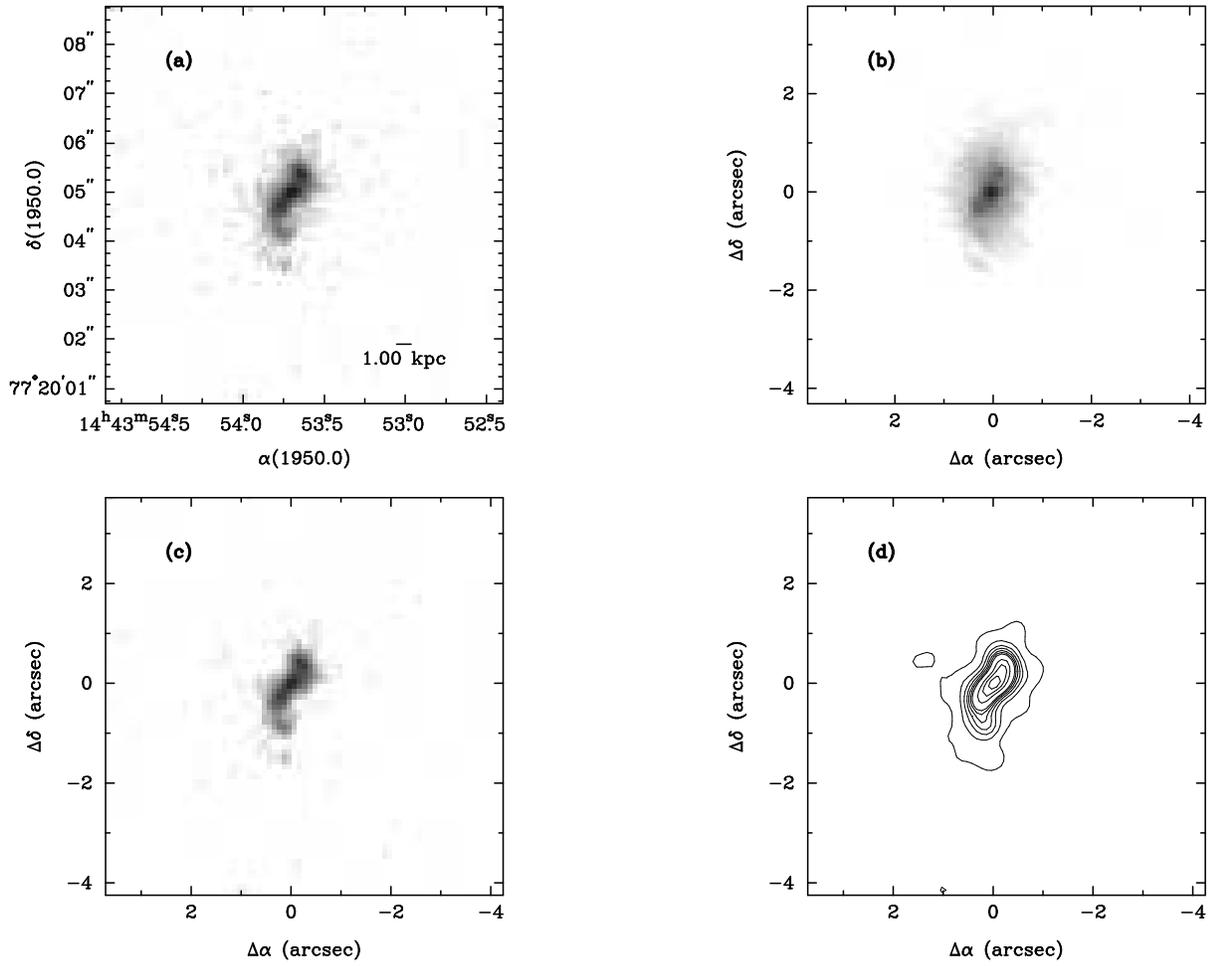}}
\caption{3C303.1 (Optical Montage) - Starting at the upper left, going clockwise: LRF Image; Broadband Image; Continuum Subtracted image [Contours]; Continuum Subtracted Image [Grey scale]}
\label{3C303.1-montage}
\end{figure}

\begin{figure}
\includegraphics[scale=.5]{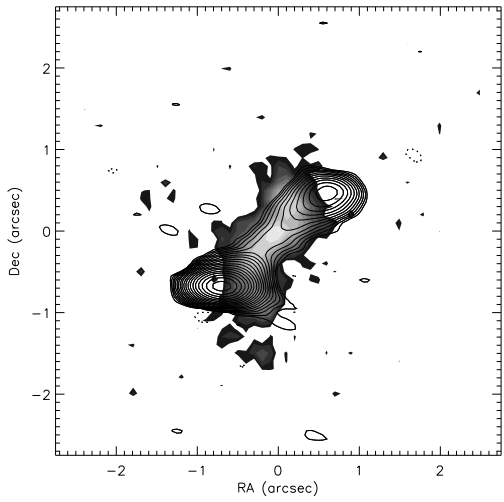}
\caption{3C303.1 (Radio/Optical Overlay) - Radio is shown in contours. Optical is shown in grey scale. Radio levels are (0.228, 0.322, 0.455, 0.644, 0.910, 1.288, 1.821, 2.575, 3.642, 5.150, 7.283, 10.300, 14.566, 20.600, 29.133, 41.200, 58.266, 82.400, 116.531) mJy. Optical levels are (46.3, 65.5, 92.5, 130.9, 185.2, 261.9, 370.3, 523.7, 740.7, 1047.4, 1481.3) * {$10^{-14}$} erg $s^{-1}$ $cm^{-2}$. }
\label{3C303.1-overlay}
\end{figure}

\clearpage

\begin{figure}
\rotatebox{-90}{\includegraphics[scale=.8]{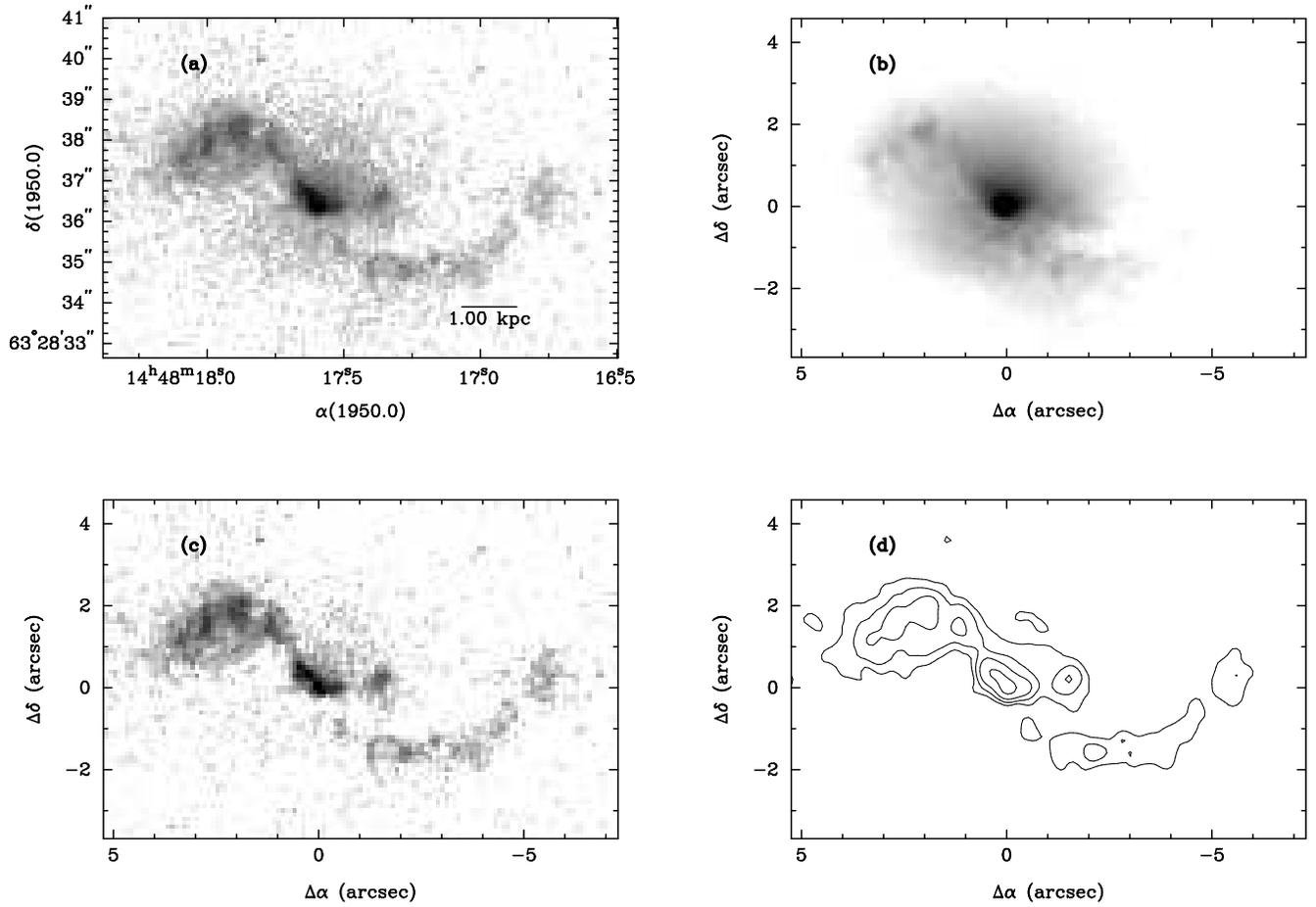}}
\caption{3C305 (Optical Montage) - Starting at the upper left, going clockwise: LRF Image; Broadband Image; Continuum Subtracted image [Contours]; Continuum Subtracted Image [Grey scale]}
\label{3C305-montage}
\end{figure}

\begin{figure}
\includegraphics[scale=.5]{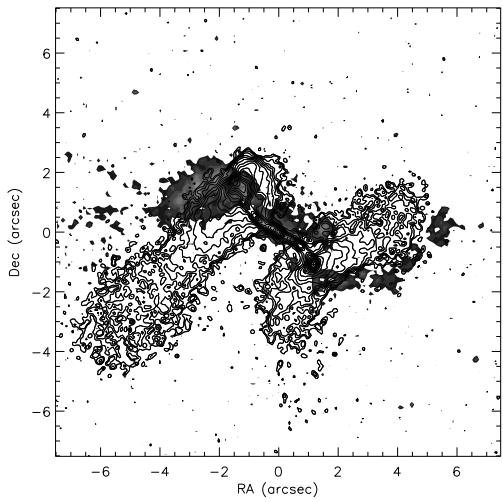}
\caption{3C305 (Radio/Optical Overlay) - Radio is shown in contours. Optical is shown in grey scale. Radio levels are (0.236, 0.334, 0.473, 0.669, 0.946, 1.337, 1.891, 2.675, 3.782, 5.349, 7.565, 10.698, 15.130, 21.396, 30.259, 42.793, 60.518, 85.586) mJy. Optical levels are (166.3, 235.2, 332.6, 470.4, 665.3, 940.9, 1330.6, 1881.7, 2661.2, 3763.5, 5322.4) * {$10^{-14}$} erg $s^{-1}$ $cm^{-2}$. The radio map was provided by J. P. Laing (Image obtained from the 3CRR Atlas (http://www.jb.man.ac.uk/atlas)).}
\label{3C305-overlay}
\end{figure}

\clearpage

\begin{figure}
\rotatebox{-90}{\includegraphics[scale=.8]{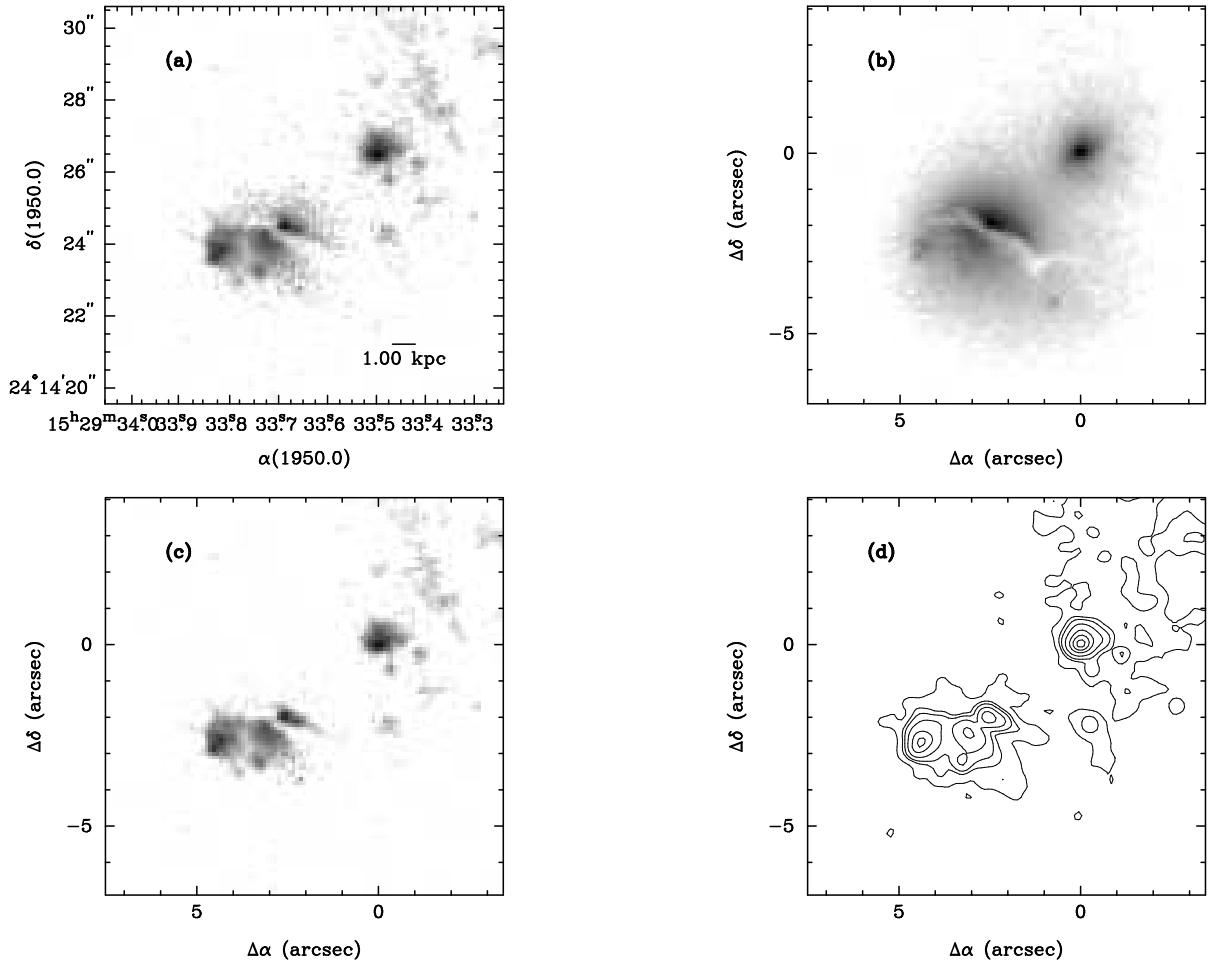}}
\caption{3C321 (Optical Montage) - Starting at the upper left, going clockwise: LRF Image; Broadband Image; Continuum Subtracted image [Contours]; Continuum Subtracted Image [Grey scale]}
\label{3C321-montage}
\end{figure}

\begin{figure}
\includegraphics[scale=.5]{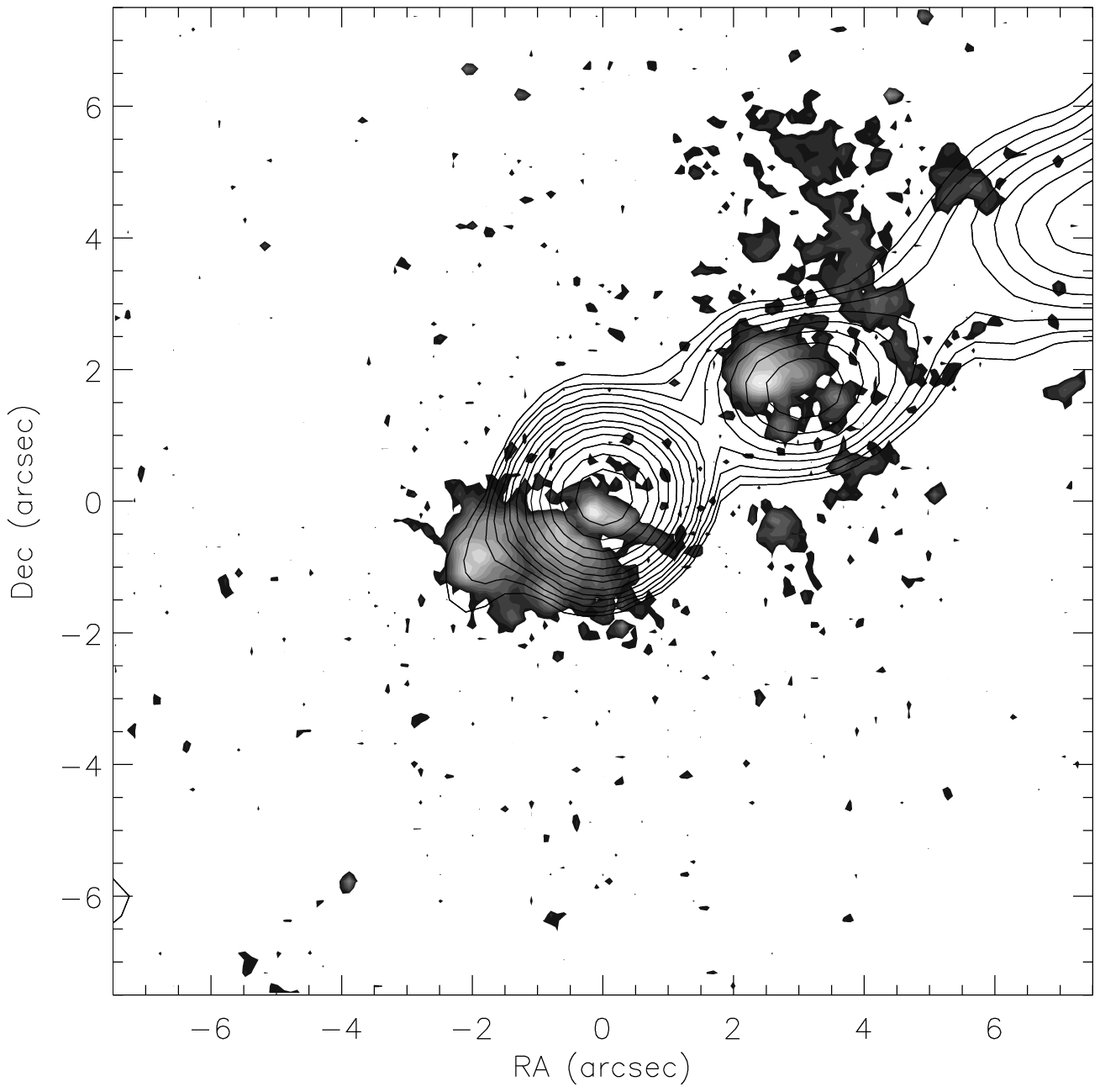}
\hfill
\includegraphics[scale=.5]{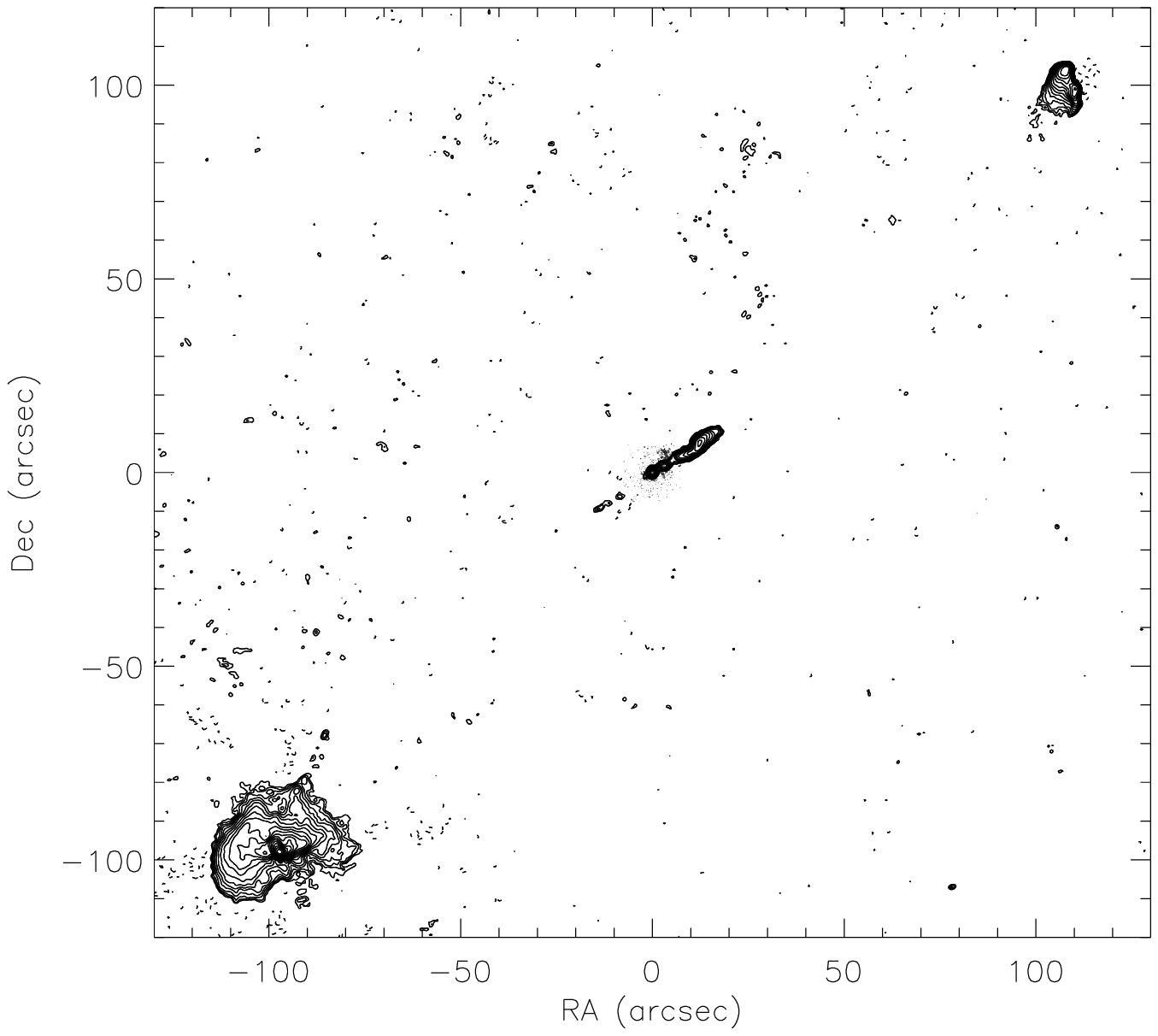}
\caption{3C321 (Radio/Optical Overlay) - Radio is shown in contours. Optical is shown in grey scale. Left: Closeup of the core. Right: view of the overall radio source. Radio levels are (0.368, 0.520, 0.736, 1.040, 1.471, 2.081, 2.942, 4.161, 5.885, 8.322, 11.770, 16.645, 23.539, 33.289, 47.078, 66.579, 94.157, 133.158, 188.314, 266.316) mJy. Optical levels are (81.0, 114.5, 162.0, 229.1, 232.9, 458.1, 647.9, 916.3, 1295.8, 1832.6, 2591.6, 3665.1) * {$10^{-14}$} erg $s^{-1}$ $cm^{-2}$. Both images have the same levels. The radio map was provided courtesy of David Floyd (private communication).}
\label{3C321-overlay}
\end{figure}

\clearpage

\begin{figure}
\includegraphics[scale=.5]{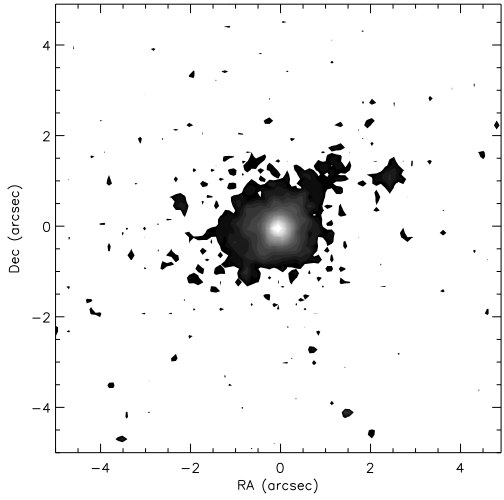}
\caption{3C323.1 (Narrow band Image) - Optical levels are (94.1, 133.0, 188.1, 266.1, 376.3, 532.2, 752.6, 1064.3, 1505.2, 2128.6, 3010.4, 4257.3, 6020.7, 8514.6, 12041.4, 17029.2, 24082.9, 34058.3, 48165.8, 68116.7) * {$10^{-14}$} erg $s^{-1}$ $cm^{-2}$.}
\label{3C323.1}
\end{figure}

\clearpage

\begin{figure}
\rotatebox{-90}{\includegraphics[scale=.8]{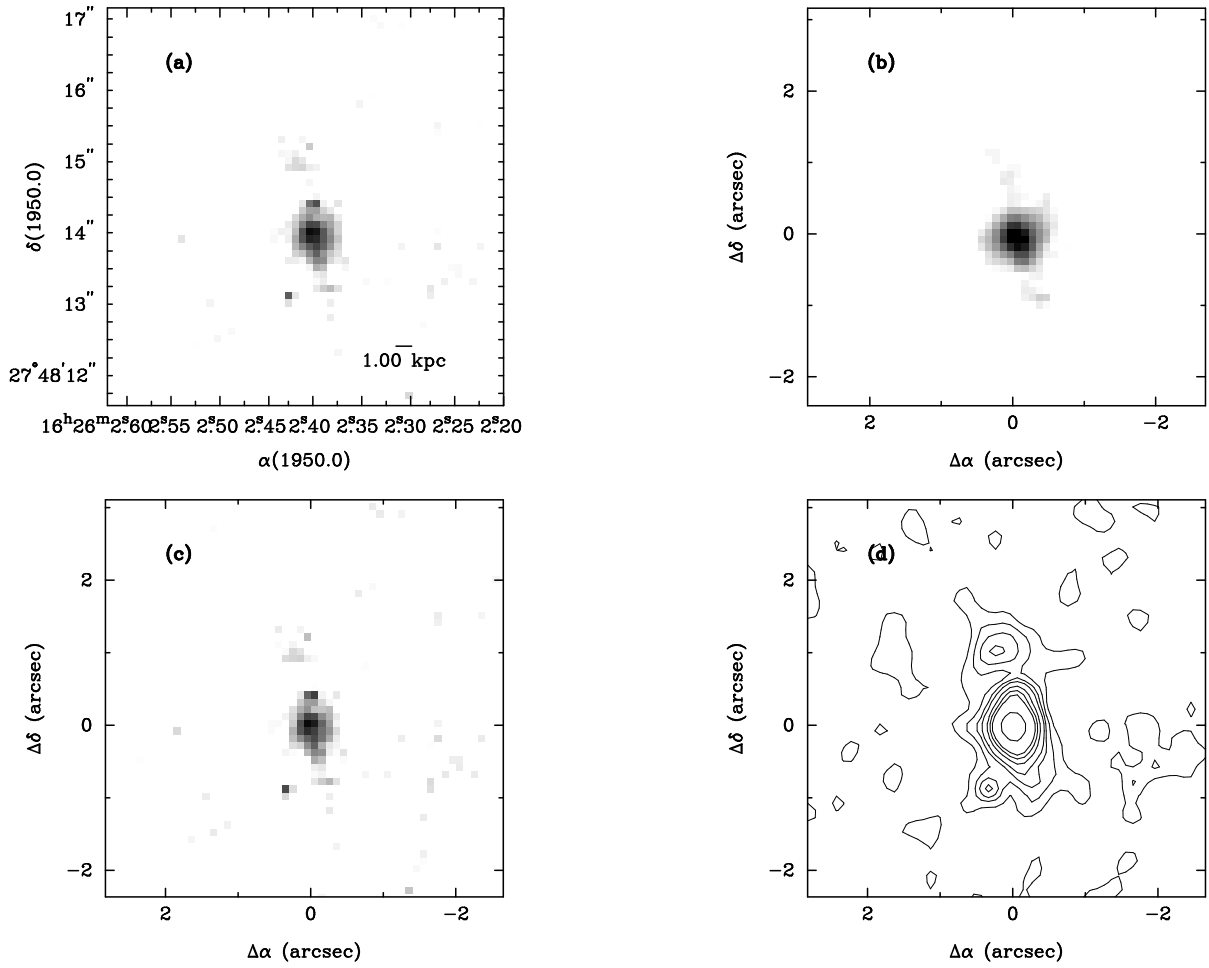}}
\caption{3C341 (Optical Montage) - Starting at the upper left, going clockwise: LRF Image; Broadband Image; Continuum Subtracted image [Contours]; Continuum Subtracted Image [Grey scale]}
\label{3C341-montage}
\end{figure}

\begin{figure}
\includegraphics[scale=.5]{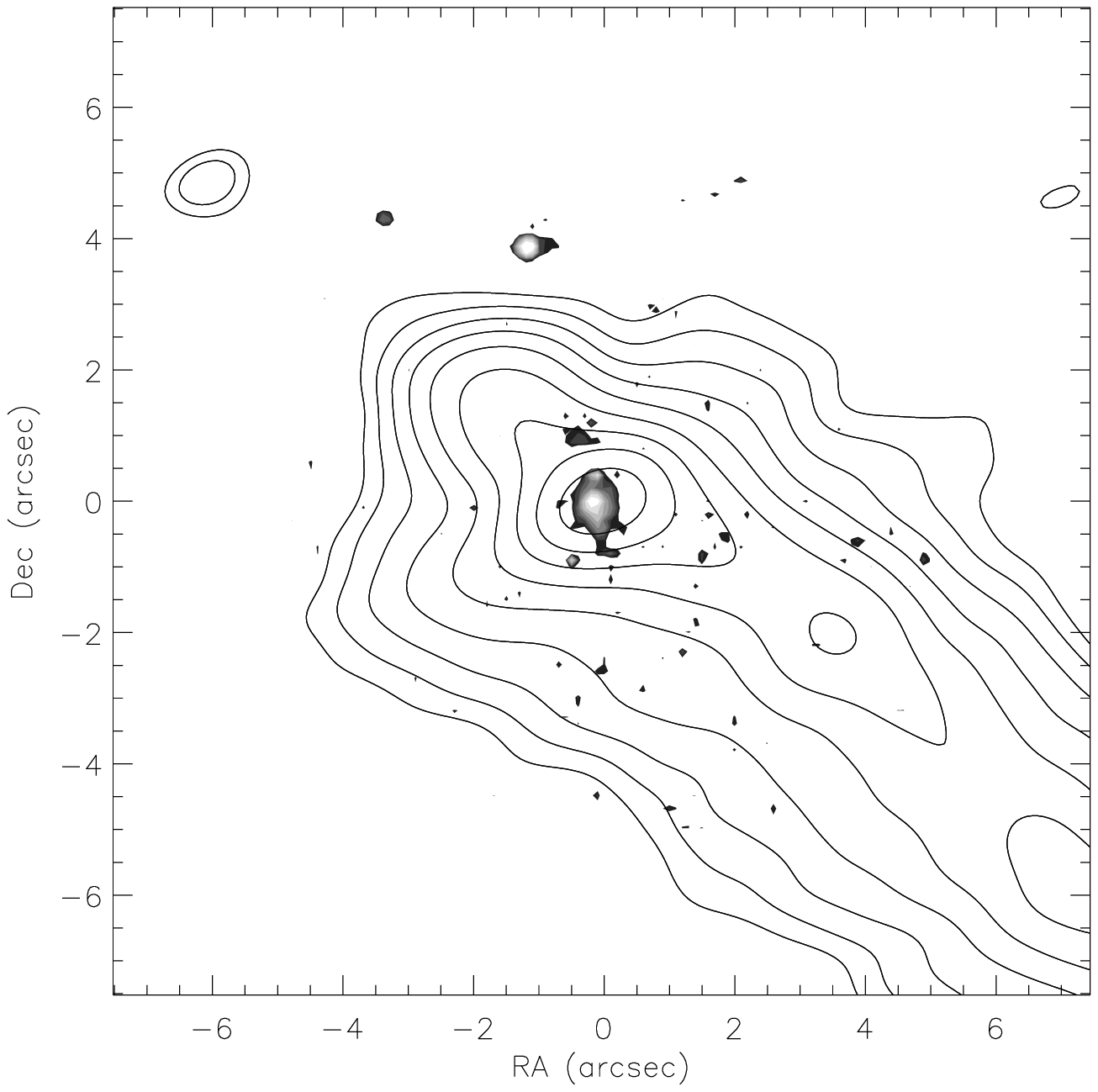}
\hfill
\includegraphics[scale=.5]{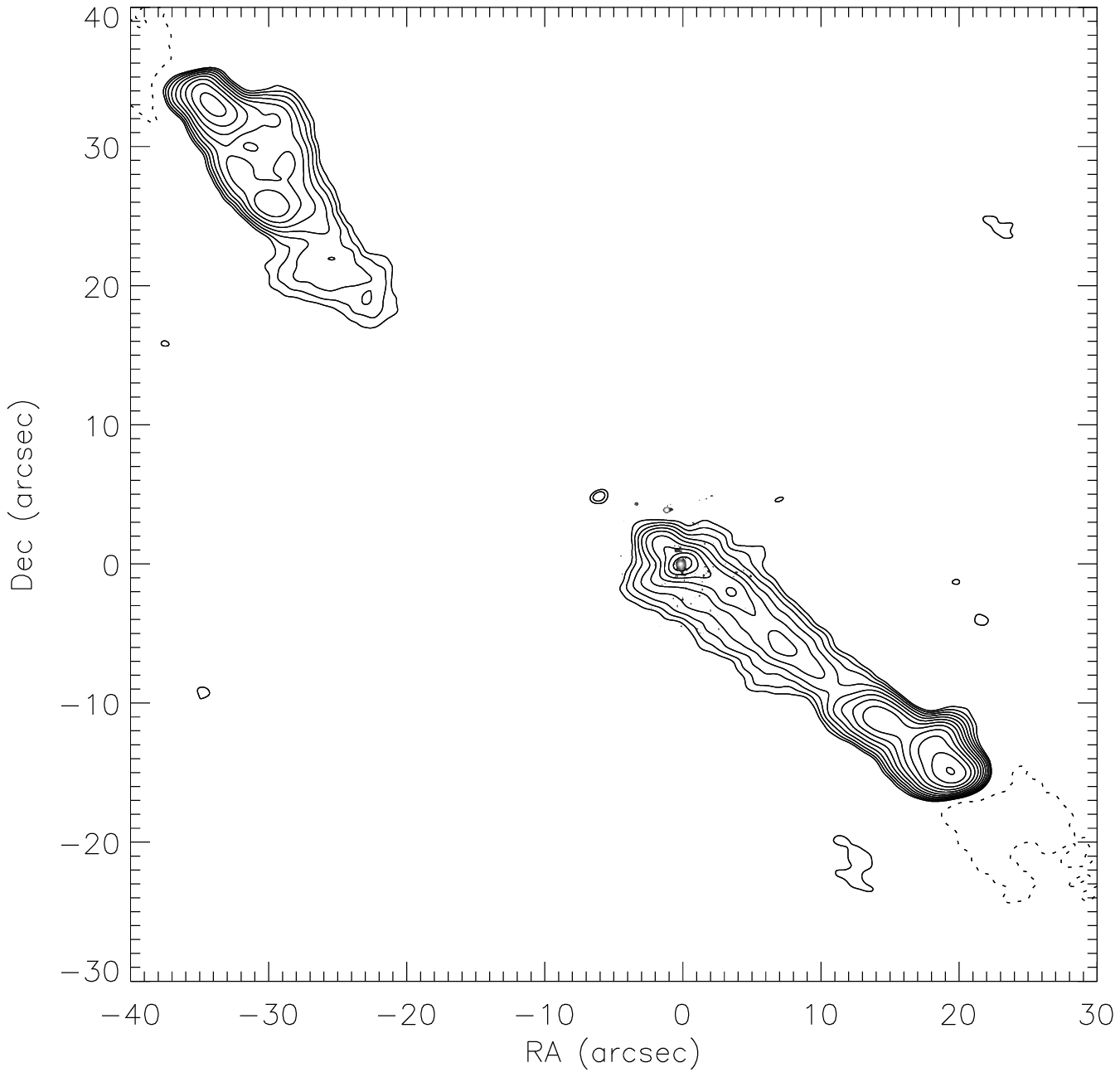}
\caption{3C341 (Radio/Optical Overlay) - Radio is shown in contours. Optical is shown in grey scale. Left: Closeup of the core. Right: View of the overall radio source. Radio levels are (0.288, 0.407, 0.576, 0.815, 1.152, 1.629, 2.304, 3.258, 4.608, 6.517, 9.216, 13.033) mJy. Optical levels are (58.0, 82.0, 115.9, 163.9, 231.9, 327.9, 463.7, 655.8) * {$10^{-14}$} erg $s^{-1}$ $cm^{-2}$. Both images have the same levels. }
\label{3C341-overlay}
\end{figure}

\clearpage

\begin{figure}
\rotatebox{-90}{\includegraphics[scale=.8]{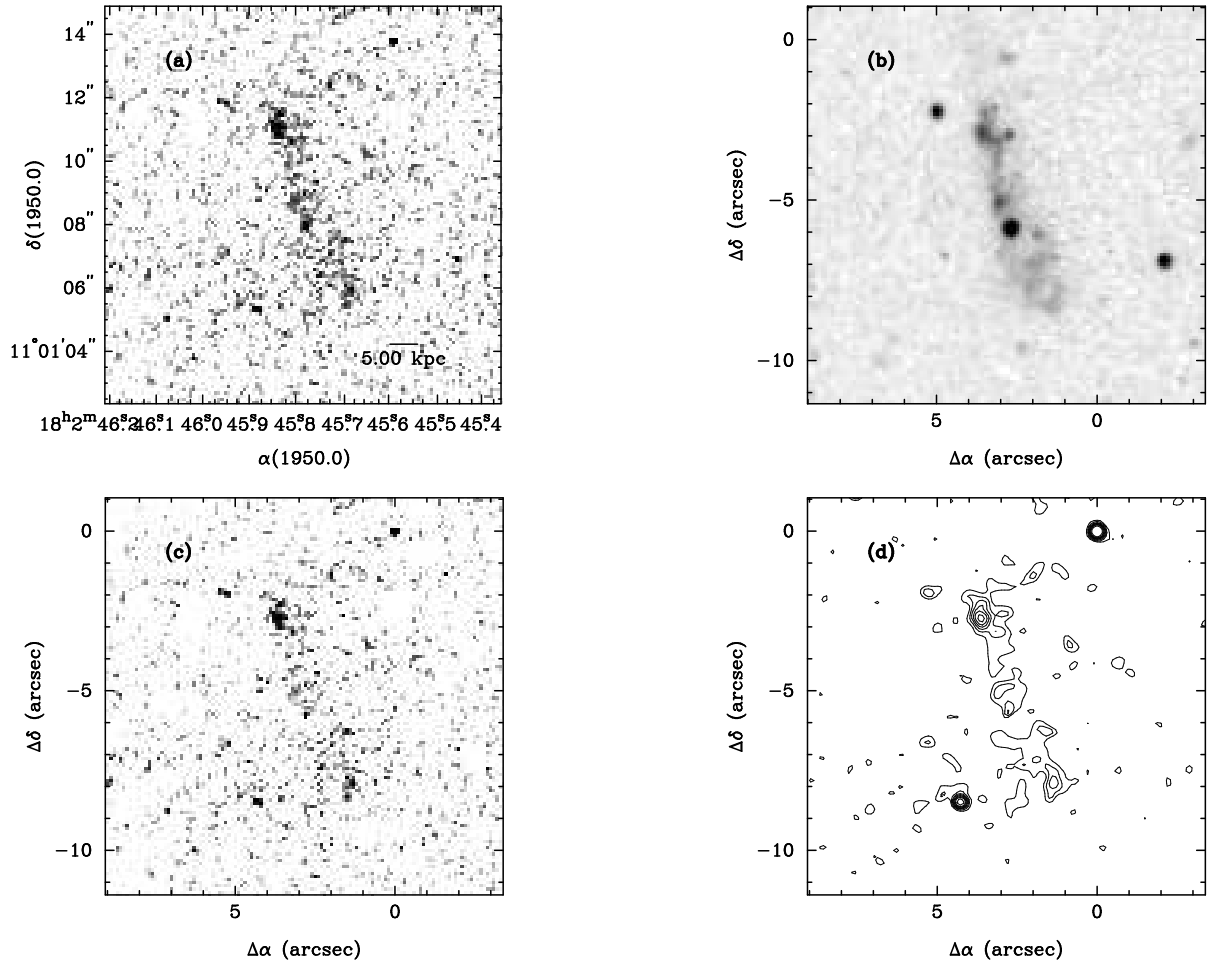}}
\caption{3C368 (Optical Montage) - Starting at the upper left, going clockwise: LRF Image; Broadband Image; Continuum Subtracted image [Contours]; Continuum Subtracted Image [Grey scale]}
\label{3C368-montage}
\end{figure}

\begin{figure}
\includegraphics[scale=.5]{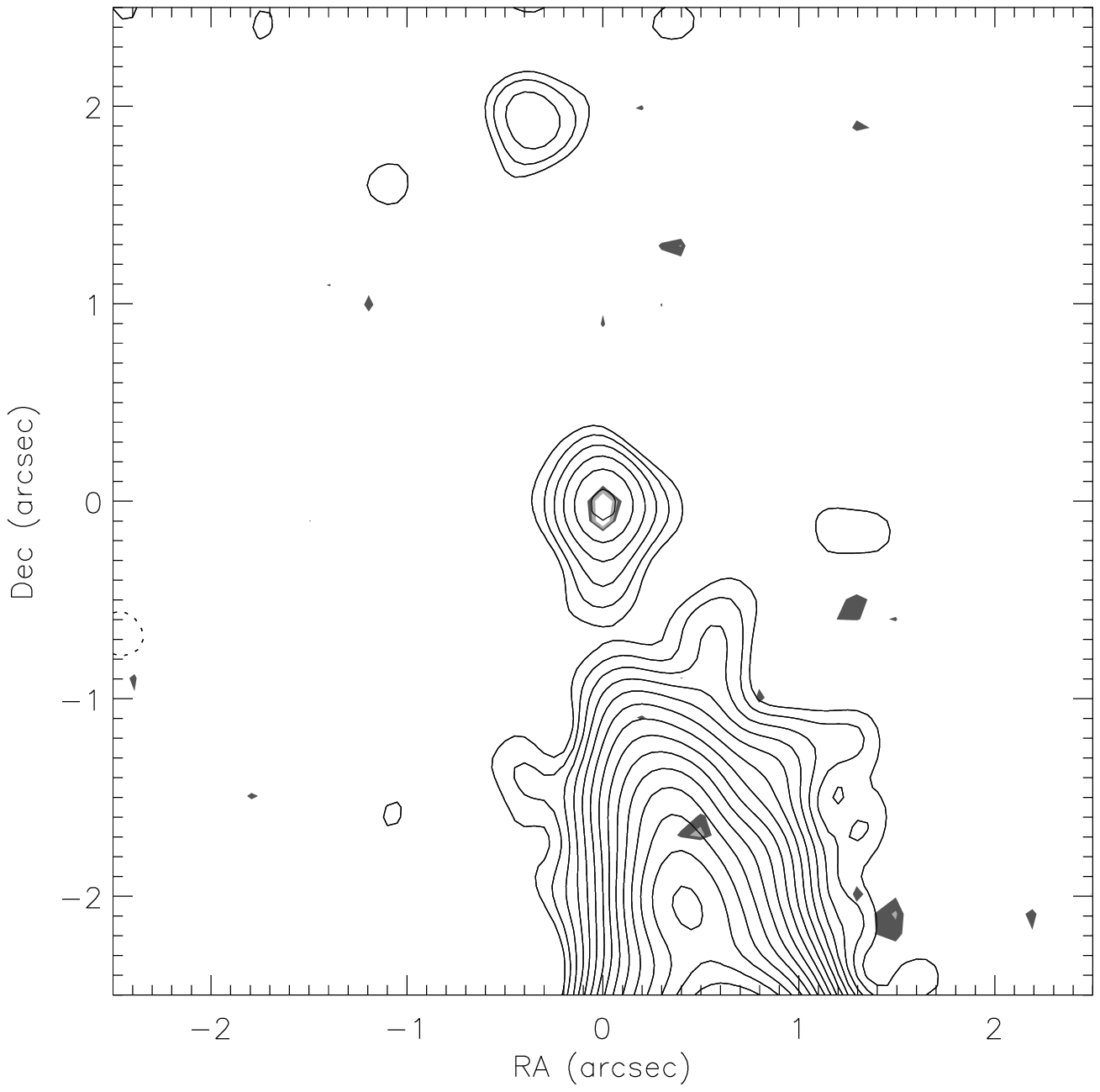}
\includegraphics[scale=.5]{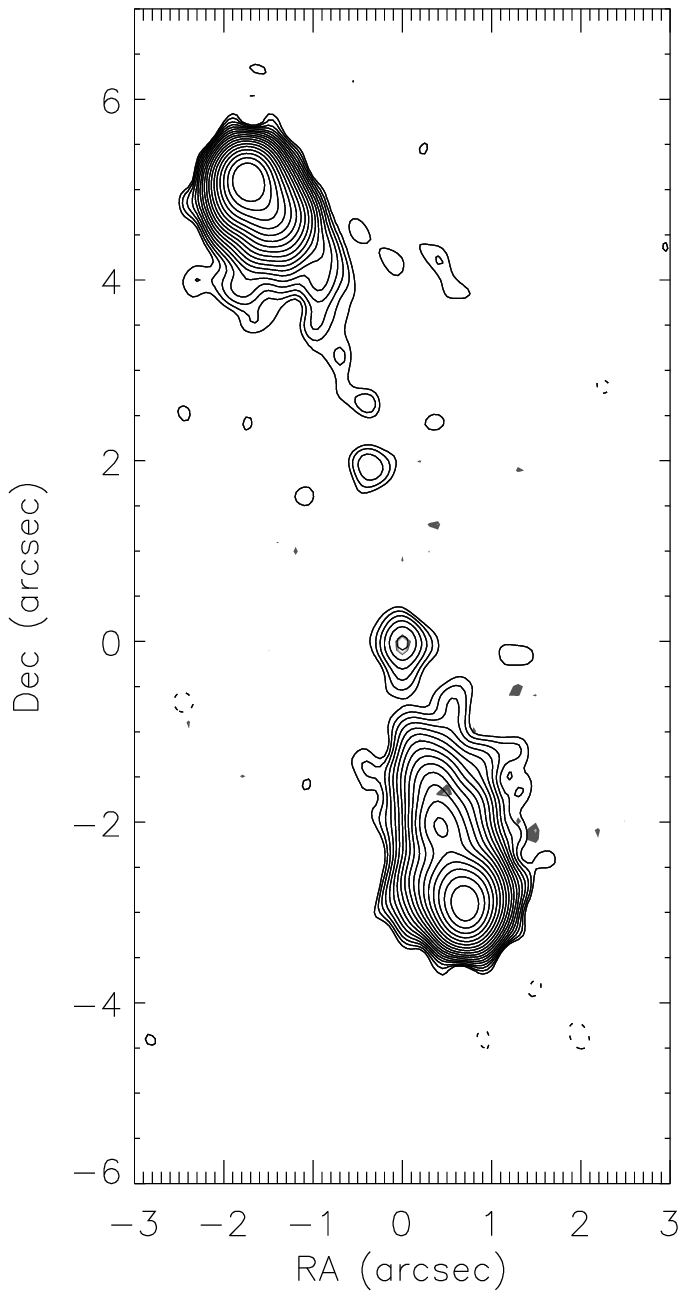}
\caption{3C368 (Radio/Optical Overlay) - Radio is shown in contours. Optical is shown in grey scale. Left: Closeup of the core. Right: View of the overall radio source. Radio levels are (0.098, 0.138, 0.195, 0.276, 0.391, 0.553, 0.782, 1.105, 1.563, 2.211, 3.126, 4.421, 6.253, 8.843, 12.506, 17.686, 25.011, 35.371) mJy. Optical levels are (68.7, 97.2, 137.5) * {$10^{-14}$} erg $s^{-1}$ $cm^{-2}$. Both images have the same levels. }
\label{3C368-overlay}
\end{figure}

\clearpage

\begin{figure}
\rotatebox{-90}{\includegraphics[scale=.8]{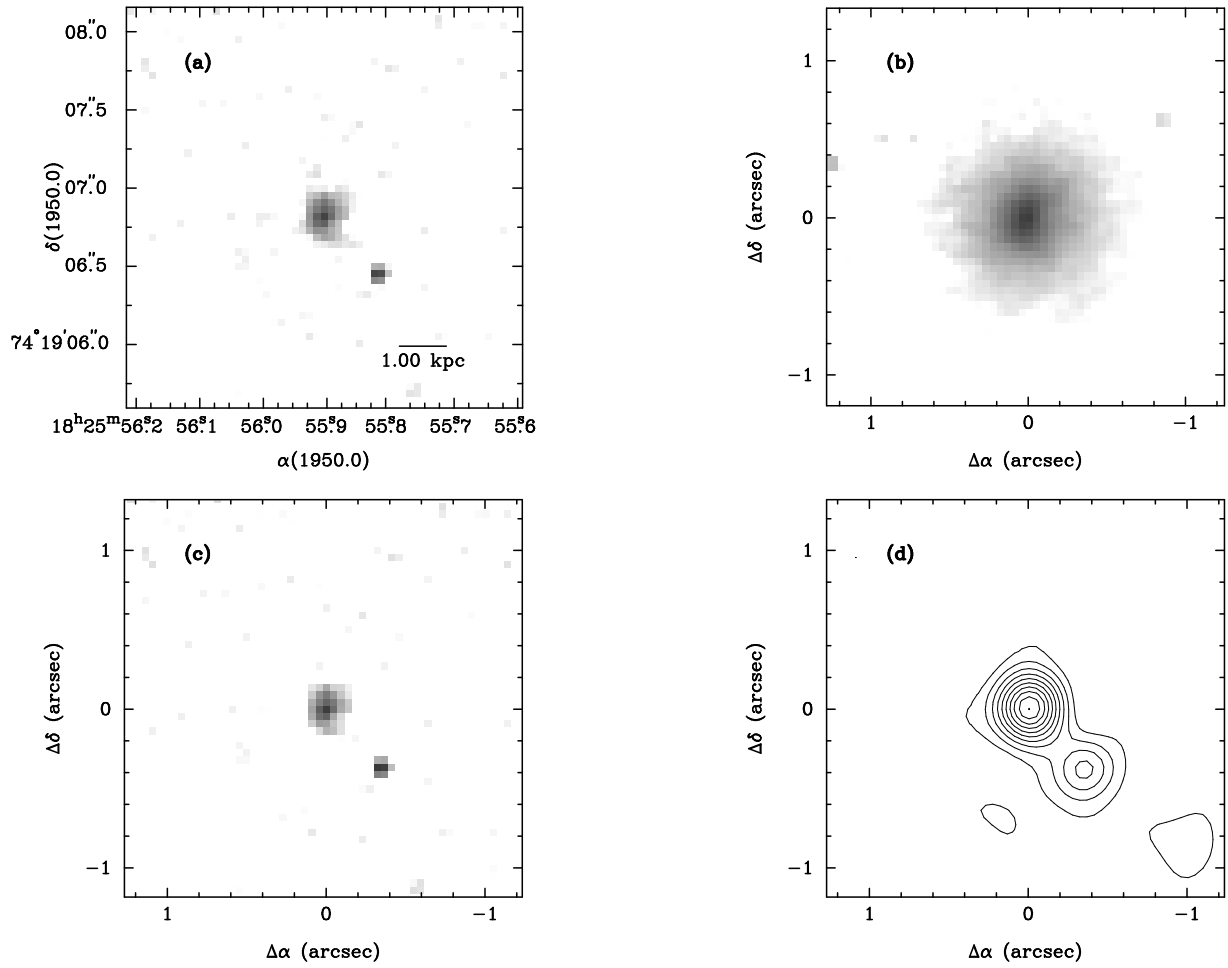}}
\caption{3C379.1 (Optical Montage) - Starting at the upper left, going clockwise: LRF Image; Broadband Image; Continuum Subtracted image [Contours]; Continuum Subtracted Image [Grey scale]}
\label{3C379.1-montage}
\end{figure}

\begin{figure}
\includegraphics[scale=.5]{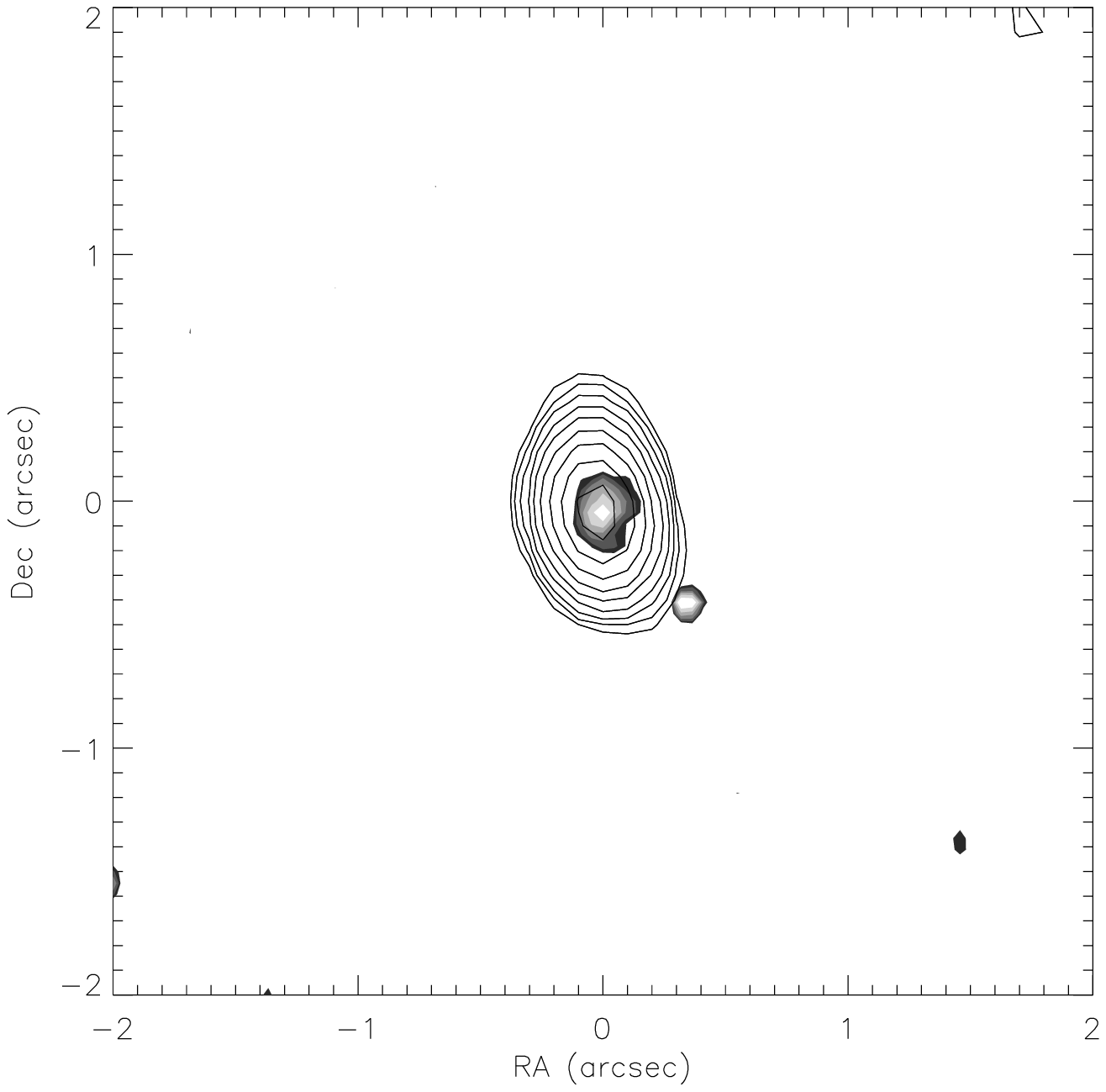}
\hfill
\includegraphics[scale=.5]{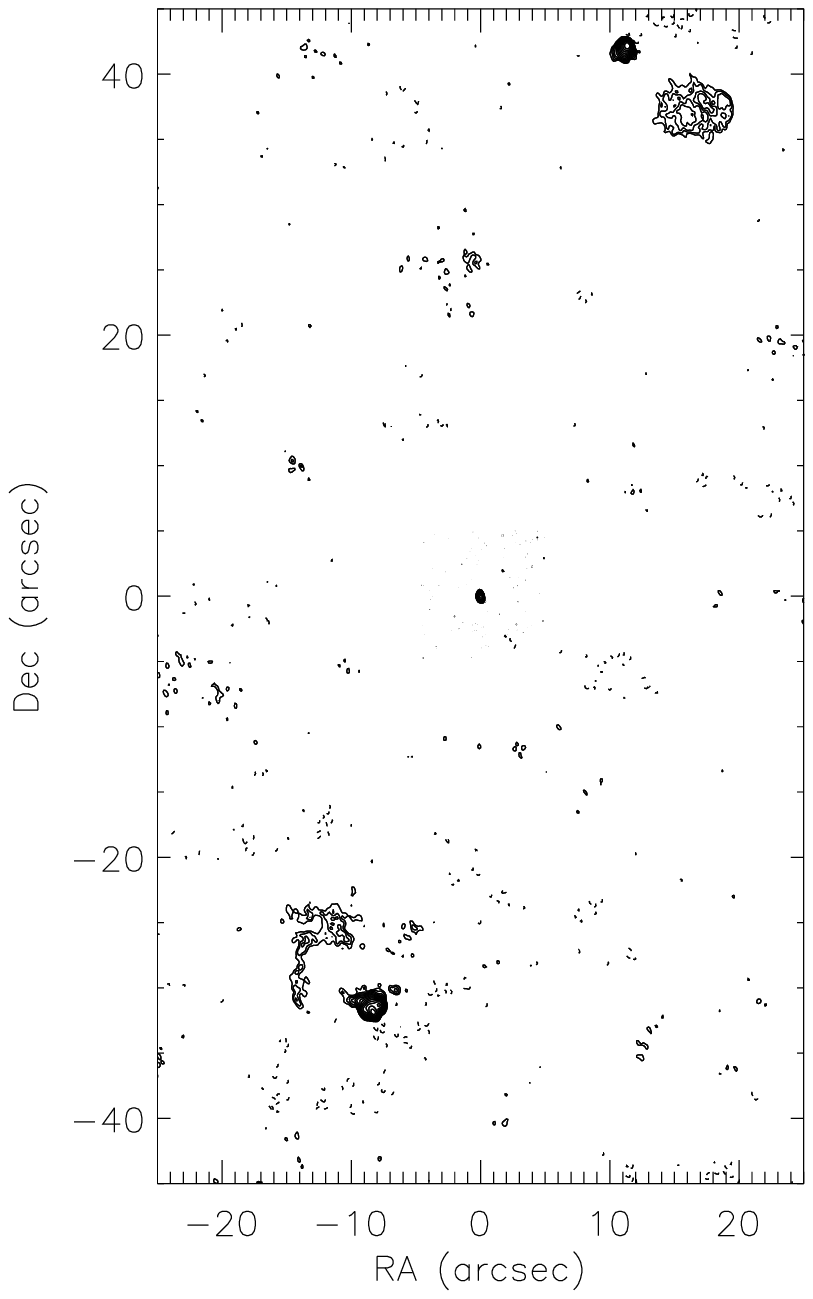}
\caption{3C379.1 (Radio/Optical Overlay) - Radio is shown in contours. Optical is shown in grey scale. Left: Closeup of the core. Right: View of the overall radio source. Radio levels are (0.225, 0.318, 0.449, 0.635, 0.898, 1.271, 1.797, 2.541, 3.594, 5.082, 7.187, 10.164, 14.374, 20.328) mJy. Optical levels are (186.8, 264.2, 373.7, 528.4, 747.3, 1056.9) * {$10^{-14}$} erg $s^{-1}$ $cm^{-2}$. Both images have the same levels. }
\label{3C379.1-overlay}
\end{figure}

\clearpage

\begin{figure}
\rotatebox{-90}{\includegraphics[scale=.8]{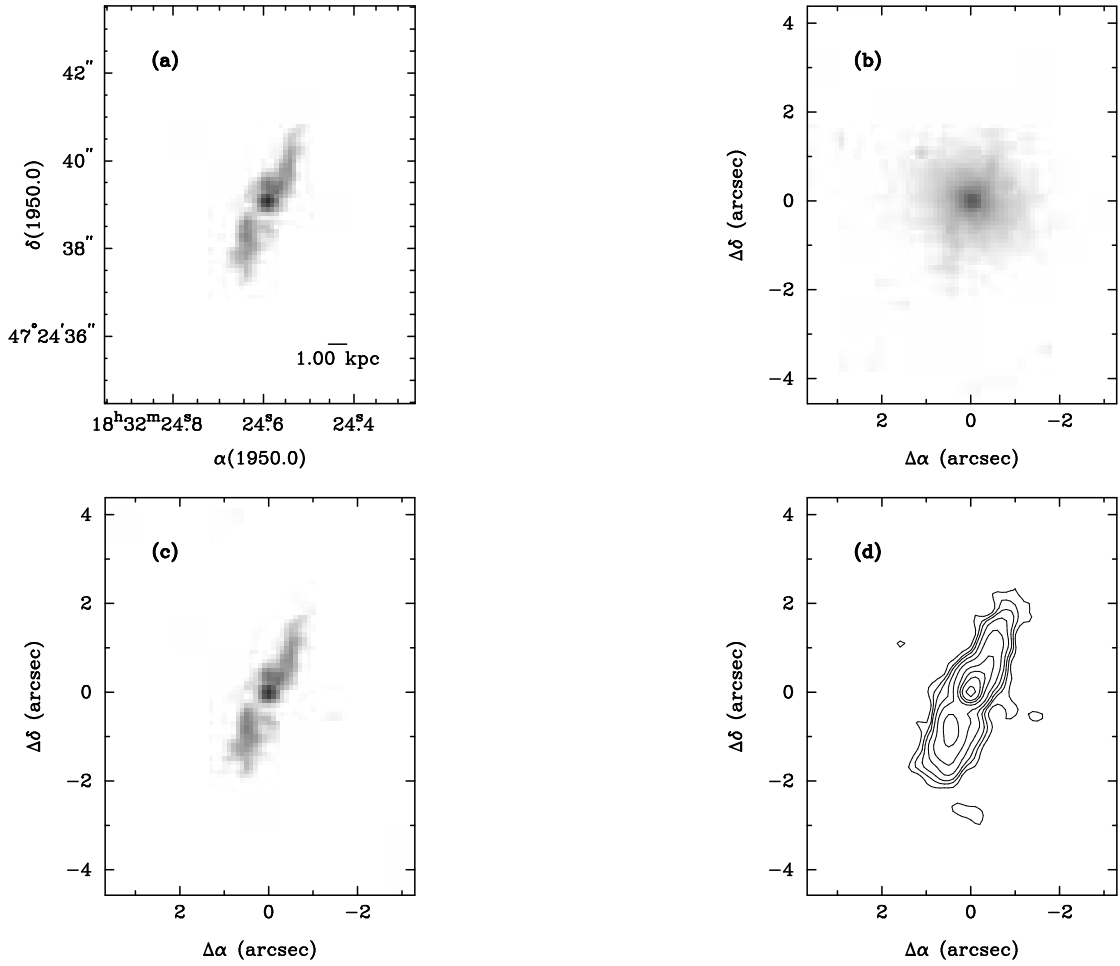}}
\caption{3C381 (Optical Montage) - Starting at the upper left, going clockwise: LRF Image; Broadband Image; Continuum Subtracted image [Contours]; Continuum Subtracted Image [Grey scale]}
\label{3C381-montage}
\end{figure}

\begin{figure}
\includegraphics[scale=.5]{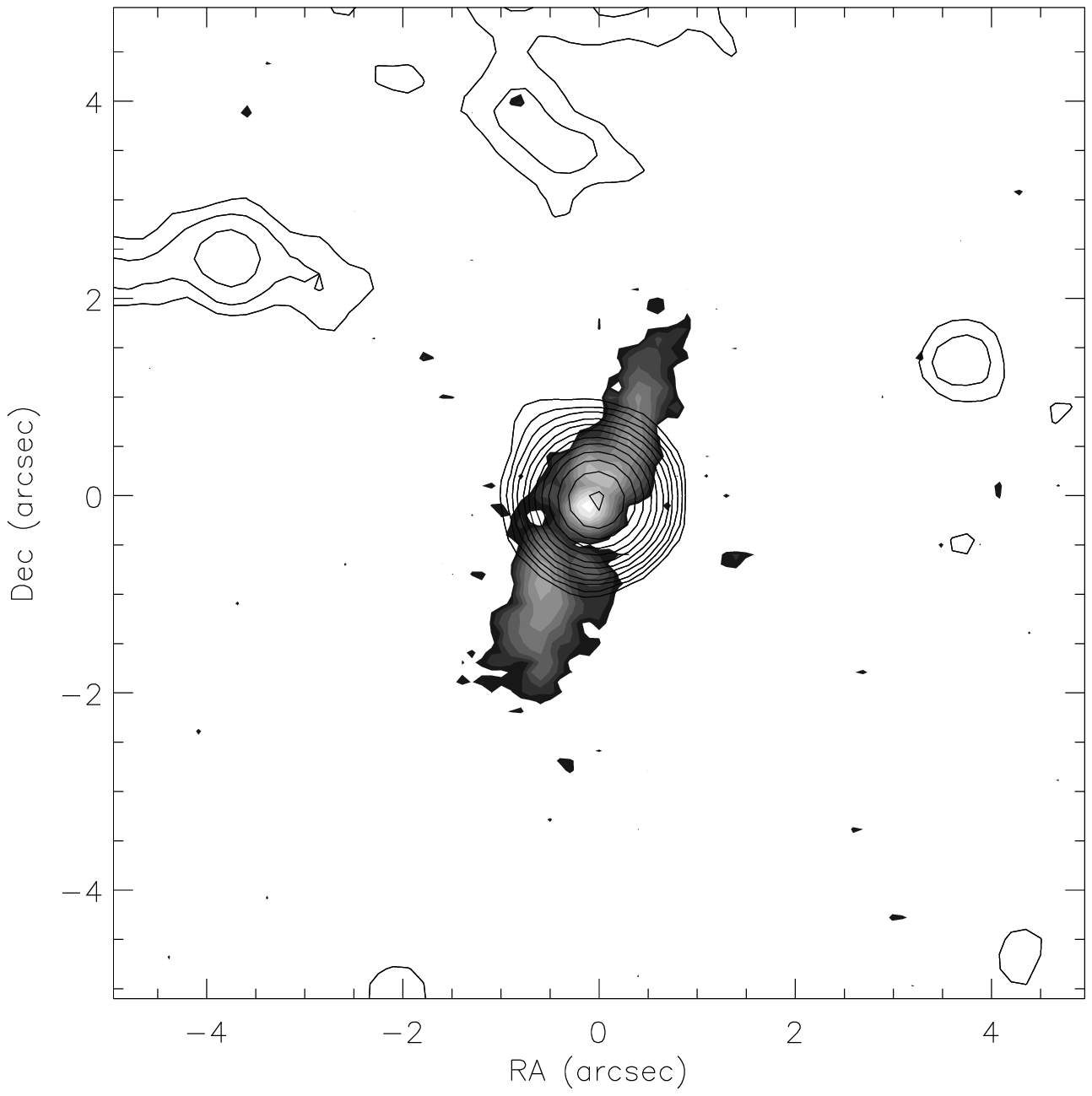}
\hfill
\includegraphics[scale=.5]{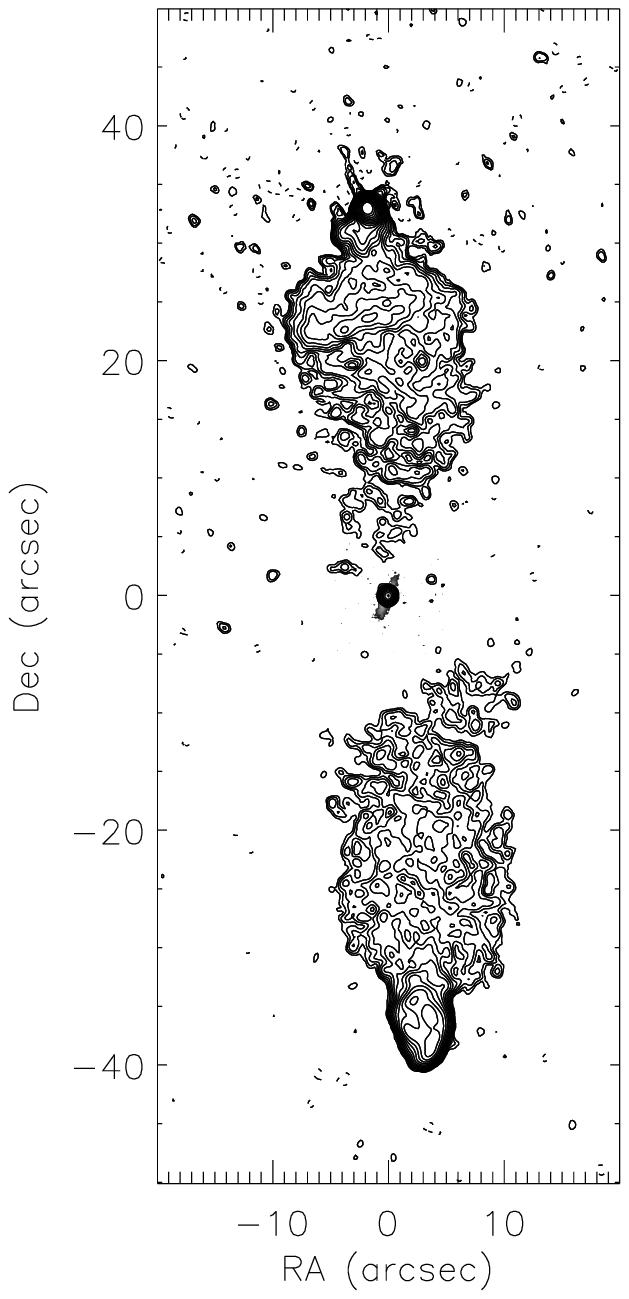}
\caption{3C381 (Radio/Optical Overlay) - Radio is shown in contours. Optical is shown in grey scale. Left: Closeup of the core. Right: View of the overall radio source. Radio levels are (0.102, 0.144, 0.203, 0.288, 0.407, 0.575, 0.814, 1.151, 1.627, 2.301, 3.254, 4.602, 6.509, 9.205, 13.018, 18.410, 26.035, 36.819, 52.070, 73.639) mJy. Optical levels are (87.8, 124.2, 175.7, 248.4, 351.4, 496.9, 702.7, 993.8, 1405.5, 1987.6, 2811.0) * {$10^{-14}$} erg $s^{-1}$ $cm^{-2}$. Both images have the same levels. The radio map was provided by David Floyd (Private Communication).}
\label{3C381-overlay}
\end{figure}

\clearpage

\begin{deluxetable}{lrrrr}
\tablecaption{Radio Observation Parameters}
\tablehead{
\colhead{Name} & \colhead{Date} & \colhead {Frequency} & \colhead{Bandwidth} &
\colhead{RMS Noise} \\
\colhead{} & \colhead{} & \colhead{} & \colhead{} & \colhead{(*10$^{-4}$ Jy/Beam)}}
\startdata
3C124 & 1987-08-17 & 4.885 GHz. & 0.1 GHz & 0.63 \\
3C135 & 1990-05-25 & 8.465 GHz. & 0.025 GHz &  0.50 \\
3C284 & 1985-01-11 & 4.885 GHz. & 0.1 GHz & 5.76 \\
3C303.1 & 1991-08-11 & 8.415 GHz. & 0.1 GHz & 0.78 \\
3C341 & 1996-12-14 & 8.435 GHz. & 0.1 GHz & 6.87 \\
3C368 & 1995-07-31 & 4.535 GHz. & 0.1 GHz & 0.32 \\
3C379.1 & 1987-10-10 & 4.873 GHz. & 0.05 GHz & 0.78 \\
3C382 & 1983-09-20 & 4.885 GHz. & 0.1 GHz & 3.55 \\
\enddata
\label{Table:VLA_obs_params}
\end{deluxetable}

\clearpage

\begin{deluxetable}{crrcccc}
\tablecaption{Optical Observations}
\tablehead{
\colhead{Source} & \colhead{z} &
\colhead{$M_{v}$} & \colhead{WFPC2 Chip} &
\colhead{Emission} & \colhead{Detected?} & \colhead{Continuum}\\
\colhead{} & \colhead{} & \colhead{} & \colhead{} & \colhead{Line} & \colhead{} & \colhead{Exposure (s)}
} 
\startdata
3C6.1&0.840&\nodata&WF2&[OII]&Yes & 300\\
3C17&0.219&-22.137&WF2&[OIII]&Yes & 2x140\\
3C31&0.016&\nodata&WF2&H$\alpha$,[NII]&Yes & 2x140 \\
3C36&1.301&\nodata&WF4&[OII]&Yes & 300\\
3C42&0.395&\nodata&WF3&[OIII]&Yes & 300\\
3C46&0.437&\nodata&WF4&[OIII]&Yes & 300\\
3C49&0.621&\nodata&WF2&[OII]&Marginal Detection & 300\\
3C55&0.734&\nodata&WF3&[OII]&No & 300\\
3C63&0.175&\nodata&PC1 (F588N)&[OIII]&Yes & 2x140\\
3C78&0.028&\nodata&PC1 (F673N)&H$\alpha$,[NII]&Yes & 2x140 \\
3C83.1&0.025&\nodata&PC1 (F673N)&H$\alpha$,[NII]&Yes & 2x140\\
3C84&0.017&\nodata&WF2&H$\alpha$,[NII]&Yes & 2x140\\
3C93&0.358&-23.247&WF2&[OIII]&Yes & 2x140\\
3C93.1&0.244&-21.423&WF3&[OIII]&Yes & 300\\
3C98&0.030&-20.155&PC1&[OIII]&Yes & \nodata\\
3C103&0.330&\nodata&WF2&[OIII]&Yes & 300\\
3C109&0.305&-22.971&WF3&[OIII]&Yes & 2x140\\
3C111&0.048&-18.564&WF3&[OIII]&No & 2x140\\
3C124&1.083&\nodata&WF4&[OII]&Yes & 300\\
3C135&0.127&-21.848&WF2&[OIII]&Yes & 2x140\\
3C136.1&0.064&\nodata&WF4&[OIII]&Yes & 2x140 \\
3C138&0.759&-24.524&PC1 (F656N)&[OII]&No & 2x140\\
3C169.1&0.633&\nodata&WF2&[OII]&Yes & 300\\
3C171&0.238&-21.282&WF2&[OIII]&Yes & 300\\
3C172&0.519&\nodata&PC1&[OIII]&Yes & 300\\
3C175&0.768&-26.795&PC1 (F658N)&[OII]&Yes & 2x140\\
3C184.1&0.118&-21.676&WF2&[OIII]&Yes & 2x140\\
3C192&0.059&-19.980&WF3&[OIII]&No & 2x140\\
3C196.1&0.198&\nodata&WF4&[OIII]&Yes & 2x140\\
3C207&0.684&-24.934&PC1 (F631N)&[OII]&Yes & 2x140\\
3C213.1&0.194&\nodata&WF4&[OIII]&Yes & 300\\
3C219&0.174&-22.158&PC1 (F588N)&[OIII]&Yes & 2x140\\
3C220.1&0.620&\nodata&WF2&[OII]&Yes & 300\\
3C223&0.136&-21.950&WF2&[OIII]&Yes & 2x140\\
3C227&0.086&-21.610&WF4&[OIII]&Yes & 2x140\\
3C234&0.184&-22.463&WF4&[OIII]&Yes & 2x140\\
3C244.1&0.428&-23.187&WF4&[OIII]&Yes & 300\\
3C249.1&0.311&-25.310&PC1 (F656N)&[OIII]&Yes & 2x140\\
3C263&0.646&-26.612&WF2&[OII]&Yes & 2x140\\
3C264&0.020&-20.697&WF2&H$\alpha$,[NII]&Yes & 2x140\\
3C268.1&0.973&\nodata&WF4&[OII]&No & 300\\
3C268.2&0.362&\nodata&WF2&[OIII]&Yes & 300\\
3C268.3&0.371&-21.479&WF2&[OIII]&Yes & 300\\
3C268.4&1.400&-26.602&PC1&[OII]&No & 2x140\\
3C273&0.158&-26.516&WF3&[OIII]&Yes & \nodata\\
3C277.1&0.320&-23.172&PC1&[OIII]&Yes & 2x140\\
3C277.3&0.085&\nodata&WF4&H$\alpha$,[NII]&Yes & 2x140 \\
3C280&0.997&-22.101&WF3&[OII]&No & 300\\
3C284&0.239&\nodata&WF2&[OIII]&Yes & 2x140\\
3C289&0.967&\nodata&WF4&[OII]&No & \nodata\\
3C297&1.406&-24.533&PC1&[OII]&Yes & 300\\
3C299&0.367&\nodata&WF2&[OIII]&Yes & 300\\
3C303.1&0.267&-22.147&WF3&[OIII]&Yes & 300\\
3C305&0.041&-22.521&WF3&[OIII]&Yes & 2x140\\
3C305.1&1.132&\nodata&WF3&[OII]&Yes & 2x140\\
3C321&0.096&-22.194&WF3&[OIII]&Yes & 2x140\\
3C323.1&0.264&-23.929&WF3&[OIII]&Yes & \nodata\\
3C330&0.550&-21.504&WF3&[OII]&Yes & 300\\
3C332&0.151&-23.258&WF3&[OIII]&Yes & 2x140\\
3C338&0.029&\nodata&PC1 (F673N)&H$\alpha$,[NII]&Yes & 2x140 \\
3C341&0.448&\nodata&WF4&[OIII]&Yes & 300\\
3C343.1&0.750&-22.632&WF3&[OII]&Yes & 300\\
3C349&0.205&\nodata&WF2&[OIII]&Yes & 300\\
3C352&0.805&\nodata&PC1 (F673N)&[OII]&Yes & 300\\
3C356&1.079&-22.815&WF4&[OII]&No & 300\\
3C368&1.130&\nodata&WF3&[OII]&Yes & 300\\
3C379.1&0.256&\nodata&PC1 (F631N)&[OIII]&Yes & 2x140\\
3C380&0.691&-26.302&WF3&[OII]&Yes  & 2x140\\
3C381&0.160&-22.157&WF3&[OIII]&Yes & 2x140 \\
3C382&0.058&-21.651&WF3&[OIII]&Yes & 2x140 \\
3C390.3&0.056&-21.582&WF3&[OIII]&Yes & 2x140 \\
3C402&0.025&\nodata&PC1 (F673N)&H$\alpha$,[NII]&No & 2x140 \\
3C433&0.101&-20.512&WF3&[OIII]&Yes & 2x140 \\
3C436&0.215&\nodata&WF2&[OIII]&Yes & 2x140 \\
3C445&0.056&-21.192&WF3&[OIII]&No & 2x140 \\
3C449&0.017&\nodata&WF2&H$\alpha$,[NII]&Yes & 2x140 \\
3C452&0.081&-21.243&WF4&[OIII]&Yes & 2x140 \\
3C454.3&0.860&-27.601&PC1&[OII]&Yes & 2x140 \\
3C458&0.290&\nodata&WF3&[OIII]&Yes & 300\\
3C465&0.029&-22.190&PC1 (F673N)&H$\alpha$,[NII]&Yes & 2x140 \\
\enddata
\label{Table:Optical_obs}
\tablecomments{Col 2 gives the redshift of the source. Col 3 is the V-Band host magnitude from \citet{Veron03}. Col 3 is the WFPC2 chip upon which the source was observed. When using the LRF, the chip on which a source is imaged depends on the filter's central wavelength. If the source was not imaged with the LRF, both the WFPC2 chip used and the narrow filter used are given. Col 4 lists the emission line observed in the narrow-band images. Col 5 states if a source was detected in the narrow-band images.}
\end{deluxetable}

\clearpage

\begin{deluxetable}{crrrrcrrr}
\rotate
\tablecaption{Optical Properties}
\tablehead{
\colhead{Source} & \colhead{$\theta$} &
\colhead{PA} & \colhead{Line Flux} &
\colhead{L$_{H\alpha}$} & \colhead{$\theta_{cone}$} &
\colhead{Morphology} & \colhead{Host PA} &\colhead{Ref} \\
\colhead{} & \colhead{('')} & \colhead{(deg)} & 
\colhead{(x10$^{-15}$ erg s$^{-1}$ cm$^{-2}$)} &
\colhead{(erg/s)} & \colhead{(deg)} & \colhead{} & \colhead{(deg)}}
\startdata
3C6.1&\nodata&\nodata&\nodata&\nodata&\nodata&Partially Res.&\nodata &   \\
3C17&\nodata&41&0.35&41.2&\nodata&Partially Res.&42  & 2   \\
3C31&\nodata&\nodata&1.01&38.4&\nodata&Partially Res.&144 & 4      \\
3C42&\nodata&\nodata&0.06&41.0&\nodata&Partially Res.&153  & 2     \\
3C46&2.13&26&0.29&41.8&17&Extended&177 & 2 \\
3C83.1&\nodata&\nodata&\nodata&\nodata&\nodata&Partially Res.&166 & 4      \\
3C84&15.67&see notes&59.14&40.3&180&Extended&100  & 4 \\
3C93.1&\nodata&65&0.35&41.3&\nodata&Partially Res.&132 & 2  \\
3C93&\nodata&95&0.19&41.4&\nodata&Partially Res.&\nodata &       \\
3C103&0.87&49&0.01&39.9&61&Partially Res.&34 & 2   \\
3C109&2.34&25&1.72&42.2&39&Extended&145 & 1 \\
3C124&1.08&4&0.13&41.8&15&Extended&\nodata  &    \\
3C135&3.71&54&1.05&41.1&35&Extended&141 & 2 \\
3C136.1&0.53&175&0.00013&36.6&35&Partially Res.&117 & 4     \\
3C169.1&\nodata&\nodata&0.33&41.7&\nodata&Partially Res.&\nodata &       \\
3C171&3.44&85&1.38&41.8&23&Extended&165 & 1 \\
3C172&0.73&16&0.07&41.4&77&Partially Res.&\nodata &       \\
3C184.1&2.64&24&0.01&38.7&\nodata&Partially Res.&40 & 2    \\
3C196.1&0.67&53&0.06&40.3&15&Partially Res.&57  & 1 \\
3C213.1&\nodata&\nodata&0.08&40.4&\nodata&Partially Res.&161 & 2   \\
3C220.1&\nodata&\nodata&0.06&40.9&\nodata&Partially Res.&\nodata &        \\
3C223&2.1&128&2.29&41.5&43&Extended&50  & 1 \\
3C234&2.37&79&9.85&42.4&32&Extended&80  & 2 \\
3C244.1&3.47&133&0.66&42.1&25&Partially Res.&76 & 2 \\
3C249.1&2.08&80&2.27&42.3&112&Extended&\nodata  & \\
3C264&\nodata&\nodata&0.49&38.3&\nodata&Partially Res.&152 & 1     \\
3C268.2&1.8&15&0.44&41.8&56&Extended&164   & 2     \\
3C268.3&3.2&146&0.42&41.8&26&Extended&144  & 2    \\
3C277.1&1.85&124&1.4&42.1&52&Partially Res.&126  & 3 \\
3C277.3&\nodata&\nodata&0.21&39.2&0&Partially Res.&170 & 4  \\
3C284&2.9&74&0.56&41.4&30&Extended&151 & 2  \\
3C299&4.24&62&1.19&42.2&70&Extended&47 & 2  \\
3C303.1&1.56&140&0.95&41.8&38&Extended&169 &2     \\
3C305&6.37&46&13.06&41.2&35&Extended&74   & 1 \\
3C305.1&\nodata&33&0.05&41.4&0&Partially Res.&24 & 3       \\
3C321&6.21&108&5.8&41.6&42&Extended&18  & 1 \\
3C330&1&103&0.06&40.8&47&Partially Res.&\nodata & \\
3C341&1.03&172&0.27&41.8&55&Extended&17 & 2\\
3C349&\nodata&36&0.16&40.7&61&Partially Res.&14 & 2 \\
3C352&\nodata&8&0.05&41.1&0&Partially Res.&\nodata &     \\
3C368&6.74&24&0.59&42.5&24&Extended&\nodata   &  \\
3C379.1&0.72&43&0.06&40.6&15&Extended&\nodata &  \\
3C380&\nodata&\nodata&0.56&42.0&0&Partially Res.&131 &3   \\
3C381&5.69&150&2.35&41.7&31&Extended&156    & 2    \\
3C382&2.14&112&13.16&41.5&49&Partially Res.&85 & 4 \\
3C390.3&0&62&11.34&41.4&0&Partially Res.&97   & 1  \\
3C433&5.87&135&1.49&41.1&55&Partially Res.&147 & 1 \\
3C436&\nodata&\nodata&0.26&41.0&0&Partially Res.&3  & 2     \\
3C449&1.7&174&0.37&38.0&44&Partially Res.&1   & 4  \\
3C452&1.58&125&0.71&40.5&30&Partially Res.&101  & 4 \\
3C454.3&\nodata&\nodata&0.86&42.2&\nodata&Partially Res.&\nodata &       \\
3C458&\nodata&75&0.07&40.7&\nodata&Partially Res.&\nodata   &    \\

\enddata
\label{Table:Optical_data}
\tablecomments{Col 2 lists the measured angular size in arcsec for the continuum-subtracted emission line image. Col 3 is the position angle (east of north) for the ELR. Col 4 lists the emission line flux. Col 5 is the log of the equivalent H$\alpha$ luminosity calculated from Col 4 and conversion factors derived from \citet{Koski78}. Col 6 lists a cone angle in which the line-emission was contained. Col 7 specifies the morphological category. Col 8 is the host galaxy position angle. Col 9 gives the reference for the host galaxy position angle. Reference key: 1 - \citet{Baum89a}. 2 - \citet{DeKoff96}. 3 - \citet{deVries97}. 4 - {\citet{Martel98}.}}
\end{deluxetable}

\clearpage

\begin{deluxetable}{ccrcrcr}
\tablecaption{Radio Properties}
\tablehead{
\colhead{Source} & \colhead{Type} &
\colhead{Angular Size} & 
\colhead{PA} & \colhead{Ref} &
\colhead{Log Luminosity} \\
\colhead{} & \colhead{} & \colhead{('')} &\colhead{(deg)} &
\colhead{} & \colhead{(erg/s/Hz)}}
\startdata
3C6.1   &FR II  &25.8   &26     &13      &35.7      \\
3C17    &FR II  &30.0   &147    &5      &34.4      \\
3C31    &FR I   &1833.0 &159    &11      &32.0     \\
3C36    &FR II  &9.0    &20     &13      &35.9      \\
3C42    &FR II  &28.0   &132    &5      &34.8      \\
3C46    &FR II  &163.0  &68     &5      &34.8      \\
3C49    &CSS    &0.98   &84     &6      &35.2 \\
3C55    &FR II  &69.0   &94     &13      &35.7       \\
3C63    &FR II  &22.0   &34     &5      &34.2       \\
3C78    &FR I   &210.0  &51     &11      &32.5       \\
3C83.1  &FR I   &680.0  &96     &11      &32.6     \\
3C84    &FR I   &492.0  &162    &11      &32.6   \\
3C93    &FR II  &34.7   &44     &17,99     &34.8      \\
3C93.1  &CSS    &0.3    &165    &4,5      &34.2   \\
3C103   &FR II  &88     &159    &5,8      &35.0  \\
3C109   &FR II  &96.0   &143    &5      &34.8     \\
3C111   &FR II  &220.0  &62     &11      &33.5     \\
3C124   &CSS    &1.3    &7      &13      &35.8  \\
3C135   &FR II  &130.0  &70     &5      &33.9     \\
3C136.1 &FR II  &522.0  &108    &11      &33.1  \\
3C138   &CSS    &0.6    &70     &1        &35.8  \\
3C169.1 &FR II  &38.0   &137    &8      &35.1   \\
3C171   &FR II (dtsb)   &30.0   &99     &2      &34.5       \\
3C172   &FR II  &103.0  &37     &13      &35.2     \\
3C175   &FR II  &48.0   &55     &10,99   &35.7        \\
3C184.1 &FR II  &167.0  &157    &3      &33.7  \\
3C192   &FR II  &192.0  &124    &3      &33.2    \\
3C196.1 &CSS    &4.0    &43     &5      &34.3 \\
3C207   &FR II  &11.4   &90     &14        &35.4        \\
3C213.1 &FR II  &43.0   &162    &5      &33.8   \\
3C219   &FR II  &184.0  &40     &5      &34.5     \\
3C220.1 &FR II  &29.7   &79     &13      &35.4    \\
3C223   &FR II  &300.0  &164    &5      &33.8    \\
3C227   &FR II  &246.0  &86     &11      &33.7     \\
3C234   &FR II  &110.0  &64     &5      &34.5     \\
3C244.1 &FR II  &52.0   &168    &5      &35.1   \\
3C249.1 &FR II  &12.7   &54     &99     &34.5   \\
3C263   &FR II  &43.1   &120    &19       &35.4        \\
3C264   &FR I   &91.0   &40     &11      &32.4       \\
3C268.1 &FR II  &46.0   &83     &13      &36.0    \\
3C268.2 &FR II  &96.0   &21     &5      &34.6    \\
3C268.3 &CSS    &1.0    &161    &5      &34.7      \\
3C268.4 &FR II  &10.2   &42     &2,18   &36.1 \\
3C277.1 &CSS    &1.5    &131    &6,99   &34.4     \\
3C277.3 &FR II  &29.0   &158    &11      &33.2   \\
3C280   &FR II  &12.9   &82     &1,10    &36.1        \\
3C284   &FR II  &176.0  &101    &5      &34.3    \\
3C297   &\nodata\tablenotemark{1}&4.0    &167    &13      &36.1    \\
3C299   &FR II  &12.0   &60     &5      &34.7      \\ 
3C303.1 &CSS    &1.9    &130    &6      &34.2      \\
3C305   &CSS    &14.0   &44     &10      &32.8       \\
3C305.1 &FR II  &7.9    &11     &13      &35.5     \\
3C321   &FR II  &309.0  &131    &3      &33.5    \\
3C330   &FR II  &62.0   &62     &13      &35.5      \\
3C332   &FR II  &81.0   &20     &5      &33.8      \\
3C338   &FR I   &115.0  &85     &11      &32.9      \\
3C341   &FR II  &71.0   &50     &5      &34.9      \\
3C343.1 &CSS    &1.0    &97     &13      &35.5       \\
3C349   &FR II  &82.0   &142    &5      &34.2     \\
3C352   &FR II  &10.2   &164    &13      &35.5     \\
3C356   &FR II  &72.0   &162    &13      &35.9     \\
3C368   &FR II  &8.5    &17     &7,13    &36.0    \\
3C379.1 &FR II  &76.0   &161    &5      &34.2   \\ 
3C380   &CSS    &1.0    &145    &2,9    &36.1      \\
3C381   &FR II  &69.0   &4      &5      &34.1       \\
3C382   &FR II  &179.0  &50     &11      &33.2     \\
3C390.3 &FR II  &231.0  &145    &3      &33.5  \\
3C402   &FR I   &528.0  &163    &11      &32.1     \\
3C433   &FR II  &58.0   &172    &5      &34.2     \\
3C436   &FR II  &105.0  &172    &5,10    &34.4      \\
3C445   &FR II  &576.0  &171    &11      &33.3    \\
3C449   &FR I   &1742.0 &10     &11      &31.9     \\
3C452   &FR II  &277.0  &79     &11      &33.9     \\
3C454.3 &CSS    &10.0   &128    &2,16     &35.7   \\
3C458   &FR II  &161    &75     &12,15        &34.6        \\
3C465   &FR I   &375.0  &125    &11      &32.9  \\

\tablenotetext{1}{We were unable to find good maps or a FR classification for this source.}

\enddata
\label{Table:Radio_data}
\tablecomments{Table of radio data from the literature and our maps. Note that the 4 sources eliminated from our sample due to the lack of a continuum image (see $\S$ \ref{sec:Observations}) are not included in this table. Col 1 is the 3CR source name. Col 2 is the source classification, given as Fanaroff-Riley Type \citep{Fanaroff74}, Compact Steep Spectrum (CSS) source (from \citet{deVries97}), or other morphology. Col 3 gives the angular size in arcseconds. Col 4 is the position angle of the radio source, generally measured from hotspot to hotspot. Col 5 is the reference for the angular size. Reference key: 1 - \citet{Akujor95}. 2 - \citet{Allington84}. 3 - \citet{Baum89a}. 4 \citet{Dallacasa95}. 5 - \citet{DeKoff96}. 6 - \citet{deVries97}. 7 - \citet{Dunlop93}. 8 - \citet{Leahy86}. 9 - \citet{Lister05}. 10 - \citet{Mackay69}. 11 - \citet{Martel99}. 12 - \citet{McCarthy95}. 13 - \citet{McCarthy97}. 14 - \citet{Mullin06}. 15 - \citet{Nilsson93}. 16 - \citet{Perley82}. 17 - \citet{Price93}. 18 - \citet{Reid95}. 19 - \citet{Swarup84}. 99 - this paper. Col 6 is the Logarithm of the Radio Luminosity, calculated at 178 MHz.}
\end{deluxetable}

\clearpage

\begin{deluxetable}{lccccccc}
\rotate
\tablecaption{Statistical Correlations.}
\tablehead{
\colhead{Relationship} & \colhead{Kendall's Tau} & \colhead{P$_{tau}$} & 
\colhead{Spearman's rho} & \colhead{P$_{rho}$}}
\startdata
Emission Line Luminosity vs. Delta Pa&-0.6577&0.040&-0.474&0.0044 \\
Emission Line Luminosity vs. Delta Pa ($z < 0.6$)&-0.5417&0.0258&-0.404&0.0222 \\
Emission Line Luminosity vs Delta pa ($0.1 < z < 0.6$)& -0.6338 & 0.0219 & \nodata & \nodata \\
Emission Line Luminosity vs. Host Mag.&-0.8734&0.0003&-0.611&0.0005 \\
Emission Line Luminosity vs. Radio Luminosity&1.095&0.0000&0.740&0.0000 \\
Emission Line Nebulae Size vs. Radio Luminosity&0.7177&0.0039&0.487&0.0067 \\
Emission Line Nebulae Size vs. Radio Luminosity ($z < 0.6$)&0.6483&0.0119&0.442&0.0172 \\
Emission Line Nebulae Size vs. z&0.6976&0.0050&0.499&0.0055 \\
Emission Line Nebulae Size vs. z ($z < 0.6$)&0.6253&0.0152&0.458&0.0136 \\
Delta PA vs. Radio Size&0.6050& 0.0105&0.455&0.0080 \\
Delta PA vs. Size Ratio&1.0099&0.0001&0.684&0.0003 \\
\enddata
\tablecomments{The probabilities given are the probabilities that a correlation is not present.}
\label{Table:Correlations}
\end{deluxetable}

\clearpage

\begin{deluxetable}{lccccc}
\rotate
\tablecaption{Statistical Non-Correlations.}
\tablehead{\colhead{Relationship} & \colhead{Kendall's Tau} & \colhead{P$_{tau}$} 
& \colhead{Spearman's rho} & \colhead{P$_{rho}$}}
\startdata
Delta PA vs. z&-0.3874&0.0913&-0.238&0.1540 \\
Delta PA vs. Emission Line Nebulae Asymmetry&-0.3126&0.2243&-0.222&0.2327 \\
Emission Line Nebulae Luminosity vs Radio Size&-0.3624&0.0423&-0.280&0.0330\\
Emission Line Nebulae Size vs Radio Size&-0.3484&0.1686&-0.283&0.1216 \\
\enddata
\tablecomments{The probabilities given are the probabilities that a correlation is not present.}
\label{Table:Non-Correlations}
\end{deluxetable}

\clearpage

\begin{deluxetable}{lccc}
\tablecaption{Relative Alignment Statistical Results}
\tablehead{\colhead{Delta PA Category} & \colhead{Data} & \colhead{Max ABS Discrepancy} & \colhead{Significance} \\ & \colhead{Points} & & \emph{(\%)}}

\startdata
This Paper (HST), $z<0.1$ & 7 & 0.169 & 98.8\\
This Paper (HST), $0.3<z<0.6$ & 13 & 0.157 & 90.5\\
This Paper (HST), $0.1<z<0.6$ & 26 & 0.152 & 58.8 \\
\citet{McCarthy95}, $z<0.1$ & 15 & 0.171 & 77.0\\
\citet{McCarthy95}, $0.3<z<0.6$ & 12 & 0.44 & 1.90 \\
\citet{McCarthy95}, $0.1<z<0.6$ & 30 & 0.343 & 0.168 \\
\enddata
\tablecomments{KS Test results comparing the distribution of Delta PAs (see Figure \ref{all_delta_pas}) with that of a uniform distribution. Col. 3 indicates the likliehood that the two distributions are the same. Note that the KS test works best for at least 40 points. Our study has fewer points and the KS test results may be unreliable for some categories.}
\label{Table:KS-results}
\end{deluxetable}

\clearpage

\appendix
\section{Individual Source Notes}
\label{Appendix}

3C46 (Figs. \ref{3C46-montage} \& \ref{3C46-overlay}) is a large FR II radio source. The emission line nebula and the radio source are not well aligned. The optical continuum image seems to show a double structure. However, the line emission is predominantly in the southern portion.

3C49 (Figs. \ref{3C49-montage} \& \ref{3C49-overlay}) was a marginal detection. A smoothed image was used for the overlay and the lowest contour is 2-$\sigma$.  

3C84 (Figs. \ref{3C84-montage} \& \ref{3C84-overlay}) has a diffraction cross in the emission line image. A sufficiently deep 5 or 8 GHz radio map was not available and could not be found in archival data, so an overlay is not shown. The line emission is symmetrically distributed within the host galaxy so there is no meaningful position angle. 

3C98 (Fig \ref{3C98}) did not have a continuum image so only the narrow band image is presented here.

3C109 (Figs \ref{3C109-montage} \& \ref{3C109-overlay}) has an emission line region which is not aligned with the radio source. The ELR is in a disk which is perpendicular to the radio jet.

3C124 (Figs \ref{3C124-montage} \& \ref{3C124-overlay}) has an emission line region which is very closely aligned with the radio source. The exact center of the host galaxy is not well determined for our images.

3C135 (Figs \ref{3C135-montage} \& \ref{3C135-overlay}) features an emission line region which is closely aligned with the overall radio source. The radio source is very large compared to the emission line region. The emission line region is elongated along the radio jet axis and features a disconnected region of emission towards the Northwest hotspot.

3C171 (Figs \ref{3C171-montage} \& \ref{3C171-overlay}) The emission line region traces the jets, linking the core to the hotspots. There is emission around the east hotspot. The ELR structure has been discussed in detail in \citet{Tilak05}.

3C223 (Figs. \ref{3C223-montage} \& \ref{3C223-overlay}) does not have an aligned emission line nebula. However, the core shows slight extension in the direction of the emission line nebulae.

3C234 (Figs. \ref{3C234-montage} \& \ref{3C234-overlay}) is aligned with the radio source and the core shows slight extension along the major axis of the nebula. There are also filaments extending almost perpendicular to the radio axis. The broad lines in 3C234 are highly polarized so it is not surprising the properties of this source are similar to that of NLRGs \citep{Antonucci84}.

3C249.1 (Figs \ref{3C249.1-montage} \& \ref{3C249.1-overlay}) has an ELR which does not have a preferential alignment with the radio source. It features two regions of emission which are disconnected from the nuclear emission at the radio core.

3C268.2 (Figs. \ref{3C268.2-montage} \& \ref{3C268.2-overlay}) shows an emission line nebula which is closely aligned with the radio source. The ELR is asymmetrical around the center of the host galaxy (measured from the broadband continuum image). There is little to no line emission detected from the nucleus in our images. The core was not detected in the radio maps, however the location was inferred using the WCS coordinates given on NED. As the core was not detected in the radio map, we are unable to examine spatial coincidence of the ELR and the radio core, but are still able to measure the alignment of the ELR and radio source. 

3C268.3 (Figs. \ref{3C268.3-montage} \& \ref{3C268.3-overlay}) is an aligned CSS source \citep{deVries99}. The emission extends beyond the radio source. Registration of the radio core and the emission line image was problematic due to the lower resolution of the optical image. 3C268.3 shows broad lines, indicating this source may not be in the plane of the sky.

3C273 (Fig \ref{3C273}) did not have a continuum image, however the narrow band image is presented here.

3C284 (Figs. \ref{3C284-montage} \& \ref{3C284-overlay}) has and ELR which is composed of three separate components in our images. The PA along the overall structure is not aligned with the source. However, taking an ``inner PA'' of the central structure gives the alignment with the radio source, so we consider this ELR aligned with the radio source.

3C299 (Figs. \ref{3C299-montage} \& \ref{3C299-overlay}) has aligned emission-line and radio morphology. Both the emission line region and the radio source are asymmetric about the center of the host galaxy.

3C303.1 (Figs. \ref{3C303.1-montage} \& \ref{3C303.1-overlay}) has an emission line nebula which is of similar size in our WFPC images as the radio source. They are closely aligned. The emission line region is symmetrical about the nucleus and is elongated along the radio axis.

3C305 (Figs. \ref{3C305-montage} \& \ref{3C305-overlay}) The emission line region is aligned with the jets and hotspots. However, the emission extends beyond the extent of the radio source, limiting the probability of shock ionization as a primary mechanism for transferring energy to ambient gas \citep{Jackson95}.

3C321 (Figs. \ref{3C321-montage} \& \ref{3C321-overlay}) Due to the dust lane (noted in \citet{Martel98}, determining the proper center of the host galaxy is problematic. The ELR is aligned with the radio source.

3C323.1 (Fig \ref{3C323.1}) did not have a continuum image for subtraction, but the LRF image is presented here.

3C341 (Figs. \ref{3C341-montage} \& \ref{3C341-overlay}) does not show alignment between the emission line nebula and the radio source. The size of the ELR is comparable to a resolution element in the radio map, making examination of spatial coincidence problematic.

3C368 (Figs. \ref{3C368-montage} \& \ref{3C368-overlay}) shows close alignment between the ELR and the radio source.

3C379.1 (Figs. \ref{3C379.1-montage} \& \ref{3C379.1-overlay}) shows an ELR which is not aligned with the radio source. The ELR is smaller than the resolution of our radio image.

3C381 (Figs. \ref{3C381-montage} \& \ref{3C381-overlay}) does not show alignment between the emission line nebula and the radio source.


\end{document}